\documentclass[reqno,a4paper]{amsart}
\usepackage{times}
\usepackage{mathrsfs}
\usepackage{latexsym,amsopn,amssymb,amsmath}
\usepackage{bm} 
\usepackage{amsthm}
\usepackage{amsfonts,amsbsy,amscd,stmaryrd}
\usepackage{paralist}
\usepackage{geometry}
\oddsidemargin 
\evensidemargin 
\newcommand{\mbc}{\mathbb{C}}

\newcommand{\mbr}{\mathbb{R}}

\newcommand{\mbz}{\mathbb{Z}}
\newcommand{\ca}{\mathcal{A}}
\newcommand{\cb}{\mathcal{B}}
\newcommand{\cc}{\mathcal{C}}

\newcommand{\ce}{\mathcal{E}}
\newcommand{\cf}{\mathcal{F}}
\newcommand{\cg}{\mathcal{G}}
\newcommand{\ch}{\mathcal{H}}
\newcommand{\ci}{\mathcal{I}}
\newcommand{\cj}{\mathcal{J}}
\newcommand{\ck}{\mathcal{K}}

\newcommand{\cp}{\mathcal{P}}

\newcommand{\cs}{\mathcal{S}}
\newcommand{\ct}{\mathcal{T}}

\newcommand{\cv}{\mathcal{V}}
\newcommand{\cw}{\mathcal{W}}
\newcommand{\cx}{\mathcal{X}}
\def\rond{\mathscr}
\newcommand{\ra}{\rond{A}}
\newcommand{\rb}{\rond{B}}
\newcommand{\rc}{\rond{C}}
\newcommand{\rd}{\rond{D}}
\newcommand{\re}{\rond{E}}

\newcommand{\ri}{\rond{I}}

\newcommand{\rk}{\rond{K}}
\newcommand{\rl}{\rond{L}}
\newcommand{\rM}{\rond{M}}
\newcommand{\mr}{\rond{M}}
\newcommand{\rn}{\rond{N}}

\newcommand{\rp}{\rond{P}}
\newcommand{\rr}{\rond{R}}
\newcommand{\rs}{\rond{S}}
\newcommand{\rt}{\rond{T}}

\def\rmb{\mathrm{b}}
\def\rmc{\mathrm{c}}
\def\rmd{\text{d}}
\def\rme{\mathrm{e}}
\def\rmo{\mathrm{o}}
\def\rmu{\mathrm{u}}

\newcommand{\Cbu}{\cc_{\rmb}^{\rmu}}

\newcommand{\Co}{\cc_{\rmo}}
\newcommand{\Cc}{\cc_{\rmc}}
\renewcommand{\proof}{\noindent{\bf Proof: }}

\def\cchi{\raisebox{.45 ex}{$\chi$}} 

\def\bra#1{\langle{#1}|}
\def\ket#1{|{#1}\rangle}
\def\braket#1#2{\langle{#1}|{#2}\rangle}

\def\jap#1{\langle {#1} \rangle}
\def\spe{\mathrm{Sp_{ess}}}
\def\sp{\mathrm{Sp}}

\def\rarrow{\rightarrow}

\def\what{\widehat}
\def\what#1{\widehat{ #1\,}}
\def\wtilde{\widetilde}

\def\nin{\notin}

\def\supp{\mbox{\rm supp\! }}

\def\nin{\notin}
\def\pprod{\textstyle \prod}

\def\ccup{\textstyle{\bigcup}}
\def\ccap{\textstyle{\bigcap}}

\def\qed{\hfill \raisebox{0.5ex}{\framebox[1.6ex]{
                                       \rule[0ex]{0ex}{0.3ex} }}}

\def\build#1_#2^#3{\mathrel{\mathop{\kern 0pt#1}\limits_{#2}^{#3}}}
\newcounter{PAR}[section]

\newtheorem{theorem}{Theorem}[section]
\newtheorem{lemma}[theorem]{Lemma}
\newtheorem{proposition}[theorem]{Proposition}
\newtheorem{corollary}[theorem]{Corollary}

\newtheorem*{acknowledgement}{Acknowledgement}
{\theoremstyle{remark}
\newtheorem{definition}[theorem]{\bf Definition}
\newtheorem{remark}[theorem]{\bf Remark}

\newtheorem{example}[theorem]{\bf Example}

}
\long\def\symbolfootnote[#1]#2{\begingroup%
\def\thefootnote{\fnsymbol{footnote}}\footnote[#1]{#2}\endgroup} 
\begin{document}

\title{On the spectral analysis of many-body systems} 

\author{Mondher Damak}
\address[Mondher Damak]{University of Sfax 3029 Sfax, Tunisia}
   \email{mondher.damak@fss.rnu.tn}

\author{Vladimir Georgescu} 
\address[Vladimir Georgescu]{CNRS and 
  University of Cergy-Pontoise 95000 Cergy-Pontoise, France}
\email{vlad@math.cnrs.fr} 

\date{\today}

\begin{abstract}
\noindent
We describe the essential spectrum and prove the Mourre estimate for
quantum particle systems interacting through $k$-body forces and
creation-annihilation processes which do not preserve the number of
particles. For this we compute the ``Hamiltonian algebra'' of the
system, i.e. the $C^*$-algebra $\rc$ generated by the Hamiltonians
we want to study, and show that, as in the $N$-body case, it is
graded by a semilattice. Hilbert $C^*$-modules graded by
semilattices are involved in the construction of $\rc$.
For example, if we start with an $N$-body system whose Hamiltonian
algebra is $\rc_N$ and then we add field type couplings between
subsystems, then the many-body Hamiltonian algebra $\rc$ is the
imprimitivity algebra of a graded Hilbert $\rc_N$-module.
\end{abstract}

\maketitle 

{\fontsize{10pt}{10pt}
\selectfont
\parskip=1pt
\setcounter{tocdepth}{1}
\tableofcontents}

\section{Introduction}
\label{s:intro}
\protect\setcounter{equation}{0}

\subsection{}
The quantum systems studied in this paper are obtained by coupling a
certain number (finite or infinite) of $N$-body systems. A
(standard) $N$-body system consists of a fixed number $N$ of
particles which interact through $k$-body forces which preserve $N$
(arbitrary $1\leq k \leq N$). The many-body type interactions
include forces which allow the system to make transitions between
states with different numbers of particles. These transitions are
realized by creation-annihilation processes as in quantum field
theory.

The Hamiltonians we want to analyze are rather complex objects and
standard Hilbert space techniques seem to us inefficient in this
situation. Our approach is based on the observation that the
$C^*$-algebra $\rc$ generated by a class of physically interesting
Hamiltonians often has a quite simple structure which allows one to
describe its quotient with respect to the ideal of compact operators
in rather explicit terms \cite{GI1,GI2}.  From this one can deduce
certain important spectral properties of the Hamiltonians.  We refer
to $\rc$ as the \emph{Hamiltonian algebra} (or $C^*$-algebra of
Hamiltonians) of the system. 

The main difficulty in this algebraic approach is to isolate the
correct $C^*$-algebra. This is especially problematic in the present
situations since it is not a priori clear how to define the
couplings between the various $N$-body systems but in very special
situations.  It is rather remarkable that the $C^*$-algebra
generated by a small class of elementary and natural Hamiltonians
will finally prove to be a fruitful choice.  These elementary
Hamiltonians are analogs of the Pauli-Fierz Hamiltonians.

The purpose of the preliminary Section \ref{s:euclid} is to present
this approach in the simplest but physically important case when the
configuration spaces of the $N$-body systems are Euclidean
spaces. We start with a fundamental example, the standard $N$-body
case. Then we describe the many-body formalism in the Euclidean case
and we state our main results on the spectral analysis of the
corresponding Hamiltonians.

There is one substantial simplification in the Euclidean case: each
subspace has a canonical supplement, the subspace orthogonal to
it. This plays a role in the way we present the framework in Section
\ref{s:euclid}.  However, the main constructions and results do not
depend on the existence of a supplement but to see this requires
more sophisticated tools from the theory of crossed product
$C^*$-algebras and Hilbert $C^*$-modules which are not apparent in
this introductory part.  In the rest of the paper we consider
many-body type couplings of systems whose configuration space is an
arbitrary abelian locally compact group.  One of the simplest
nontrivial physically interesting cases covered by this framework is
that when the configuration spaces of the $N$-body systems are
discrete groups, e.g.  discretizations $\mbz^D$ of $\mbr^D$.

\subsection{}
We summarize now the content of the paper.  Section \ref{s:euclid}
starts with a short presentation of the standard \mbox{$N$-body}
formalism, the rest of the section being devoted to a rather
detailed description of our framework and main results in the case
when the configuration spaces of the $N$-body subsystems are
Euclidean spaces. These results are proven in a more general and
natural setting in the rest of the paper.  In Section \ref{s:grad}
we recall some facts concerning $C^*$-algebras graded by a
semilattice $\cs$ (we take here into account the results of Athina
Mageira's thesis \cite{Ma}) and then we present some results on
$\cs$-graded Hilbert $C^*$-modules. This notion, due to Georges
Skandalis \cite{Sk}, proved to be very natural and useful in our
context: thanks to it many results can be expressed in a simple and
systematic way thus giving a new and interesting perspective to the
subject (this is discussed in more detail in \cite{DG4}). The heart
of the paper is Section \ref{s:grass}, where we define the many-body
Hamiltonian algebra $\rc$ in a general setting and prove that it is
naturally graded by a certain semilattice $\cs$. In Section
\ref{s:id} we give alternative descriptions of the components of
$\rc$ which are important for the affiliation criteria presented in
Section \ref{s:af}, where we point out a large class of self-adjoint
operators affiliated to the many-body algebra.  The $\cs$-graded
structure of $\rc$ gives then an HVZ type description of the
essential spectrum for all these operators. The main result of
Section \ref{s:mou} is the proof of the Mourre estimate for
nonrelativistic many-body Hamiltonians. Finally, an Appendix is
devoted to the question of generation of some classes of
$C^*$-algebras by "elementary" Hamiltonians.

\subsection{Notations} 
\label{ss:inotations}
We recall some notations and terminology.  If $\ce,\cf$ are normed
spaces then $L(\ce,\cf)$ is the space of bounded operators
$\ce\rarrow\cf$ and $K(\ce,\cf)$ the subspace consisting of compact
operators. If $\cg$ is a third normed space and $(e,f)\mapsto ef$ is
a bilinear map $\ce\times\cf\to\cg$ then $\ce\cf$ is the linear
subspace of $\cg$ generated by the elements $ef$ with
$e\in\ce,f\in\cf$ and $\ce\cdot\cf$ is its closure. If $\ce=\cf$
then we set $\ce^2=\ce\cdot\ce$.  Two unusual abbreviations are
convenient: by \emph{lspan} and \emph{clspan} we mean ``linear
span'' and ``closed linear span'' respectively.  If $\ca_i$ are
subspaces of a normed space then $\sum^\rmc_i\ca_i$ is the clspan of
$\cup_i\ca_i$.  If $X$ is a locally compact topological space then
$\Co(X)$ is the space of continuous complex functions which tend to
zero at infinity and $\Cc(X)$ the subspace of functions with compact
support.

By \emph{ideal} in a $C^*$-algebra we mean a closed self-adjoint
ideal. A $*$-homomorphism between two \mbox{$C^*$-algebras} will be
called \emph{morphism}.  We write $\ra\simeq\rb$ if the
$C^*$-algebras $\ra,\rb$ are isomorphic and $\ra\cong\rb$ if they
are canonically isomorphic (the isomorphism should be clear from the
context).

A self-adjoint operator $H$ on a Hilbert space $\ch$ is
\emph{affiliated} to a $C^*$-algebra $\ra$ of operators on $\ch$ if
$(H+i)^{-1}\in\ra$; then $\varphi(H)\in\ra$ for all
$\varphi\in\Co(\mbr)$. If $\ra$ is the closed linear span of the
elements $\varphi(H)A$ with $\varphi\in\Co(\mbr)$ and $A\in\ra$, we
say that $H$ is \emph{strictly affiliated to $\ra$}.  The
$C^*$-algebra generated by a set $\re$ of self-adjoint operators is
the smallest $C^*$-algebra such that each $H\in\re$ is affiliated to
it.

We now recall the definition of $\cs$-graded $C^*$-algebras
following \cite{Ma2}. Here $\cs$ is a \emph{semilattice}, i.e. a set
equipped with an order relation $\leq$ such that the lower bound
$\sigma\wedge\tau$ of each couple of elements $\sigma,\tau$ exists.
We say that a subset $\ct$ of $\cs$ is a \emph{sub-semilattice} of
$\cs$ if $\sigma,\tau\in\ct\Rightarrow\sigma\wedge\tau\in\ct$.  The
set $\rs$ of all closed subgroups of a locally compact abelian group
is a semilattice for the order relation given by set inclusion.  The
semilattices which are of main interest for us are (inductive limits
of) sub-semilattices of $\rs$.

A $C^*$-algebra $\ra$ is called \emph{$\cs$-graded} if a linearly
independent family $\{\ra(\sigma)\}_{\sigma\in\cs}$ of
$C^*$-subalgebras of $\ra$ has been given such that
$\sum^\rmc_{\sigma\in\cs}\ra(\sigma)=\ra$ and
$\ra(\sigma)\ra(\tau)\subset\ra(\sigma\wedge\tau)$ for all
$\sigma,\tau$. The algebras $\ra(\sigma)$ are the \emph{components
  of $\ra$}. It is useful to note that some of the algebras
$\ra(\sigma)$ could be zero. If $\ct$ is a sub-semilattice of
$\cs$ and $\ra(\sigma)=\{0\}$ for $\sigma\nin\ct$ we say that $\ra$
is \emph{supported} by $\ct$; then $\ra$ is in fact
$\ct$-graded. Reciprocally, any $\ct$-graded $C^*$-algebra becomes
$\cs$-graded if we set $\ra(\sigma)=\{0\}$ for $\sigma\nin\ct$.

\subsection{Note}
The preprint \cite{DG4} is a preliminary version of this paper. We
decided to change the title because the differences between the two
versions are rather important: the preliminaries concerning the
theory of Hilbert $C^*$-modules and the role of the imprimitivity
algebra of a Hilbert $C^*$-module in the spectral analysis of
many-body systems are now reduced to a minimum; on the other hand,
the Euclidean case and the spectral theory of the corresponding
Hamiltonians are treated in more detail.

\vspace{3mm}

\begin{acknowledgement}{\rm
The authors thank Georges Skandalis for very helpful suggestions and
remarks.}
\end{acknowledgement}

\section{Euclidean framework: main results}
\label{s:euclid}
\protect\setcounter{equation}{0}

\subsection{The Hamiltonian algebra of a standard $N$-body system}
\label{ss:inb}

Consider a system of $N$ particles moving in the physical space
$\mbr^d$. In the nonrelativistic case the Hamiltonian is of the form
\begin{equation}\label{eq:inonrel}
H={\textstyle\sum_{j=1}^N P_j^2/2m_j + \sum_{j=1}^N V_j(x_j)+
\sum_{j<k} V_{jk}(x_j-x_k)}
\end{equation}
where $m_1,\dots,m_N$ are the masses of the particles,
$x_1,\dots,x_N\in\mbr^d$ their positions, and $P_j=-i\nabla_{x_j}$
their momenta. In the simplest situation the potentials $V_j,V_{jk}$
are real continuous functions with compact support on $\mbr^d$. The
state space of the system is the Hilbert space $L^2(X)$ with
$X=(\mbr^d)^N$.

Let $P=(P_1,\dots,P_N)$, this is a set of commuting self-adjoint
operators on $L^2(X)$ and so $h(P)$ is a well defined self-adjoint
operator for any real Borel function $h$ on $X$. In what follows we
replace the kinetic energy part $\sum_j P_j^2/2m_j$ in
\eqref{eq:inonrel} by an operator $h(P)$ with $h$ continuous and
divergent at infinity. Denote $x=(x_1,\dots,x_N)$ the points of $X$
and let us consider the linear subspaces of $X$ defined as follows:
$X_j=\{x\in X\mid x_j=0\}$ if $1\le j \le N$ and $X_{jk}=\{x\in
X\mid x_j=x_k\}$ for $j<k$.  Let $\pi_j$ and $\pi_{jk}$ be the
natural maps of $X$ onto the quotient spaces $X/X_j$ and $X/X_{jk}$
respectively (the Euclidean structure of $X$ allows us to identify
these abstract spaces with the subspaces of $X$ orthogonal to $X_j$
and $X_{jk}$, but this is irrelevant here). Then $H$ may be written
in the form
\begin{equation}\label{eq:inonr}
  H=h(P) + {\textstyle\sum_j v_j\circ\pi_j(x)+
    \sum_{j<k} v_{jk}\circ\pi_{jk}(x)}
\end{equation}  
for some real functions $v_j\in\Cc(X/X_j)$ and
$v_{jk}\in\Cc(X/X_{jk})$. Thus Hamiltonians of the form
\eqref{eq:inonr} are natural objects in the $N$-body problem.  Note
that there should be no privileged origin in the momentum space, so
if we accept $h(P)$ as an admissible kinetic energy operator then
$h(P+p)$ should also be admissible for any
$p=(p_1,\dots,p_N)\in(\mbr^d)^N$.

Let $\rs\equiv\rs(X)$ be the set of linear subspaces of $X$ equipped
with the order relation $Y\leq Z \Leftrightarrow Y\subset Z$. Then
$\rs$ is a semilattice with $Y\wedge Z =Y\cap Z$.  If $Y\in\rs$ then
we realize $\Co(X/Y)$ as a $C^*$-algebra of operators on $L^2(X)$ by
associating to $v$ the operator of multiplication by $v\circ\pi_Y$,
where $\pi_Y:X\to X/Y$ is the canonical surjection. The following
fact is easy to prove: \emph{if $\cs\subset\rs$ is finite then
$\cc(\cs):=\sum_{Y\in\cs} \Co(X/Y)$ is a direct topological sum and
is an algebra if and only if $\cs$ is a sub-semilattice of $\rs$; in
this case, $\cc(\cs)$ is an $\cs$-graded $C^*$-algebra}.

In the next proposition $\cs$ is the semilattice of subspaces of $X$
generated by the $X_j$ and $X_{jk}$, i.e. the set of subspaces of
$X$ obtained by taking arbitrary intersections of subspaces of the
form $X_j$ and $X_{jk}$.  We denote $\rt_X$ the $C^*$-algebra of
operators on $L^2(X)$ of the from $\varphi(P)$ with
$\varphi\in\Co(X)$.

\begin{proposition}\label{pr:mad}
  Let $h:X\to\mbr$ be continuous with $h(x)\to\infty$ if
  $x\to\infty$ and let $H_p$ be the self-adjoint operator
  \eqref{eq:inonr} with $h(P)$ replaced by $h(P+p)$. Then the
  $C^*$-algebra generated by the operators $H_p$ when $p$ runs over
  $(\mbr^d)^N$ and $ v_j$ and $ v_{jk}$ run over the set of real
  functions in $\Cc(X/X_j)$ and $\Cc(X/X_{jk})$ respectively is
  $\rc=\cc(\cs)\cdot\rt_X=\sum_{Y\in\cs}
  \Co(X/Y)\cdot\rt_X$. Moreover, $\rc$ is $\cs$-graded by this
  decomposition.
\end{proposition}

This has been proved in \cite{DG1}. More general results of this
nature are presented in Appendix \ref{s:appb}.  Observe that we decided
to fix the function $h$ which represents the kinetic energy but not
the potentials $v_j,v_{jk}$. However, as a consequence of
Proposition \ref{pr:mad}, if we allow $h$ to vary we get the same
algebra.

Proposition \ref{pr:mad} provides a basic example of ``Hamiltonian
algebra''.  We mention that $\rc$ is the crossed product of the
$C^*$-algebra $\cc(\cs)$ by the natural action of the additive group
$X$, so it is a natural mathematical object.  We shall see in a more
general context that the set of self-adjoint operators affiliated to
it is much larger than expected (cf. Theorem \ref{th:iaff} for
example).

If we are in the nonrelativistic case and $V_j=0$ for all $j$ then
the center of mass of the system moves freely and it is more
convenient to eliminate it and to take as origin of the reference
system the center of mass of the $N$ particles. Then the
configuration space $X$ is the set of points
$x=(x_1,\dots,x_N)\in(\mbr^d)^N$ such that $\sum_km_k x_k=0$.
Proposition \ref{pr:mad}  remains valid if $\cs$ is conveniently
defined, see \S \ref{ss:cexample}.

The following ``generalized'' class of $N$-body systems is suggested
by results from \cite{Ma,Ma3}.

\begin{definition}\label{df:nb}
  An \emph{$N$-body structure on a locally compact abelian group
    $X$} is a set $\cs_X$ of closed subgroups such that
  $X\in\cs_X$ and such that for all $Y,Z\in\cs_X$ the following
  three conditions are satisfied: (i)~ $Y\cap Z \in \cs$; (ii) the
  subgroups $Y,Z$ of $X$ are compatible; (iii) if $Y\supsetneq Z$
  then $Y/Z$ is not compact.
\end{definition}
$X$ must be thought as configuration space of the system. The notion
of compatible subgroups is defined in Subsection \ref{ss:compat} (if
$X$ is a $\sigma$-compact topological space this means that $Y+Z$ is
a closed subgroup).  We shall see that the Hamiltonian algebra
associated to such an $N$-body system is an interesting object:
\begin{equation}\label{eq:gnbh}
\rc_X(\cs_X):= \cc(\cs_X)\cdot\rt_X \cong \cc(\cs_X)\rtimes X
\quad\text{where}\quad  \cc(\cs_X)=
\textstyle{\sum^\rmc_{Y\in\cs_X}} \Co(X/Y).
\end{equation} 

Here $\rt_X\cong\Co(X^*)$ is the group $C^*$-algebra of $X$ and
$\rtimes$ means crossed product.

\begin{example}\label{ex:}
  This framework covers an interesting extension of the standard
  $N$-body setting. Assume that $X$ is a finite dimensional real
  vector space.  In the standard framework the semilattice $\cs$
  consists of linear subspaces of $X$ but here we allow them to be
  closed additive subgroups. The closed additive
  subgroups of $X$ are of the form $Y=E+L$ where $E$ is a vector
  subspace of $X$ and $L$ is a lattice in a vector subspace $F$ of
  $X$ such that $E\cap F=\{0\}$.  More precisely, $L=\sum_k\mbz f_k$
  where $\{f_k\}$ is a basis in $F$. Thus $F/L$ is a torus and if
  $G$ is a third vector subspace such that $X=E\oplus F\oplus G$
  then the space $X/Y\simeq (F/L)\oplus G$ is a cylinder with $F/L$
  as basis.
\end{example}

\subsection{The Euclidean many-body algebra}
\label{ss:sintro}

We introduce here an abstract framework which allows us to study
couplings between several $N$-body systems of the type considered
above. A concrete and physically interesting example may be found in
\S\ref{ss:cexample}.

Let $\cx$ be a real prehilbert space. Let $\rs(\cx)$ be the set of
finite dimensional subspaces of $\cx$ equipped with the order
relation given by set inclusion. This is a lattice with $X\wedge
Y=X\cap Y$ and $X\vee Y=X+Y$, but only the semilattice structure is
relevant for what follows.

Each finite dimensional subspace
$X\subset \cx$ is equipped with the Euclidean structure induced by
$\cx$ hence the Hilbert space $\ch_X= L^2(X)$ and the $C^*$-algebras
$\rl_X=L(\ch_X)$ and $\rk_X=K(\ch_X)$ are well defined.  The group
algebra $\rt_X$ is defined as the closure in $\rl_X$ of the set of
operators of convolution with functions of class $\Cc(X)$.  If
$O=\{0\}$ then $\ch_O=\mbc$ and $\rl_O=\rk_O=\rt_O=\mbc$ by
convention. 

We denote $h(P)$ the operator on $\ch_X$ given by
$\cf_X^{-1}M_h\cf_X$, where $\cf_X$ is the Fourier transformation
and $M_h$ is the operator of multiplication by the function
$h:X\to\mbc$. Then $\rt_X=\{\psi(P)\mid\psi\in\Co(X)\}$. We use the
notation $P=P_X$  if the space $X$ has to be specified. 

If $X,Y$ are finite dimensional subspaces of $\cx$ we set
$\rl_{XY}=L(\ch_Y,\ch_X)$ and $\rk_{XY}=K(\ch_Y,\ch_X)$.  We define
a closed subspace $\rt_{XY}\subset\rl_{XY}$ as follows. If
$\varphi\in\Cc(X+Y)$ then one may easily check that
$(T_{XY}(\varphi)f)(x)=\int_Y\varphi(x-y)f(y) \rmd y$ defines a
continuous operator $\ch_Y\to\ch_X$. Let
\begin{equation}\label{eq:txy}
\rt_{XY}= \text{ norm closure of the set of operators} \hspace{2mm}
T_{XY}(\varphi)  \hspace{2mm} \text{with} \hspace{2mm} 
\varphi\in\Cc(X+Y).
\end{equation}
Clearly $\rt_{XX}=\rt_X$. The space $\rt_{XY}$ is a ``concrete''
realization of the Hilbert $C^*$-module introduced by Philip Green
to show the Morita equivalence of the crossed products
$\Co(Z/Y)\rtimes X$ and $\Co(Z/X)\rtimes Y$ where $Z=X+Y$ (this has
been noticed by Georges Skandalis, see Remark \ref{re:rief} for
more details).

Now we fix a sub-semilattice $\cs\subset\rs$, i.e. we assume that
$X\cap Y\in\cs$ if $X,Y\in\cs$.  This set completely determines the
many-body system and the class of Hamiltonians that we intend to
study. For each $X\in\cs$ the Hilbert space $\ch_X$ is thought as
the state space of an $N$-body system with $X$ as configuration
space.  We define the state space of the many-body system as the
Hilbertian direct sum
\begin{equation}\label{eq:hs}
\ch\equiv\ch_\cs=\oplus_{X\in\cs} \ch_X.
\end{equation}
We have a natural embedding $\rl_{XY}\subset L(\ch)$ for all
$X,Y\in\cs$. Let $\rl\equiv\rl_\cs$ be the closed linear span of the
subspaces $\rl_{XY}$. Clearly $\rl$ is a $C^*$-subalgebra of
$L(\ch)$ which is equal to $L(\ch)$ if and only if $\cs$ is finite.
We will be interested in subspaces $\rr$ of $\rl$ constructed as
follows: for each couple $X,Y$ we are given a closed subspace
$\rr_{XY}\subset\rl_{XY}$ and $\rr\equiv(\rr_{XY})_{X,Y\in\cs}=
{\textstyle\sum^\rmc_{X,Y\in\cs}}\rr_{XY}$ where $\sum^\rmc$ means
closure of the sum. Note that
$\rk\equiv\rk_\cs=(\rk_{XY})_{X,Y\in\cs}=K(\ch)$.

\begin{theorem}\label{th:C}
  Let $\rt\equiv\rt_\cs=(\rt_{XY})_{X,Y\in\cs}$. Then $\rt$ is a
  closed self-adjoint subspace of $\rl$ and $\rc\equiv\rc_\cs=\rt^2$
  is a non-degenerate $C^*$-algebra of operators on $\ch$.
\end{theorem}

We say that \emph{$\rc$ is the Hamiltonian algebra of the many-body
  system $\cs$}.  This terminology will be justified later on: we
shall see that physically interesting many-body Hamiltonians are
self-adjoint operators affiliated to $\rc$. Moreover, in a quite
precise way, $\rc$ is the smallest $C^*$-algebra with this
property. For the purposes of this paper \emph{we define a
many-body Hamiltonian as a self-adjoint operator affiliated to
$\rc$}.

We now equip $\rc$ with an $\cs$-graded $C^*$-algebras
structure. This structure will play a central role in the spectral
analysis of self-adjoint operators affiliated to $\rc$.  We often
say ``graded'' instead of $\cs$-graded.

To define the grading we need new objects.  If $Y\not\subset X$ we
set $\cc_X(Y)=\{0\}$. If $Y\subset X$ then we define $\cc_X(Y)$ as
the set of continuous functions on $X$ which are invariant under
translations in the $Y$ directions and tend to zero in the $Y^\perp$
directions.  This is a $C^*$-algebra of bounded uniformly continuous
functions on $X$ canonically isomorphic with $\Co(X/Y)$ where $X/Y$
is the orthogonal of $Y$ in $X$. Thus
\begin{equation}\label{eq:2cxy}
\cc_X(Y)\cong\Co(X/Y) \text{ if } Y\subset X \quad\text{and}\quad
\cc_X(Y)=\{0\} \text{ if } Y\not\subset X.
\end{equation}
Let $\cc_X\equiv\cc_X(\cs):=\sum^\rmc_{Y\in\cs}\cc_X(Y)$, this is a
$C^*$-algebra of bounded uniformly continuous functions on $X$.  We
embed it in $\rl_X$ by identifying a function $\varphi$ with the
operator on $\ch_X$ of multiplication by $\varphi$. Then
\begin{equation}\label{eq:imp0}
\cc\equiv\cc_\cs=\oplus_{X\in\cs}\cc_X
\end{equation}
is a $C^*$-algebra of operators on $\ch$ included in $\rl$.  For
each $Z\in\cs$ we define a $C^*$-subalgebra of $\cc$ by
\begin{equation}\label{eq:impz}
\cc(Z)\equiv\cc_\cs(Z)=\oplus_X\cc_X(Z)=\oplus_{X\supset Z}\cc_X(Z).
\end{equation}
It is easy to see that the family $\{\cc(Z)\}_{Z\in\cs}$ defines a
graded $C^*$-algebra structure on $\cc$.

\begin{theorem}\label{th:CG}
  We have $\rc=\rt\cdot\cc=\cc\cdot\rt$.  For each $Z\in\cs$ the
  space $\rc(Z)=\rt\cdot\cc(Z)=\cc(Z)\cdot\rt$ is a $C^*$-subalgebra
  of $\rc$ and the family $\{\rc(Z)\}_{Z\in\cs}$ defines a graded
  $C^*$-algebra structure on $\rc$.
\end{theorem}

In particular we get $\rc=(\rc_{XY})_{X,Y\in\cs}$ with (the second
and fourth equalities are not obvious):
\begin{align}
\rc_{XY} 
&=\rt_{XY}\cdot\cc_Y 
={\textstyle\sum^\rmc_{Z\subset X\cap Y}}\rt_{XY}\cdot\cc_Y(Z)
\label{eq:icxy1} \\
&=\cc_X\cdot\rt_{XY}= {\textstyle\sum^\rmc_{Z\subset X\cap
    Y}}\cc_X(Z)\cdot\rt_{XY}. \label{eq:icxy2}
\end{align}
The $C^*$-algebras $\cc(Z)$ and $\rc(Z)$ ``live'' in the closed
subspace $\ch_{\geq Z}=\oplus_{X\supset Z}\ch_X$ of $\ch$.  More
precisely, they leave invariant $\ch_{\geq Z}$ and their restriction
to its orthogonal subspace is zero. Moreover, if we denote
$\rl_{\geq Z}=(\rl_{X,Y})_{X,Y\supset Z}\subset L(\ch_{\geq
  Z})$ then clearly $\cc(Z)$ and $\rc(Z)$ are subalgebras of
$\rl_{\geq Z}$.

\begin{remark}\label{re:nbody}{\rm 
The diagonal element $\rc_{XX}\equiv\rc_X$ of the ``matrix'' $\rc$
is given  by
\begin{equation}\label{eq:NB}
\rc_X=\cc_X\cdot\rt_X=
{\textstyle\sum^\rmc_{Z\subset X}\Co(X/Z)\cdot\rt_X}
\subset\rl_X.
\end{equation}
This $C^*$-algebra is \emph{the Hamiltonian algebra of the
  (generalized) $N$-body system} associated to the semi-lattice
$\cs_X=\{Z\in\cs\mid Z\subset X\}$ of subspaces of $X$,
cf. \eqref{eq:gnbh}. The non-diagonal elements $\rc_{XY}$ are
Hilbert $C^*$-bimodules which define the coupling between the
$N$-body type systems $X$ and $Y$.  }\end{remark}

\begin{remark}\label{re:bigc}{\rm Note that if we take $\cs$ equal
    to the set of all finite dimensional subspaces of $\cx$ then we
    get a graded $C^*$-algebra $\rc$ canonically associated to the
    prehilbert space $\cx$. According to a remark in
    \S\ref{ss:inotations}, any other choice of $\cs$ would give us a
    graded $C^*$-subalgebra of this one.  }\end{remark}

\begin{remark}\label{re:sym}
  The algebra $\rc$ is not adapted to symmetry considerations, in
  particular in applications to physical systems consisting of
  particles one has to assume them distinguishable. The Hamiltonian
  algebra for systems of identical particles interacting through
  field type forces (both bosonic and fermionic case) is constructed
  in \cite{Geo}. We mention that a quantum field model without
  symmetry considerations corresponds to the case when $\cs$ is a
  distributive relatively ortho-complemented lattice.
\end{remark}

\subsection{Particle systems with conserved total mass}
\label{ss:cexample}

We give now a physically interesting example of the preceding abstract
construction.  We shall describe the many-body system associated to
$N$ ``elementary particles'' of masses $m_1,\dots,m_N$ moving in the
physical space $\mbr^d$ without external fields. We shall get
\emph{a system in which the the total mass is conserved but not the
  number of particles.}

We go back the framework of \S \ref{ss:inb} but assume that 
the particles interact only through 2-body forces. Then
\begin{equation}\label{eq:nonrel}
H={\textstyle\sum_{j=1}^N P_j^2/2m_j +\sum_{j<k} V_{jk}(x_j-x_k)}.
\end{equation}
and the center of mass of the system moves freely so it is
convenient to eliminate it. This is a standard procedure that we
sketch now, cf.  \cite{ABG,DeG1} for a detailed discussion of the
formalism.  We take as origin of the reference system the center of
mass of the $N$ particles so the configuration space $X$ is the set
of points $x=(x_1,\dots,x_N)\in(\mbr^d)^N$ such that $\sum_km_k
x_k=0$. We equip $X$ with the scalar product
$\braket{x}{y}=\sum_{k=1}^N2m_k x_k y_k$. The advantage is that the
reduced Hamiltonian, the operator acting in $L^2(X)$ naturally
associated to the expression \eqref{eq:nonrel}, is
$\Delta_X+\sum_{j<k}V_{jk}(x_j-x_k)$ where $\Delta_X$ is the
Laplacian associated to this scalar product. We denote by the same
symbol $H$ this reduced operator.

The first step is to describe the $C^*$-algebra generated by these
Hamiltonians, i.e. to get the analog of Proposition \ref{pr:mad} in
the present context. Thus we have to describe the semilattice of
subspaces of $X$ generated by the $X_{(jk)}:=X_{jk}\cap X$. We give
the result below and refer to the Appendix \S\ref{ss:anbody} for
proofs.

A partition $\sigma$ of the set $\{1,\dots,N\}$ is also called
cluster decomposition. Then the sets of the partition are called
clusters. A cluster $a\in\sigma$ is thought as a ``composite
particle'' of mass $m_a=\sum_{k\in a}m_k$.  Let $|\sigma|$ be the
number of clusters of $\sigma$. We interpret $\sigma$ as a system of
$|\sigma|$ particles with masses $m_a$ hence its configuration space
should be the set of $x=(x_a)_{a\in\sigma}\in(\mbr^d)^{|\sigma|}$
such that $\sum_a m_a x_a=0$ equipped with a scalar product similar
to that defined above.

Let $X_\sigma$ be the set of $x\in X$ such that $x_i=x_j$ if $i,j$
belong to the same cluster and let us equip $X_\sigma$ with the
scalar product induced by $X$. Then there is an obvious isometric
identification of $X_\sigma$ with the configuration space of the
system $\sigma$ as defined above. The advantage now is that all the
spaces $X_\sigma$ are isometrically embedded in the same $X$.  We
equip the set $\mathfrak{S}$ of partitions with the order relation:
$\sigma\leq\tau$ if and only if ``$\tau$ is finer than $\sigma$''
(this is opposite to the usual convention). Then $\sigma\leq\tau$ is
equivalent to $X_\sigma\subset X_\tau$ and $X_\sigma\cap
X_\tau=X_{\sigma\wedge\tau}$.  Thus we see that $\mathfrak{S}$ is
isomorphic as semilattice with the set $\cs=\{X_\sigma\mid
\sigma\in\mathfrak{S}\}$ of subspaces of $X$ with inclusion as order
relation. Now it is easy to check that \emph{$\cs$ coincides with
  the semilattice of subspaces of $X$ generated by the $X_{(jk)}$}.

We abbreviate $\ch_\sigma=\ch_{X_\sigma}=L^2(X_\sigma)$.  According
to the identifications made above, this is the state space of a
system of $|\sigma|$ particles with masses $(m_a)_{a\in\sigma}$.

Now we may apply our construction to $\cs$. We get a system whose
state space is $\ch=\oplus_\sigma \ch_{\sigma}$. If the system is in
a state $u\in\ch_{\sigma}$ then it consists of $|\sigma|$ particles
of masses $m_a$. Note that $\min\mathfrak{S}$ is the partition
consisting of only one cluster $\{1,\dots,N\}$ with mass
$M=m_1+\dots+m_N$. Since there are no external fields and we decided
to eliminate the motion of the center of mass, this system must be
the vacuum. And its state space is indeed
$\ch_{\min\mathfrak{S}}=\mbc$. The algebra $\rc$ in this case
predicts usual inter-cluster interactions associated, for examples,
to potentials defined on $X^\sigma=X/X_\sigma$, but also
interactions which force the system to make a transition from a
``phase'' $\sigma$ to a ``phase'' $\tau$. In other terms, the system
of $|\sigma|$ particles with masses $(m_a)_{a\in\sigma}$ is
transformed into a system of $|\tau|$ particles with masses
$(m_b)_{b\in\tau}$. Thus the number of particles varies from $1$ to
$N$ but the total mass is constant and equal to $M$.

\subsection{Natural morphisms and essential spectrum}
\label{ss:specan} 

We return to the general case.  Sub-semilattices $\ct$ of $\cs$
define many-body type subsystems (this is discussed in more detail
in\S\ref{ss:subsys}). The spectral properties of the total many-body
Hamiltonian are described in terms of a special class of such
subsystems.

Each $X\in\cs$ determines a new many-body system $\cs_{\geq
  X}=\{Y\in\cs\mid Y\supset X\}$ whose state space is $\ch_{\geq
  X}$. Let $\rc_{\ge X}$ be the corresponding Hamiltonian algebra
$\rc_{\cs_{\geq X}}$. It is easy to see that $\rc_{\geq
  X}=\sum^\rmc_{Y\supset X}\rc(Y)$.  Thus $\rc_{\geq X}$ is a
$C^*$-subalgebra of $\rc$ which lives and is non-degenerate on the
subspace $\ch_{\geq X}$ of $\ch$. We mention one fact:
if $\Pi_{\ge X}$ is the orthogonal projection $\ch\to\ch_{\ge X}$,
then $\Pi_{\ge X}\rc\Pi_{\ge X}$ is an $\cs$-graded
$C^*$-subalgebra of $\rc_\cs$ and we have $\rc_{\geq
  X}\subset\Pi_{\ge X}\rc\Pi_{\ge X}$ strictly in general.

Then the general theory of graded
$C^*$-algebras implies that \emph{there is a unique linear
  continuous projection $\rp_{\geq X}:\rc\to\rc_{\geq X}$ such that
  $\rp_{\geq X}(T)=0$ if $T\in\rc(Y)$ with $Y\not\supset X$ and this
  projection is a morphism}. These are the \emph{natural morphisms}
  of the graded algebra $\rc$.

This extends to unbounded operators as follows: if $H$ is a
self-adjoint operator on $\ch$ strictly affiliated to $\rc$ then
there is a unique self-adjoint operator $H_{\ge X}\equiv\rp_{\ge
  X}(H)$ on $\ch_{\ge X}$ such that $\rp_{\ge
  X}(\varphi(H))=\varphi(H_{\ge X})$ for all $\varphi\in\Co(\mbr)$.
If $H$ is only affiliated to $\rc$ then $H_{\ge X}$ could be not
densely defined.

Assume that the semilattice $\cs$ has a smallest element
$\min\cs$. Then $X\in\cs$ is called \emph{atom} if the only element
of $\cs$ strictly included in $X$ is $\min\cs$.  Let $\cp(\cs)$ be
the set of atoms of $\cs$. We say that $\cs$ is \emph{atomic} if
each of its elements distinct from $\min\cs$ contains an atom.  The
following HVZ type theorem is an immediate consequence Theorem
\ref{th:ga}. The symbol $\overline{\cup}$ means ``closure of
union''.

\begin{theorem}\label{th:imp4}
If $H$ is a self-adjoint operator on $\ch$ strictly affiliated to
$\rc$ then for each $X\in\cs$ there is a unique self-adjoint
operator $H_{\geq X}\equiv\rp_{\geq X}(H)$ on $\ch_{\geq X}$ such
that $\rp_{\geq X} (\varphi(H))=\varphi(H_{\geq X})$ for all
$\varphi\in\Co(\mbr)$. The operator $H_{\geq X}$ is strictly
affiliated to $\rc_{\geq X}$.  If $O\in\cs$ and $\cs$ is atomic then
\begin{equation}\label{eq:esi}
\spe(H)=\overline{\ccup}_{X\in\cp(\cs)}\sp(H_{\geq X}).
\end{equation} 
\end{theorem}

The theorem remains valid for operators which are only affiliated to
$\rc$ but then we must allow them to be non-densely defined.

\subsection{Subsystems and subhamiltonians}
\label{ss:subsys}

If $\ct$ is an arbitrary subset of $\cs$ then the Hilbert space
$\ch_\ct=\oplus_{X\in\ct}\ch_X$ is well defined and naturally
embedded as a closed subspace of $\ch_\cs$. Let $\Pi_\ct$ be the
orthogonal projection of $\ch$ onto $\ch_\ct$. Note that the
definition of $\rt_\cs$ makes sense for any set $\cs$ of finite
dimensional subspaces , in particular we may replace $\cs$ by
$\ct$. Then we have:
\begin{align}
& \rt_\ct =\Pi_\ct\rt_\cs\Pi_\ct, 
\quad
\cc_\cs^\ct := \Pi_\ct\cc_\cs\Pi_\ct= \oplus_{X\in\ct}\cc_X(\cs), 
\label{eq:subs1}\\
& \rc_\cs^\ct:= \Pi_\ct\rc_\cs\Pi_\ct
=\Pi_\ct\rt_\cs\cdot\cc_\cs\Pi_\ct=
\rt_\ct\cdot\cc_\cs^\ct \label{eq:subs2}
\end{align}
From this we easily get that $\rc_\cs^\ct=
{\textstyle\sum_{X,Y\in\ct}^\rmc}\rc_{XY}$ is an $\cs$-graded
$C^*$-subalgebra of $\rc$ supported by the ideal generated by $\ct$
in $\cs$ (an ideal is a subset $\cj$ of $\cs$ such that $X\subset
Y\in\cj\Rightarrow X\in\cj$).  The operators affiliated to
$\rc_\cs^\ct$ are affiliated to $\rc_\cs$, so are many-body
Hamiltonians in our sense. 

The case when $\ct$ is a sub-semilattice of $\cs$ is interesting.
Indeed, then $\ct$ defines a many-body system whose Hamiltonian
algebra $\rc_\ct$ is a $\ct$-graded $C^*$-algebra of operators on
the Hilbert space $\ch_\ct$. We emphasize that \emph{this algebra
  does not coincide with $\rc_\cs^\ct$}.  We always have
$\rc_\ct\subset\rc_\cs^\ct$ but \emph{the inclusion is strict unless
  $\ct$ is an ideal of $\cs$}.  This is clear because
$\cc_\ct\subset\cc_\cs^\ct$ strictly in general.

The simplest sub-semilattices are the chains (totally ordered
subsets). Then $\ch_\ct$ has a structure analogous to a Fock
space. Nonrelativistic Hamiltonians affiliated to such
$C^*$-algebras $\rc_\cs^\ct$ have been studied before in \cite{SSZ}.
If we take $\ct=\{X\}$ for an arbitrary $X\in\cs$ then the
associated subsystem has $\ch_X$ as state space and its Hamiltonian
algebra is just $\rc_X=\cc_X(\cs)\cdot\rt_X$, the Hamiltonian
algebra of the (generalized) $N$-body system determined by the
semi-lattice $\cs_X$. We refer to \S\ref{ss:T} and to the Example
\ref{ex:fried} for other simple but instructive examples of
subsystems.

\subsection{Intrinsic descriptions}
\label{ss:id}

We give two explicit descriptions of $\rc(Z)$ as an algebra of
operators on $\ch_{\geq Z}$. Since $\rc$ is the closure of the sum
of the algebras $\rc(Z)$, these descriptions allow one to check
rather easily whether a self-adjoint operator is affiliated to $\rc$
or not. Both theorems are consequences of more general results in
Section \ref{s:id}.

For any vector $a\in\cx$ and any finite dimensional subspace $X$ of
$\cx$ we define two unitary operators in $\ch_X$ by
$(U_af)(x)=f(x+a_X)$ and $(V_a f)(x)=\rme^{i\braket{a}{x}} f(x)$
where $a_X$ is the orthogonal projection of $x$ on $X$. Then
$\{U_a\}_{a\in\cx}$ and $\{V_a\}_{a\in\cx}$ are strongly continuous
representations of (the additive group) $\cx$ on $\ch_X$ such that
$U_a=1 \Leftrightarrow V_a=1 \Leftrightarrow a\perp X$. The direct
sum over $X\in\cs$ of these representations give representations of
$\cx$ on $\ch$ for which we use the same notations.

\begin{theorem}\label{th:CZU}
  $\rc(Z)$ is the set of $T\in\rl_{\geq Z}$ such that
\begin{compactenum}
\item[{\rm(i)}] $U_a^*T U_a=T$ for all $a\in Z$ and $\|T(V_a-1)\|\to
  0$ if $a\to 0$ in $Z^\perp$ ,
\item[{\rm(ii)}] $\|T(U_a-1)\|\to 0$ and $\|V^*_a T V_a-T\|\to 0$ if
  $a\to 0$ in $\cx$.
\end{compactenum}
\end{theorem}

Let $\cs/Z$ be the set of subspaces of $\cx/Z=Z^\perp$ of the form
$X/Z$ with $X\in\cs, X\supset Z$. Clearly $\cs/Z$ is a semilattice
of finite dimensional subspaces of $\cx/Z$ so the Hilbert space
$\ch_{\cs/Z}$ and the corresponding algebra of compact operators
$\rk_{\cs/Z}$ are well defined.  If $X\supset Z$ then $X=Z\oplus
X/Z$ so we have a canonical factorization
$\ch_X=\ch_Z\otimes\ch_{X/Z}$. Thus $\ch_{\geq
  Z}=\ch_Z\otimes\ch_{\cs/Z}$.

\begin{theorem}\label{th:CZ}
$\rc(Z)=\rt_Z\otimes \rk_{\cs/Z}$ relatively to the factorization
$\ch_{\geq Z}=\ch_Z\otimes\ch_{\cs/Z}$.
\end{theorem}

\subsection{Factorization properties}
\label{ss:fact}

For $Z\subset X\cap Y$ we have $X=Z\oplus (X/Z)$ and $Y=Y\oplus
(Y/Z)$ hence we have canonical factorizations
\begin{equation}\label{eq:xyzint}
\ch_X=\ch_Z\otimes\ch_{X/Z}
\quad\text{and}\quad 
\ch_Y=\ch_Z\otimes\ch_{Y/Z}.
\end{equation}
Relatively to these factorizations, we get from Theorem \ref{th:CZ}:
\begin{equation}\label{eq:CZ}
\rc_{XY}(Z)=\rt_Z\otimes \rk_{X/Z,Y/Z} \cong \Co(Z^*;\rk_{X/Z,Y/Z}).
\end{equation}
The tensor product (and those below) is in the category of Hilbert
modules, cf.  \S\ref{ss:ha}. We have written $Z^*$ above in spite of
the canonical Euclidean isomorphism $Z^*\cong Z$ in order to stress
that we consider functions of momentum not of position. For any
$X,Y$ we set
\begin{equation}\label{eq:xony}
X/Y=X/(X\cap Y)= X\ominus (X\cap Y)
\end{equation}
and so we have 
\begin{equation}\label{eq:xyonz}
X/Z=X/(X\cap Y)\oplus (X\cap Y)/Z = X/Y\oplus (X\cap Y)/Z 
\end{equation}
and similarly for $Y/Z$. Then from \eqref{eq:CZ} and
\eqref{eq:comtens} we get the finer factorization:
\begin{equation}\label{eq:xyzeptens}
\rc_{XY}(Z)=\rt_Z\otimes \rk_{(X\cap Y)/Z}
\otimes \rk_{X/Y,Y/X}. 
\end{equation}
In particular, we get
\begin{equation}\label{eq:xyzeint}
\rc_{XY}=\rc_{X\cap Y}\otimes \rk_{X/Y,Y/X}
\end{equation}
relatively to the tensor factorizations 
\begin{equation}\label{eq:xyeht}
\ch_X=\ch_{X\cap Y}\otimes\ch_{X/Y} \quad\text{and}\quad 
\ch_Y=\ch_{X\cap Y}\otimes\ch_{Y/X}.
\end{equation}
Since $\rk_{X/Y,O}\cong\ch_{X/Y}$
in the special case $Z\subset Y\subset X$ we have 
\begin{equation}\label{eq:xby}
\rc_{XY}=\rc_{Y}\otimes\ch_{X/Y}
\quad\text{and}\quad
\rc_{XY}(Z)=\rt_Z\otimes \rk_{Y/Z} \otimes \ch_{X/Y}.
\end{equation}

\begin{example}\label{ex:fried}
  These factorizations give us the possibility of expressing quite
  explicitly the Hamiltonian algebra of some subsystems. We refer to
  \S\ref{ss:T} for more general situations and consider here
  sub-semilattices of the form $\ct=\{X,Y\}$ with $X\supset Y$.
  This is a toy model, an $N$-body system coupled to one of its
  subsystems, and can be nicely formulated in a purely abstract
  setting, cf. Proposition \ref{pr:toymodel}.  We have
  $\ch_\ct=\ch_X\oplus\ch_Y$ with $\ch_X=\ch_Y\otimes\ch_{X/Y}$.
  From \eqref{eq:xby} we have
\[
\rc^\ct_\cs=
\begin{pmatrix}
\rc_X & \rc_{Y}\otimes\ch_{X/Y}\\
\rc_{Y}\otimes\ch^*_{X/Y} & \rc_Y
\end{pmatrix}.
\]
where $\ch^*_{X/Y}$ has a natural meaning (see
\S\ref{ss:affil}). The grading is defined for $Z\in\cs_X$
by
\begin{compactenum}
\item
If  $Z\subset Y$ then
\[
\rc^\ct_\cs(Z)=
\begin{pmatrix}
\rc_X(Z) & \rc_{Y}(Z)\otimes\ch_{X/Y}\\
\rc_{Y}(Z)\otimes\ch^*_{X/Y} & \rc_Y(Z)
\end{pmatrix}.
\]
\item   \label{p:i2ex}
If  $Z\not\subset Y$ then
\[
\rc^\ct_\cs(Z)=
\begin{pmatrix}
\rc_X(Z) & 0\\
0 & 0
\end{pmatrix}.
\]
\end{compactenum}
If $\ct=\{X,O\}$ we get a version of the Friedrichs model: an
$N$-body system coupled to the vacuum. The case when $\ct$ is an
arbitrary chain (a totally ordered subset of $\cs$) is very similar.
The case $\ct=\{X,Y\}$ with not comparable $X,Y$ is more complicated
and is treated in \S\ref{ss:T} in a more general setting.
\end{example}

\subsection{Examples of many-body Hamiltonians}
\label{ss:ex}

Here we use Theorems \ref{th:CZ} and \ref{th:CZU} to construct
self-adjoint operators strictly affiliated to $\rc$. For simplicity,
\emph{in this and the next subsections $\cs$ is assumed finite}. If
$\cs$ is infinite then an assumption of the same nature as the
non-zero mass condition in quantum field theory models is needed to
ensure that the kinetic energy operator $K$ is affiliated to $\rc$.

The Hamiltonians will be of the form $H=K+I$ where the self-adjoint
operator $K$ is the kinetic energy and $I$ is an
interaction term bounded in form sense by $K$. More precisely, $I$
is a symmetric sesquilinear form on the domain of $|K|^{1/2}$ which
is continuous, i.e. satisfies
\begin{equation}\label{eq:formbound}
\pm I\leq \mu|K+ia| \hspace{2mm}\text{for some real numbers } \mu, a.
\end{equation}
$H$ and $K$ are matrices of operators, e.g. $H=(H_{XY})_{X,Y\in\cs}$
where $H_{XY}$ is defined on a subspace of $\ch_Y$ and has values in
$\ch_X$ and the relation $H_{XY}^*=H_{YX}$ holds at least
formally. By construction $K$ is given by a diagonal matrix, so
$K_{XY}=0$ if $X\neq Y$, and we set $K_X=K_{XX}$. The interaction
will be a matrix of sesquilinear forms. Then $H_{XX}=K_{X}+I_{XX}$
will be an $N$-body type Hamiltonian, i.e. a self-adjoint operator
affiliated to $\rc_X$, cf. Remark \ref{re:nbody}. The non-diagonal
elements $H_{XY}=I_{XY}$ define the interaction between the systems
$X$ and $Y$.  We give now a rigorous construction of such
Hamiltonians.

{\bf(a)} For each $X$ we choose a kinetic energy operator
$K_X=h_X(P)$ for the system having $X$ as configuration space. The
function $h_X:X\to\mbr$ is continuous and such that
$h_X(x)\to\infty$ if $x\to\infty$.  We stress that there are no
relations between the kinetic energies of the systems corresponding
to different $X$. Denote $\cg^2_X$ the domain of $K_X$ equipped with
the graph norm and let $\cg^s_X$ ($s\in\mbr$) be the scale of
Hilbert spaces associated to it, e.g.  $\cg^0_X=\ch_X$,
$\cg^1_X=D(K_X^{1/2})$ is the form domain of $K_X$, and $\cg^{-1}_X$
its adjoint space.

{\bf(b)} The total kinetic energy of the system is by definition
$K=\oplus_X K_X$. We call this a \emph{standard kinetic energy
  operator}.  Then the spaces $\cg^s$ of the scale determined by the
domain $\cg^2$ of $K$ can be identified with direct sums
$\cg^s\simeq\oplus_X\cg^s_X$. In particular this holds for the form
domain $\cg^1=D(K^{1/2})$ and for its adjoint space $\cg^{-1}$.
Note that we may also introduce the operators $K_{\geq
  X}=\oplus_{Y\supset X} K_Y$ and the associated spaces $\cg^s_{\ge
  X}$. If $s>0$ we have $\cg^s_{\ge X}=\cg^s\cap\ch_{\ge X}$.

{\bf(c)} The simplest type of interactions that we may consider are
given by symmetric elements $I$ of the multiplier algebra of
$\rc$. Then $H=K+I$ is strictly affiliated to $\rc$ and $\rp_{\geq
  X}(H)=K_{\geq X}+\rp_{\geq X} (I)$ where $\rp_{\geq X}$ is
extended to the multiplier algebras as explained in
\cite[p. 18]{La}.

{\bf(d)} In order to cover singular interactions (form bounded but
not necessarily operator bounded by $K$) we assume that the
functions $h_X$ are equivalent to regular weights. This is a quite
weak assumption, cf. page \pageref{p:regw}.  For example, it
suffices that $c'|x|^{\alpha}\leq h_X(x)\leq c''|x|^{\alpha}$ for
large $x$ where $c',c'',\alpha>0$ are numbers depending on $X$.
Then $U_a,V_a$ induce continuous operators in each of the spaces
$\cg^s_X$, $\cg^s$, $\cg^s_{\ge X}$.

{\bf(e)} The interaction will be of the form $I=\sum_{Z\in\cs} I(Z)$
where the $I(Z)$ are continuous symmetric sesquilinear forms on
$\cg^1$ such that $ I(Z) \geq -\mu_Z K -\nu$ for some positive
numbers $\mu_Z$ and $\nu$ with $\sum_Z\mu_Z<1$. Then the form sum
$K+I$ defines a self-adjoint operator $H$ on $\ch$.

{\bf(f)} We identify $I(Z)$ with a symmetric operator
$\cg^1\to\cg^{-1}$ and we assume that $I(Z)$ is supported by the
subspace $\ch_{\ge Z}$.  In other terms, $I(Z)$ is the sesquilinear
form on $\cg^1$ associated to an operator $I(Z):\cg^1_{\ge
  Z}\to\cg^{-1}_{\ge Z}$.  Moreover, we assume that this last
operator satisfies
\begin{equation}\label{eq:iaff}
U_a I(Z)=I(Z) U_a \text{ if } a\in Z, \ 
I(Z)(V_a-1)\to 0 \text{ if } a\to 0 \text{ in } Z^\perp, \ 
V^*_a I(Z) V_a\to I(Z) \text{ if } a\to 0
\end{equation}
where the limits hold in norm in $L(\cg^2_{\ge Z},\cg^{-1}_{\ge Z})$.

Note that the first part of condition {\bf(f)}, saying that $I(Z)$
is supported by $\ch_{\ge Z}$, is equivalent to an estimate of the
form $\pm I(Z)\le\mu K_{\geq Z}+ \nu\Pi_{\geq Z}$ for some positive
numbers $\mu,\nu$. See also Remark \ref{re:iaff}.

\begin{theorem}\label{th:iaff}
  The Hamiltonian $H$ is a self-adjoint operator strictly affiliated
  to $\rc$, we have   
  $H_{\geq X}=K_{\geq X}+\sum_{Z\geq X} I(Z)$, 
  and $\spe(H)=\ccup_{X\in\cp(\cs)}\sp(H_{\geq X})$.
\end{theorem}

\begin{remark}\label{re:pauli}
  We required the $h_X$ to be bounded from below only for the
  simplicity of the statements. Moreover, a simple extension of the
  formalism allows one to treat particles with arbitrary
  spin. Indeed, if $E$ is a complex Hilbert then Theorem \ref{th:CG}
  remains true if $\rc$ is replaced by $\rc^E=\rc\otimes K(E)$ and
  the $\rc(Z)$ by $\rc(Z)\otimes K(E)$. If $E$ is the spin space
  then it is finite dimensional and one obtains $\rc^E$ exactly as
  above by replacing the $\ch(X)$ by $\ch(X)\otimes
  E=L^2(X;E)$. Then one may consider instead of scalar kinetic
  energy functions $h$ self-adjoint operator valued functions
  $h:X^*\to L(E)$. For example, we may take as one particle kinetic
  energy operators the Pauli or Dirac Hamiltonians.
\end{remark}

\begin{remark}\label{re:iaff}
We give here a second, more explicit version of condition
{\bf(f)}. Since 
$I(Z)$ is a continuous symmetric operator $\cg^1\to\cg^{-1}$ we may
represent it as a matrix $I(Z)=(I_{XY}(Z))_{X,Y\in\cs}$ of
continuous operators $I_{XY}(Z):\cg^1_Y\to\cg^{-1}_X$ with
$I_{XY}(Z)^*=I_{YX}(Z)$. We take $I_{XY}(Z)=0$ if $Z\not\subset
X\cap Y$ and if $Z\subset X\cap Y$ we assume
$V^*_a I_{XY}(Z) V_a\to I_{XY}(Z) \text{ if } a\to 0 \text{ in }
X+Y$ and 
\begin{equation}\label{eq:iiaff}
U_a I_{XY}(Z)=I_{XY}(Z) U_a \text{ if }a\in Z,  \quad
I_{XY}(Z)(V_a-1)\to 0 \text{ if } a\to 0 \text{ in } Y/Z.
\end{equation}
The limits should hold in norm in $L(\cg^2_Y,\cg^{-1}_X)$.
\end{remark}

The operators $I_{XY}(Z)$satisfying \eqref{eq:iiaff} are described
in more detail in Proposition \ref{pr:zxy}. In the next example we
consider the simplest situation which is useful in the
nonrelativistic case.

If $E$ is an Euclidean space and $s$ is a real number let $\ch^s_E$
be the Sobolev space defined by the norm
\[
\|u\|_{\ch^s} = \|(1+\Delta_E)^{s/2}u\|
\] 
where $\Delta_E$ is the (positive) Laplacian associated to the
Euclidean space $E$. The space $\ch^s_E$ is equipped with two
continuous representations of $E$, a unitary one induced by
$\{U_x\}_{x\in E}$ and a non-unitary one induced by $\{V_x\}_{x\in
  E}$.  If $E=O:=\{0\}$ we define $\ch_E^s=\mbc$.

\begin{definition}\label{df:small}
  If $E,F$ are Euclidean spaces and $T:\ch^s_E\to\ch^t_F$ is a
  linear map, we say that \emph{$T$ is small at infinity} if there
  is $\varepsilon>0$ such that when viewed as a map
  $\ch^{s+\varepsilon}_E\to\ch^{t}_F$ the operator $T$ is compact.
\end{definition}
By the closed graph theorem $T$ is continuous and the compactness
property holds for all $\varepsilon>0$.  If $E=O$ or $F=O$ then we
consider that all the operators $T:\ch^s_E\to\ch^t_F$ are small at
infinity.

\begin{example}\label{ex:zxy}
  Due to assumption {\bf(d)} the form domains of $K_X$ and $K_Y$ are
  Sobolev spaces, for example $\cg^1_X=\ch^s_X$ and
  $\cg^1_Y=\ch^t_Y$. Let $I_{XY}^Z:\ch^t_{Y/Z}\to\ch^{-s}_{X/Z}$ be
  a linear small at infinity map. Then we may take
  $I_{XY}(Z)=1_Z\otimes I_{XY}^Z$ relatively to the tensor
  factorizations \eqref{eq:xyzint}.
\end{example}

We make now some comments to clarify the conditions {\bf(a)} -
{\bf(f)}.  Assume, more generally, that $\rc$ is a $C^*$-algebra of
operators on a Hilbert space $\ch$ and that $K$ is a self-adjoint
operator on $\ch$ affiliated to $\rc$. Let $I$ be a continuous
symmetric sesquilinear form on the domain of $|K|^{1/2}$. Then for
small real $\nu$ the form sum $K+\nu I$ is a self-adjoint operator
$H_\nu$. If $H_\nu$ is affiliated to $\rc$ for small $\nu$, and
since the derivative with respect to $\nu$ at zero of
$(H_\nu+i)^{-1}$ exists in norm, we get
$(K+i)^{-1}I(K+i)^{-1}\in\rc$. This clearly implies
$\jap{K}^{-2}I\jap{K}^{-2}\in\rc$.  Since
$\jap{K}^{-1/2}I\jap{K}^{-1/2}$ is a bounded operator, the map
$z\mapsto\jap{K}^{-z}I\jap{K}^{-z}$ is holomorphic on $\Re{z}>1/2$
hence we get
\begin{equation}\label{eq:ent}
\jap{K}^{-\alpha}I\jap{K}^{-\alpha}\in\rc \text{ if } \alpha>1/2. 
\end{equation}
Reciprocally, if $K$ is strictly affiliated to $\rc$ (and $K$ as
defined at (b) has this property) then Theorem 2.8 from \cite{DG3}
says that $\jap{K}^{-1/2}I\jap{K}^{-\alpha}\in\rc$ suffices to
ensure that $H=K+I$ is strictly affiliated to $\rc$ under a quite
general condition needed to make this operator well defined (this is
the role of assumption (e) above).  Condition (f) is formulated such
as to imply $\jap{K}^{-1/2}I\jap{K}^{-1}\in\rc$. To simplify the
statement we added condition (d) which implies that the spaces
$\cg^s$ are stable under the group $V_a$. Formally
\[
(\jap{K}^{-1/2}I\jap{K}^{-1})_{XY}=
\jap{K_X}^{-1/2}I_{XY}\jap{K_Y}^{-1}.
\]
So this should belong to $\rc_{XY}=\sum_{Z\subset X\cap
  Y}\rc_{XY}(Z)$. Thus $I_{XY}$ must be a sum of terms $I_{XY}(Z)$
with
\[
\jap{K_X}^{-1/2}I_{XY}(Z)\jap{K_Y}^{-1}\in\rc_{XY}(Z). 
\]
Conditions (d) and (f) are formulated such as this to hold,
cf. Remark \ref{re:iaff} and Theorem \ref{th:CZU}.

\subsection{Pauli-Fierz   Hamiltonians}
\label{ss:affil} 

The next result is an a priori argument which supports our
interpretation of $\rc$ as Hamiltonian algebra of a many-body
system: we show that $\rc$ is the $C^*$-algebra generated by a
simple class of Hamiltonians which have a natural quantum field
theoretic interpretation. For simplicity we state this only for
finite $\cs$.

For each couple $X,Y\in\cs$ such that $X\supset Y$ we have
$\ch_X=\ch_Y\otimes\ch_{X/Y}$.  Then we define
$\Phi_{XY}\subset\rl_{XY}$ as the closed linear subspace consisting
of ``creation operators'' associated to states from $\ch_{X/Y}$,
i.e. operators $a^*(\theta):\ch_Y\to\ch_X$ with $\theta\in\ch_{X/Y}$
which act as $u\mapsto u\otimes\theta$.  We set
$\Phi_{YX}=\Phi_{XY}^*\subset\rl_{YX}$, this is the space of
``annihilation operators'' $a(\theta)=a^*(\theta)^*$ defined by
$\ch_{X/Y}$.  This defines $\Phi_{XY}$ when $X,Y$ are comparable,
i.e. $X\supset Y$ or $X\subset Y$, which we abbreviate by $X\sim
Y$. If $X\not\sim Y$ then we take $\Phi_{XY}=0$. Note that
$\Phi_{XX}=\mbc 1_X$, where $1_X$ is the identity operator on
$\ch_X$.  We have
\begin{equation}\label{eq:Phi}
\rt_X\cdot\Phi_{XY}=\Phi_{XY}\cdot \rt_Y=\rt_{XY} \quad
\text{if } X\sim Y.
\end{equation}

Now let $\Phi=(\Phi_{XY})_{X,Y\in\cs}\subset\rl$. This is
a closed self-adjoint linear space of bounded operators on $\ch$.  A
symmetric element $\phi\in\Phi$ will be called \emph{field
  operator}. Giving such a $\phi$ is equivalent to giving a family
$\theta=(\theta_{XY})_{X\supset Y}$ of elements
$\theta_{XY}\in\ch_{X/Y}$, the components of the operator
$\phi\equiv\phi(\theta)$ being given by:
$\phi_{XY}=a^*(\theta_{XY})$ if $X\supset Y$,
$\phi_{XY}=a(\theta_{YX})$ if $X\subset Y$, and $\phi_{XY}=0$ if
$X\not\sim Y$. 

The operators of the form $K+\phi$, where $K$ is a standard kinetic
energy operator and $\phi\in\Phi$ is a field operator, will be
called \emph{Pauli-Fierz Hamiltonians}.

\begin{theorem}\label{th:motiv}
  If $\cs$ is finite then $\rc$ is the $C^*$-algebra
  generated by the Pauli-Fierz Hamiltonians.
\end{theorem}

Thus $\rc$ is generated by a class of Hamiltonians involving only
elementary field type interactions. On the other hand, we have seen
before that the class of Hamiltonians affiliated to $\rc$ is very
large and covers \mbox{$N$-body} systems interacting between
themselves with field type interactions. We emphasize that the
$k$-body type interactions \emph{inside} each of the $N$-body
subsystems are generated by pure field interactions.

\subsection{Nonrelativistic Hamiltonians and Mourre
  estimate}
\label{ss:mouint}  

We prove the Mourre estimate only for nonrelativistic many-body systems. 
There are serious difficulties when
the kinetic energy is not a quadratic form even in the much simpler
case of $N$-body Hamiltonians, but see \cite{De1,Ger1,DG2} for some
partial results which could be extended to our setting. Note that
the quantum field case is much easier from this point of view
because of the special nature of the interactions
\cite{DeG2,Ger2,Geo}. 

Let $\cs$ be a finite semilattice of subspaces of $\cx$.  Recall
that for $X\in\cs$ we denote $\cs/X$ the set of subspaces $Y/X=Y\cap
X^\perp$ with $Y\in\cs_{\ge X}$. This is a finite semilattice of
subspaces of $\cx$ which contains $O$. Hence the Hilbert space
$\ch_{\cs/X}$ and the $C^*$-algebra $\rc_{\cs/X}$ are well defined
by our general rules and (cf. \S\ref{ss:fact}):
\begin{equation}\label{eq:f}
\ch_{\geq X}=\ch_X\otimes\ch_{\cs/X}\quad \text{and} \quad
\rc_{\geq X}=\rt_X\otimes\rc_{\cs/X} .
\end{equation}
Denote $\Delta_X$ the (positive) Laplacian associated to the Euclidean
space $X$ with the convention $\Delta_O=0$. We have
$\Delta_X=h_X(P)$ with $h_X(x)=\|x\|^2$.  We set
$\Delta\equiv\Delta_\cs=\oplus_X \Delta_X$ and define $\Delta_{\geq X}$
similarly. If $Y\supset X$ then $\Delta_Y=\Delta_X\otimes 1
+1\otimes\Delta_{Y/X}$ hence $\Delta_{\geq X}=\Delta_X\otimes 1
+1\otimes\Delta_{\cs/X}$.  The domain and form domain of the
operator $\Delta_\cs$ are given by $\ch_\cs^2$ and $\ch_\cs^1$ where
$\ch_\cs^s\equiv\ch^s=\oplus_X \ch^s(X)$ for any real
$s$.

We define nonrelativistic many-body Hamiltonian by extending to the
present setting \cite[Def. 9.1]{ABG}. We consider only strictly
affiliated operators to avoid working with not densely defined
operators. Note that the general case of affiliated operators covers
interesting physical situations (hard-core interactions).

\begin{definition}\label{df:NR}
\emph{A nonrelativistic many-body Hamiltonian of type $\cs$} is a
bounded from below self-adjoint operator $H=H_\cs$ on $\ch=\ch_\cs$
which is strictly affiliated to $\rc=\rc_\cs$ and has the following
property: for each $X\in\cs$ there is a bounded from below
self-adjoint operator $H_{\cs/X}$ on $\ch_{\geq X}$
such that 
\begin{equation}\label{eq:NR}
\rp_{\geq X}(H)\equiv H_{\geq X}=\Delta_X\otimes1+ 1\otimes H_{\cs/X}
\end{equation}
relatively to the tensor factorization $\ch_{\geq
  X}=\ch_X\otimes\ch_{\cs/X}$. 
\end{definition}

Then \emph{each $H_{\cs/X}$ is a nonrelativistic many-body
  Hamiltonian of type $\cs/X$}. Indeed, the argument from \cite[p.\
415]{ABG} extends in a straightforward way to the present situation.

\begin{remark}\label{re:maxs}
  If $X$ is a maximal element in $\cs$ then $\cs/X=\{O\}$ hence
  $\ch_{\cs/X}=\ch_O=\mbc$ and $H_O$ will necessarily be a real
  number.  Then we get $\ch_{\geq X}=\ch_X$, $\rc_{\geq X}=\rt_X$,
  and $H_{\ge X}=\Delta_X +H_O$ on $\ch_X$.
\end{remark}

\begin{remark}\label{re:mins}
  Since $\cs$ is a finite semilattice, it has a least element
  $\min\cs$. If $\cs_o=\cs/{\min\cs}$, we get
\begin{equation}\label{eq:min}
\ch_{\cs}=\ch_{\min\cs}\otimes\ch_{\cs_o}, \quad
\rc_{\cs}=\rt_X\otimes \rc_{\cs_o}, \quad  
H_{\cs}=\Delta_{\min\cs}\otimes 1 +  1\otimes H_{\cs_o}.
\end{equation}
\end{remark}

Now we give an HVZ type description of the essential spectrum of a
nonrelativistic many-body Hamiltonian. For a more detailed
statement, see the proof.

\begin{theorem}\label{th:nrhvz}
  Denote $\tau_X=\inf H_{\cs/X}$ the bottom of the spectrum of
  $H_{\cs/X}$. Then
\begin{equation}\label{eq:hvzunif}
  \spe(H)=[\tau,\infty[ \hspace{2mm}\text{with}\hspace{2mm}
  \tau=\min \{ \tau_X \mid X \text{ is minimal in } \cs\setminus
  \{O\} \}. 
\end{equation}
\end{theorem}
\proof
  From \eqref{eq:NR} we get
\begin{equation}\label{eq:speint}
\sp(H_{\geq X})=[0,\infty[\ +\, \sp(H_{\cs/X})=
[\tau_X,\infty[ \quad
\text{if } X\neq O.
\end{equation}
In particular, if $O\nin\cs$ then by taking $X=\min\cs$
in \eqref{eq:min} we get 
\begin{equation}\label{eq:mino}
\sp(H)=\spe(H)=[\inf H_{\cs_o},\infty[ .
\end{equation}
If $O\in\cs$ then Theorem \ref{th:imp4} implies 
\begin{equation}\label{eq:hvznrint}
\spe(H)=[\tau,\infty[ \hspace{2mm}\text{with}\hspace{2mm}
\tau=\min_{X\in\cp(\cs)} \tau_X.
\end{equation}
The relation \eqref{eq:hvzunif} expresses \eqref{eq:mino} and
\eqref{eq:hvznrint} in a unified way. 
\qed

For $X\in\cs$ we consider the dilation group $W_\tau=\rme^{i\tau D}$
defined on $\ch_X$ by (set $n=\dim X$):
\begin{equation}\label{eq:fed}
(W_\tau u)(x)=\rme^{n\tau/4}u(\rme^{\tau/2}x), 
\quad 
2iD=x\cdot\nabla_x+n/2= \nabla_x\cdot x-n/2. 
\end{equation}
Let $D_O=0$. We keep the same notation for the unitary operator
$\oplus_X W_\tau$ on the direct sum $\ch=\oplus_X\ch_X$ and we do
not indicate explicitly the dependence on $X$ or $\cs$ of $W_\tau$
and $D$ unless this is really needed. Note that $D$ has
factorization properties similar to that of the Laplacian,
e.g. $D_{\geq X}=D_X\otimes 1 +1\otimes D_{\cs/X}$.

We refer to Subsection \ref{ss:mest} for terminology related to the
Mourre estimate. We take $D$ as conjugate operator and we denote by
$\what\rho_H(\lambda)$ the best constant (which could be infinite)
in the Mourre estimate at point $\lambda$. The \emph{threshold set}
$\tau(H)$ of $H$ with respect to $D$ is the set where
$\what\rho_H(\lambda)\leq0$.  If $A$ is a real set then we define
$N_A:\mbr\to[-\infty,\infty[$ by $N_A(\lambda)=\sup\{ x\in A \mid
x\leq\lambda\}$ with the convention $\sup\emptyset=-\infty$.  Denote
$\mathrm{ev}(T)$ the set of eigenvalues of an operator $T$.

\begin{theorem}\label{th:thrintr}
  Let $H=H_\cs$ be a nonrelativistic many-body Hamiltonian of type
  $\cs$ and of class $C^1_\rmu(D)$. Then $\tau(H)$ is a closed
  \emph{countable} real set given by
\begin{equation}\label{eq:thrintr}
\tau(H)=\ccup_{X\neq O}\mathrm{ev}(H_{\cs/X}).
\end{equation}
The eigenvalues of $H$ which do not belong to $\tau(H)$ are of
finite multiplicity and may accumulate only to points from
$\tau(H)$. We have
$\what\rho_H(\lambda)=\lambda-N_{\tau(H)}(\lambda)$ for all real
$\lambda$.
\end{theorem}

We emphasize that if $O\nin\cs$ the threshold set
\begin{equation}\label{eq:thrinto}
\tau(H)=\ccup_{X\in\cs} \mathrm{ev}(H_{\cs/X})
\end{equation}
is very rich although the spectrum of
$H=\Delta_{\min\cs}\otimes1+1\otimes H_{\cs_o}$ is purely absolutely
continuous.

\begin{remark}\label{re:NM}
We thus see that there is no difference between nonrelativistic
$N$-body and many-body Hamiltonians from the point of view of their
channel structure.  The formulas which give the essential spectrum
and the threshold set relevant in the Mourre estimate are identical,
cf. \eqref{eq:hvznrint} and \eqref{eq:thrintr}. This is due to
the fact that both Hamiltonian algebras are graded by the same
semilattice $\cs$.
\end{remark}

\subsection{Examples of nonrelativistic many-body Hamiltonians}
\label{ss:examples}

Let $H=K+I$ with kinetic energy  $K=\Delta$. Hence
$\cg^1=\ch^1=\oplus_X \ch^1_X$ and $\cg^{-1}=\ch^{-1}=\oplus_X
\ch^{-1}_X$ with the notations of \S\ref{ss:ex}.  The interaction
term is an operator $I:\ch^1\to\ch^{-1}$ given by a sum
$I=\sum_{Z\in\cs} I(Z)$ where each $I(Z)$ is defined with the help
of the tensor factorization $\ch_{\ge Z}=\ch_Z\otimes\ch_{\cs/Z}$.

\begin{proposition}\label{pr:exnr}
  Let $I^Z:\ch^1_{\cs/Z}\to\ch^{-1}_{\cs/Z}$ be symmetric and small
  at infinity and let $I(Z):=1_Z\otimes I^Z$ which is naturally
  defined as a symmetric operator $\ch^1\to\ch^{-1}$. Assume that
  $I(Z)\geq -\mu_Z\Delta-\nu$ for some numbers $\mu_Z,\nu\ge0$
  with $\sum\mu_Z<1$. Then $H=\Delta+I$ defined in the quadratic
  form sense is a nonrelativistic many-body Hamiltonian of type
  $\cs$ and we have $H_{\geq X}=\Delta_{\geq X}+\sum_{Z\supset X}
  I(Z)$.
\end{proposition}

The first condition on $I^Z$ can be stated in terms of its
coefficients as follows: if $Z\subset X\cap Y$ then the operator
$I_{XY}^Z:\ch^1_{Y/Z}\to\ch^{-1}_{X/Z}$ is small at infinity and
such that $(I_{XY}^Z)^*=I_{YX}^Z$.  On the other hand, note that if
the operators $I^Z:\ch^1_{\cs/Z}\to\ch^{-1}_{\cs/Z}$ are compact
then they are small at infinity and for any $\mu>0$ there is a
number $\nu$ such that $\pm I(Z)\leq \mu\Delta_\cs+\nu$ for all
$Z$. The more general smallness at infinity condition covers second
order perturbations of $\Delta_\cs$.

In the next proposition we give examples of nonrelativistic
operators of class $C^1_\rmu(D)$. The operator $H$ is constructed as
in Proposition \ref{pr:exnr} but we consider only interactions which
are relatively bounded in \emph{operator} sense with respect to the
kinetic energy such as to force the domain of $H$ to be equal to the
domain of $\Delta$, hence to $\ch^2=\oplus_X\ch^2_X$.  Since
this space is stable under the action of the operators $W_\tau$,
we shall get a simple condition for $H$ to be of class
$C^1_\rmu(D)$.

\begin{proposition}\label{pr:nrm}
  For each $Z\in\cs$ assume that $I^Z:\ch^2_{\cs/Z}\to \ch_{\cs/Z}$
  is compact and symmetric as operator on $\ch_{\cs/Z}$ and that
  $[D,I^Z]: \ch^2_{\cs/Z}\to\ch^{-2}_{\cs/Z}$ is compact.  Then the
  conditions of Proposition \ref{pr:exnr} are fulfilled and each
  operator $I(Z):\ch^2\to\ch$ is $\Delta$-bounded with relative
  bound zero.  The operator $H$ is self-adjoint on $\ch^2$ and of
  class $C^1_\rmu(D)$.
\end{proposition}

So for the coefficients $I^Z_{XY}$ we ask $I^Z_{XY}=0$ if
$Z\not\subset X\cap Y$ and if $Z\subset X\cap Y$ then
$(I^{Z}_{XY})^*\supset I^Z_{YX}$ and
\begin{equation}\label{eq:3d}
I^Z_{XY}:\ch^2_{Y/Z}\to\ch_{X/Z} \text{ and } 
[D,I^Z_{XY}]: \ch^2_{Y/Z}\to\ch^{-2}_{X/Z} \text{  are  compact
  operators.} 
\end{equation}
The expression $[D,I^Z_{XY}]=D_{X/Z}I^Z_{XY}-I^Z_{XY}D_{Y/Z}$ is not
really a commutator. Indeed, if we denote $E=(X\cap Y)/Z$, so
$Y/Z=E\oplus(Y/X)$ and $X/Z=E\oplus(X/Y)$, then $\ch_{X/Z}
=\ch_E\otimes\ch_{X/Y}$ and $\ch_{Y/Z} =\ch_E\otimes\ch_{Y/X}$.
Hence the relation $D_{X/Z}=D_E\otimes 1 + 1\otimes D_{X/Y}$ and a
similar one for $Y/Z$ give
\begin{equation*}
[D,I^Z_{XY}] =[D_E,I^Z_{XY}] +D_{X/Y}I^Z_{XY} -I^Z_{XY}D_{Y/X}.
\end{equation*}
The first term above is a commutator and so is of a different nature
than the next two. Since $I^Z_{XY}D_{Y/X}$ is a restriction of
$(D_{Y/X}I^Z_{YX})^*$ it is clear that the second part of condition
\eqref{eq:3d} follows from:
\begin{equation}\label{eq:2d}
[D_E,I^Z_{XY}] \text{ and } D_{X/Y}I^Z_{XY} \text{ are compact
  operators } \ch^2_{Y/Z}\to\ch^{-2}_{X/Z} \text{ for all } X,Y,Z.
\end{equation}
We consider some simple examples of operators $I_{XY}^Z$ to clarify
the difference with respect to the $N$-body situation (see
\S\ref{ss:scs} for details and generalizations). If $E,F$ are
Euclidean spaces we denote
\begin{equation}\label{eq:ikef}
\rk^2_{FE} =K(\ch^2_E,\ch_F) \quad \text{and} \quad
\rk^{2}_E=\rk^2_{E,E}=K(\ch^2_E,\ch_E). 
\end{equation}
Denote $X \boxplus Y = X/Y\oplus Y/X$ and embed $L^2(X \boxplus
Y)\subset \rk_{X/Y,Y/X}$ by identifying a Hilbert-Schmidt operator
with its kernel. Then
\begin{equation*}
L^2(X \boxplus Y;\rk^2_E) \subset \rk^2_E\otimes \rk_{X/Y,Y/X}
\subset \rk^2_{X/Z,Y/Z}.
\end{equation*}
Thus $I_{XY}^Z \in L^2(X \boxplus Y;\rk^2_E)$ is a simple example of
operator satisfying the first part of condition \eqref{eq:3d}. Such
an $I_{XY}^Z$ acts as follows: if $u\in\ch^2_{Y/Z}\subset
L^2(Y/X;\ch^2_E)$ then
\[
I_{XY}^Z u\in \ch_{X/Z} = L^2(X/Y;\ch_E) \quad\text{is given by}\quad
(I_{XY}^Z u)(x')={\textstyle\int_{Y/X}} I_{XY}^Z(x',y')u(y') \rmd y'.
\]
Now we consider \eqref{eq:2d}.
Since $(x',y')\mapsto[D_E,I^Z_{XY}(x',y')]$ is the kernel of the
operator $[D_E,I^Z_{XY}]$, if
\[
[D_E,I^Z_{XY}]\in L^2(X \boxplus Y;K(\ch^2_E,\ch^{-2}_E)
\]
then $[D_E,I^Z_{XY}]$ is a compact operator
$\ch^2_{Y/Z}\to\ch^{-2}_{X/Z}$. For the term $D_{X/Y}I^Z_{XY}$ it
suffices to require the compactness of the operator
\[
D_{X/Y}I^Z_{XY} = 1_E\otimes D_{X/Y} \cdot I^Z_{XY}: \ch^2_{Y/Z}\to
\ch_E\otimes\ch^{-2}_{X/Y}.
\]
From \eqref{eq:fed} we see that this is a condition on the kernel
$x'\cdot\nabla_{x'}I_{XY}^Z(x',y')$. For example, it suffices that
the operator $\jap{Q_{X/Y}}I^Z_{XY}:\ch^2_{Y/Z}\to\ch_{X/Z}$ be
compact, which is a short range assumption. In summary:

\begin{example}\label{ex:hschmidt}
  For each $Z\subset X\cap Y$ let $I^Z_{XY} \in L^2(X \boxplus
  Y;\rk^2_E)$ such that the adjoint of $I^Z_{XY}(x',y')$ is an
  extension of $I^Z_{YX}(y',x')$. Assume that kernel
  $[D_E,I^Z_{XY}(x',y')]$ belongs to $L^2(X \boxplus
  Y;K(\ch^2_E,\ch^{-2}_E)$ and that the kernel
  $x'\cdot\nabla_{x'}I_{XY}^Z(x',y')$ defines a compact operator
  $\ch^2_{Y/Z}\to\ch^{-2}_{X/Z}$. Then \eqref{eq:3d} is fulfilled.
\end{example}

\begin{example}\label{ex:anc}
  Here we consider the particular case $Y\subset X$ to see the
  structure of a generalized creation operator which appears in this
  context. For each $Z\subset Y$ let $I^Z_{XY}\in
  \rk^2_{Y/Z}\otimes\ch_{X/Y}$, where the tensor product is a kind
  of weak version of $L^2(X/Y; \rk^2_{Y/Z})$ discussed in
  \S\ref{ss:ha}. Furthermore, assume that $[D_{Y/Z},I^Z_{XY}]\in
  K(\ch^2(Y/Z),\ch^{-2}_{Y/Z})\otimes\ch_{X/Y}$ and
  $D_{X/Y}I^Z_{XY}\in \rk^2_{Y/Z}\otimes\ch^{-2}_{X/Y}$. Then
  \eqref{eq:3d} holds.
\end{example}

\subsection{Boundary values of the resolvent}
\label{ss:bvr}

Theorem \ref{th:thrintr} has important consequences in the spectral
theory of the operator $H$: we shall use it together with
\cite[Theorem 7.4.1]{ABG} to show that $H$ has no singular
continuous spectrum and to prove the existence of the boundary
values of its resolvent in the class of weighted $L^2$ spaces that
we define now. Let $\ch_{s,p}=\oplus_X L^2_{s,p}(X)$ where the
$L^2_{s,p}(X)$ are the Besov spaces associated to the position
observable on $X$ (these are obtained from the usual Besov spaces
associated to $L^2(X)$ by a Fourier transformation).  Note that
$\ch_{s}=\ch_{s,2}$ is the Fourier transform of the Sobolev space
$\ch^s$.  Let $\mbc_+$ be the open upper half plane and
$\mbc^H_+=\mbc_+\cup(\mbr\setminus\tau(H))$. If we replace the upper
half plane by the lower one we similarly get the sets $\mbc_-$ and
$\mbc^H_-$. We define two holomorphic maps $R_\pm:\mbc_\pm\to
L(\ch)$ by $R_\pm(z)=(H-z)^{-1}$ and note that we have continuous
embeddings
\[
L(\ch)\subset L(\ch_{1/2,1},\ch_{-1/2,\infty}) \subset
L(\ch_{s},\ch_{-s}) \quad\text{if } s>1/2
\]
so we may consider $R_\pm$ as maps with values in
$L(\ch_{1/2,1},\ch_{-1/2,\infty})$.

\begin{theorem}\label{th:c11}
If $H$ is of class $C^{1,1}(D)$ then its singular continuous
spectrum is empty and the holomorphic maps $R_\pm:\mbc_\pm\to
L(\ch_{1/2,1},\ch_{-1/2,\infty})$ extend to weak$^*$ continuous
functions $\bar{R}_\pm$ on $\mbc^H_\pm$.  The maps
$\bar{R}_\pm:\mbc^H_\pm\to L(\ch_{s},\ch_{-s})$ are norm continuous
if $s>1/2$.
\end{theorem}

This result is optimal both with regard to the regularity of the
Hamiltonian relatively to the conjugate operator $D$ and to the
Besov spaces in which we establish the existence of the boundary
values of the resolvent. The class $C^{1,1}(D)$ will be discussed
and its optimality will be made precise in \S\ref{ss:scsc} but we
give some examples below.

We state first the simplest sufficient condition: \emph{assume that
  $H$ is as in Proposition \ref{pr:exnr} and that its domain is
  equal to $\ch^2$;if $\,[D,[D,I^Z]]\in
  L(\ch^2_{\cs/Z},\ch^{-2}_{\cs/Z})$ for all $Z$ then $H$ is of
  class $C^{1,1}(D)$}. This follows from Theorem 6.3.4 in
\cite{ABG}. The condition on $\,[D,[D,I^Z]]$ can easily be written
in terms of the coefficients $I^Z_{XY}$ by arguments similar to
those of \S\ref{ss:examples}.  Refinements allow the addition of
long range and short range interactions as in \cite[\S 9.4.2]{ABG}.

Let $\xi:\mbr\to\mbr$ be of class $C^\infty$ and such that
$\xi(\lambda)=0$ if $\lambda\le 1$ and $\xi(\lambda)=1$ if
$\lambda\ge 2$. For each Euclidean space $X$ and real $r\ge 1$ we
denote $\xi^r_X$ the operator of multiplication by the function
$x\mapsto\xi(|x|/r)$ on any Sobolev space over $X$. Then we define
$\xi^r_\cs=\oplus_{X\in\cs}\xi^r_X$ considered as operator on
$\ch^s_\cs$ for any real $s$.

\begin{definition}\label{df:slr}
  Let $T:\ch^2_\cs\to\ch_\cs$ be a symmetric operator. We say that
  $T$ is a \emph{long range interaction} if $[D,T]\in
  L(\ch^2_\cs,\ch^{-1}_\cs)$ and $\int_1^\infty \|\xi^r_\cs
  [D,T]\|_{\ch^2_\cs\to\ch^{-1}_\cs} \rmd r/r <\infty$.  We say that
  $T$ is a \emph{short range interaction} if $\int_1^\infty
  \|\xi^r_\cs [D,T]\|_{\ch^2_\cs\to\ch_\cs} \rmd r <\infty$.
\end{definition}

\begin{theorem}\label{th:BVR}
  Assume that $H=\Delta_\cs + \sum_{Z\in\cs} 1_Z\otimes I^Z$ where
  each $I^z:\ch^2_{\cs/Z}\to\ch_{\cs/Z}$ is symmetric, compact, and
  is the sum of a long range and a short range interaction. Then $H$
  is a nonrelativistic many-body Hamiltonian of class $C^{1,1}(D)$,
  hence the conclusions of Theorem \ref{th:c11} are true.
\end{theorem}

Scattering channels may be defined in a natural way in the context
of the theorem.  If the long range interactions are absent we expect
that asymptotic completeness holds.

\section{Graded Hilbert $C^*$-modules}
\label{s:grad}
\protect\setcounter{equation}{0}

\subsection{Graded $\boldsymbol{C^*}$-algebras}
\label{ss:grca}

The natural framework for the systems considered in this paper is
that of $C^*$-algebras graded by semilattices. We refer to
\cite{Ma2,Ma3} for a detailed study of this class of algebras.

Let $\cs$ be a semilattice and $\ra$ a graded $C^*$-algebra.
Following \cite{Ma2} we say that $\rb\subset\ra$ is a \emph{graded
  $C^*$-subalgebra} if $\rb$ is a $C^*$-subalgebra of $\ra$ equal to
$\sum^\rmc_\sigma\rb\cap\ra(\sigma)$. Then $\rb$ has a natural
graded $C^*$-algebra structure: $\rb(\sigma)=\rb\cap\ra(\sigma)$.
If $\rb$ is also an ideal of $\ra$ then $\rb$ is a \emph{graded
  ideal}.

A subset $\ct$ of a semilattice $\cs$ is a \emph{sub-semilattice of
  $\cs$} if $\sigma,\tau\in\ct \Rightarrow
\sigma\wedge\tau\in\ct$. We say that $\ct$ is an \emph{ideal of
  $\cs$} if $\sigma\leq\tau\in\ct \Rightarrow \sigma\in\ct$. If
$\ra$ is an $\cs$-graded $C^*$-algebra and $\ct\subset\cs$ let
$\ra(\ct)=\sum^\rmc_{\sigma\in\ct}\ra(\sigma)$ (if $\ct$ is finite
the sum is already closed).  If $\ct$ is a sub-semilattice or an
ideal then clearly $\ra(\ct)$ is a $C^*$-subalgebra or an ideal of
$\ra$ respectively.

We say that $\ra$ is \emph{supported by a sub-semilattice $\ct$} if
$\ra=\ra(\ct)$, i.e. $\ra(\sigma)=\{0\}$ for $\sigma\nin\ct$. Then
$\ra$ is also $\ct$-graded.  The smallest sub-semilattice with this
property will be called \emph{support of $\ra$}.  If $\ct$ is a
sub-semilattice of $\cs$ and $\ra$ is a $\ct$-graded algebra then
$\ra$ is $\cs$-graded: set $\ra(\sigma)=\{0\}$ for
$\sigma\in\cs\setminus\ct$.

The next result is obvious if $\cs$ is finite. For the general case,
see the proof of Proposition 3.3 in \cite{DG3}.

\begin{proposition}\label{pr:gsalg}
  Let $\ct$ be a sub-semilattice of $\cs$ such that
  $\ct'=\cs\setminus\ct$ is an ideal. Then $\ra(\ct)$ is a
  $C^*$-subalgebra of $\ra$, $\ra(\ct')$ is an ideal of $\ra$, and
  $\ra=\ra(\ct)+\ra(\ct')$ with $\ra(\ct)\cap\ra(\ct')=\{0\}$.
  In particular, the natural linear projection
  $\rp(\ct):\ra\to\ra(\ct)$ is a morphism.
\end{proposition}

If $\ct$ is a sub-semilattice then $\ct'$ is an ideal if and only if
$\ct$ is a filter
(i.e. $\sigma\ge\tau\in\ct\Rightarrow\sigma\in\ct$). Thus if $\cs$
is finite then the only sub-semilattices which have this property
are the $\cs_{\ge\sigma}$ introduced below.

The simplest sub-semilattices are the chains (totally ordered
subsets). If $\sigma\in \cs$ and
\begin{equation}\label{eq:Aa}
\cs_{\geq\sigma}=\{\tau\in \cs\mid \tau\geq\sigma\},
\quad%
\cs_{\not\geq\sigma}=
\cs'_{\geq\sigma}=\{\tau\in\cs\mid\tau\not\geq\sigma\},
\quad
\cs_{\leq\sigma}=\{\tau\in \cs\mid \tau\leq\sigma\}
\end{equation}
then $\cs_{\geq\sigma}$ is a sub-semilattice and
$\cs_{\not\geq\sigma}$ and $\cs_{\leq\sigma}$ are ideals. So
$\ra_{\ge\sigma}\equiv\ra(\cs_{\geq\sigma})$ is a graded
$C^*$-subalgebra of $\ra$ supported by $\cs_{\geq\sigma}$ and
$\ra(\cs_{\not\geq\sigma})$ is a graded ideal supported by
$\cs_{\not\geq\sigma}$ such that
\begin{equation}\label{eq:dsum}
\ra=\ra_{\ge\sigma}+\ra(\cs_{\not\geq\sigma})
\quad\text{with} \quad
\ra_{\ge\sigma}\cap\ra(\cs_{\not\geq\sigma})=\{0\}. 
\end{equation}
The projection morphism $\rp_{\ge\sigma}:\ra\to\ra_{\geq\sigma}$
defined by \eqref{eq:dsum} is the unique linear continuous map
$\rp_{\geq\sigma}:\ra\rarrow\ra$ such that $\rp_{\geq\sigma}A=A$ if
$A\in\ra(\tau)$ for some $\tau\geq\sigma$ and $\rp_{\geq\sigma}A=0$
otherwise.

$\cs$ is called \emph{atomic} if it has a smallest element
$o\equiv\min \cs$ and if each $\sigma\neq o$ is minorated by an
atom.  We denote by $\cp(\cs)$ the set of atoms of $\cs$.  If $\ct$
is an ideal of $\cs$ and $\cs$ is atomic then $\ct$ is atomic, we
have $\min\ct=\min \cs$, and $\cp(\ct)=\cp(\cs)\cap\ct$.  This next
result is also easy to prove \cite{DG3}.

\begin{theorem}\label{th:ga}  
  If $\cs$ is atomic then $\rp
  A=(\rp_{\geq\alpha}A)_{\alpha\in\cp(\cs)}$ defines a morphism
  $\rp:\ra\to\prod_{\alpha\in\cp(\cs)}\ra_{\geq\alpha}$ with
  $\ra(o)$ as kernel. This gives us a canonical embedding
\begin{equation}\label{eq:quot}
\ra/\ra(o)\subset\pprod_{\alpha\in\cp(\cs)}\ra_{\geq\alpha}.
\end{equation}
\end{theorem}

We call this ``theorem'' because it has important consequences in
the spectral theory of many-body Hamiltonians: it allows us to
compute their essential spectrum and to prove the Mourre estimate.

We assume that $\cs$ is atomic so that $\ra$ comes equipped with a
remarkable ideal $\ra(o)$. Then for $A\in\ra$ we define its
\emph{essential spectrum} (relatively to $\ra(o)$) by the formula
\begin{equation}\label{eq:eso}
\spe(A)\equiv\sp(\rp A).
\end{equation}
In our concrete examples $\ra$ is represented on a Hilbert space
$\ch$ and $\ra(o)= K(\ch)$, so we get the usual Hilbertian notion of
essential spectrum. 

In order to extend this to unbounded operators it is convenient to
define an \emph{observable affiliated to $\ra$} as a morphism
$H:\Co(\mbr)\to\ra$.  We set $\varphi(H)\equiv H(\varphi)$.  If
$\ra$ is realized on  $\ch$ then a self-adjoint
operator on $\ch$ such that $(H+i)^{-1}\in\ra$ is said to be
affiliated to $\ra$; then $H(\varphi)=\varphi(H)$ defines an
observable affiliated to $\ra$ (see Appendix A in \cite{DG3} for a
precise description of the relation between observables and
self-adjoint operators affiliated to $\ra$). The spectrum of an
observable is by definition the support of the morphism $H$:
\begin{equation}\label{eq:sp}
\sp(H)=\{\lambda\in\mbr \mid
\varphi\in\Co(\mbr),\varphi(\lambda)\neq 0 \Rightarrow
\varphi(H)\neq0\}. 
\end{equation}
Now note that $\rp H\equiv\rp\circ H$ is an observable affiliated to
the quotient algebra $\ra/\ra(o)$ so we may define the essential
spectrum of $H$ as the spectrum of $\rp H$. Explicitly, we get:
\begin{equation}\label{eq:es1}
\spe(H)=\{\lambda\in\mbr \mid 
\varphi\in\Co(\mbr),\varphi(\lambda)\neq 0 \Rightarrow
\varphi(H)\notin \ra(o)\}. 
\end{equation}
Now the first assertion of the next theorem follows immediately from
Theorem \ref{th:ga}. For the second assertion, see the proof of
Theorem 2.10 in \cite{DG2}. By $\overline{\cup}$ we denote the
closure of the union.

\begin{theorem}\label{th:gas}
Let $\cs$ be atomic. If $H$ is an observable affiliated to $\ra$
then $H_{\geq\alpha}=\rp_{\geq\alpha}H$ is an observable affiliated
to $\ra_{\geq\alpha}$ and we have:
\begin{equation}\label{eq:es2}
\spe(H)=\overline{\ccup}_{\alpha\in\cp(\cs)}\sp(H_{\geq\alpha}).
\end{equation}
If for each $A\in\ra$ the set of $\rp_{\geq\alpha}A$ with
$\alpha\in\cp(\cs)$ is compact in $\ra$ then the union in
\eqref{eq:es2} is closed.
\end{theorem}

\subsection{Hilbert $C^*$-modules}
\label{ss:preh}

Some basic knowledge of the theory of Hilbert
$C^*$-modules is useful but not indispensable for understanding our
constructions. We translate here the necessary facts in a purely
Hilbert space language.  Our main reference for the general theory
of Hilbert $C^*$-modules is \cite{La} but see also \cite{Bl,RW}.
The examples of interest in this paper are the ``concrete'' Hilbert
$C^*$-modules described below as Hilbert $C^*$-submodules of
$L(\ce,\cf)$. We recall, however, the general definition.

If $\ra$ is a $C^*$-algebra then a \emph{Banach $\ra$-module} is a
Banach space $\mr$ equipped with a continuous bilinear map
$\ra\times\mr\ni(A,M)\mapsto MA\in\mr$ such that $(MA)B=M(AB)$.  We
denote $\rM\cdot\ra$ the clspan of the elements $MA$ with $A\in\ra$
and $M\in\rM$.  By the Cohen-Hewitt theorem \cite{FD} for each
$N\in\rM\cdot\ra$ there are $A\in\ra$ and $M\in\rM$ such that
$N=MA$, in particular $\mr\cdot\ra=\mr\ra$.  Note that by module we
mean ``right module'' but the Cohen-Hewitt theorem is also valid for
left Banach modules.

Let $\ra$ be a $C^*$-algebra. A (right) \emph{Hilbert $\ra$-module}
is a Banach $\ra$-module $\mr$ equipped with an $\ra$-valued
sesquilinear map
$\braket{\cdot}{\cdot}\equiv\braket{\cdot}{\cdot}_\ra$ which is
positive (i.e. $\braket{M}{M}\geq0$) $\ra$-sesquilinear
(i.e. $\braket{M}{NA}=\braket{M}{N}A$) and such that
$\|M\|\equiv\|\braket{M}{M}\|^{1/2}$. Then $\mr=\mr\ra$.  The clspan
of the elements $\braket{M}{M}$ is an ideal of $\ra$ denoted
$\braket{\mr}{\mr}$. One says that $\mr$ is \emph{full} if
$\braket{\mr}{\mr}=\ra$.  If $\ra$ is an ideal of a $C^*$-algebra
$\rc$ then $\mr$ is equipped with an obvious structure of Hilbert
$\rc$-module. Left Hilbert $\ra$-modules are defined similarly. 

If $\rM,\rn$ are Hilbert $\ra$-modules and $(M,N)\in\mr\times\rn$
then $M'\mapsto N\braket{M}{M'}$ is a linear continuous map
$\mr\to\rn$ denoted $\ket{N}\bra{M}$ or $NM^*$.  The closed linear
subspace of $L(\mr,\rn)$ generated by these elements is denoted
$\ck(\mr,\rn)$. There is a unique antilinear isometric map $T\mapsto
T^*$ of $\ck(\mr,\rn)$ onto $\ck(\rn,\mr)$ which sends
$\ket{N}\bra{M}$ into $\ket{M}\bra{N}$. The space
$\ck(\rM)\equiv\ck(\rM,\rM)$ is a $C^*$-algebra called
\emph{imprimitivity algebra} of the Hilbert $\ra$-module $\rM$.

Assume that $\rn$ is a closed subspace of a Hilbert $\ra$-module $\mr$
and let $\braket{\rn}{\rn}$ be the clspan of the elements
$\braket{N}{N}$ in $\ra$. If $\rn$ is an $\ra$-submodule of $\mr$ then
it inherits an obvious Hilbert $\ra$-module structure from $\mr$.  If
$\rn$ is not an $\ra$-submodule of $\mr$ it may happen that there is a
$C^*$-subalgebra $\rb\subset\ra$ such that $\rn\rb\subset\rn$ and
$\braket{\rn}{\rn}\subset\rb$. Then clearly we get a Hilbert
$\rb$-module structure on $\rn$. On the other hand, it is clear that
such a $\rb$ exists if and only if $\rn\braket{\rn}{\rn}\subset\rn$
and then $\braket{\rn}{\rn}$ is a $C^*$-subalgebra of $\ra$. Under
these conditions we say that \emph{$\rn$ is a Hilbert $C^*$-submodule}
of the Hilbert $\ra$-module $\mr$. Then $\rn$ inherits a Hilbert
$\braket{\rn}{\rn}$-module structure and this defines the
$C^*$-algebra $\ck(\rn)$. Moreover, if $\rb$ is as above then
$\ck(\rn)=\ck_\rb(\rn)$.

If $\rn$ is a closed subspace of a Hilbert $\ra$-module $\rM$ then
let $\ck(\rn|\mr)$ be the closed subspace of $\ck(\mr)$ generated by
the elements $NN^*$ with $N\in\rn$. It is easy to prove that
\emph{if $\rn$ is a Hilbert $C^*$-submodule of $\mr$ then
$\ck(\rn|\mr)$ is a $C^*$-subalgebra of $\ck(\mr)$ and the map
$T\mapsto T|_\rn$ sends $\ck(\rn|\mr)$ onto $\ck(\rn)$ and is an
isomorphism of $C^*$-algebras}. Then we identify $\ck(\rn|\mr)$
with $\ck(\rn)$.

If $\ce,\cf$ are Hilbert spaces then we equip $L(\ce,\cf)$ with the
Hilbert $L(\ce)$-module structure defined as follows: the
$C^*$-algebra $L(\ce)$ acts to the right by composition and we take
$\braket{M}{N}=M^*N$ as inner product, where $M^*$ is the usual
adjoint of the operator $M$. Note that $L(\ce,\cf)$ is also equipped
with a natural left Hilbert $L(\cf)$-module structure: this time the
inner product is $MN^*$.

If $\mr\subset L(\ce,\cf)$ is a linear subspace then $\mr^*\subset
L(\cf,\ce)$ is the set of adjoint operators $M^*$ with $M\in\mr$.
Clearly $\mr_1\subset\mr_2\Rightarrow\mr_1^*\subset\mr_2^*$. If
$\cg$ is a third Hilbert spaces and $\rn\subset L(\cf,\cg)$ is a
linear subspace then $(\rn\cdot\mr)^*=\mr^*\cdot\rn^*$.  In
particular, if $\ce=\cf=\cg$, $\mr=\mr^*$, and $\rn=\rn^*$ then
$\mr\cdot\rn\subset\rn\cdot\mr$ is equivalent to
$\mr\cdot\rn=\rn\cdot\mr$.

Now let $\mr\subset L(\ce,\cf)$ be a closed linear subspace. Then
\emph{$\mr$ is a Hilbert $C^*$-submodule of $L(\ce,\cf)$ if and only
  if $\mr\mr^*\mr\subset\mr$}.

These are the ``concrete'' Hilbert $C^*$-modules we are interested
in. It is clear that $\mr^*$ will be a Hilbert $C^*$-submodule of
$L(\cf,\ce)$. We mention that $\mr^*$ is canonically identified with
the left Hilbert $\ra$-module $\ck(\mr,\ra)$ dual to $\mr$. 

\begin{proposition}\label{pr:ss}
Let $\ce,\cf$ be Hilbert spaces and let $\mr$ be a Hilbert
$C^*$-submodule of $L(\ce,\cf)$. Then $\ra\equiv\mr^*\cdot\mr$ and
$\rb\equiv\mr\cdot\mr^*$ are $C^*$-algebras of operators on $\ce$
and $\cf$ respectively and $\mr$ is equipped with a canonical
structure of $(\rb,\ra)$ imprimitivity bimodule.
\end{proposition}

For the needs of this paper the last assertion of the proposition
could be interpreted as a definition.

\begin{proposition}\label{pr:clsubmod}
Let $\rn$ be a $C^*$-submodule of $L(\ce,\cf)$ such that
$\rn\subset\mr$ and $\rn^*\cdot\rn=\mr^*\cdot\mr$,
$\rn\cdot\rn^*=\mr\cdot\mr^*$. Then $\rn=\mr$.
\end{proposition}
\proof If $M\in\mr$ and $N\in\rn$ then $MN^*\in\rb=\rn\cdot\rn^*$ and
$\rn\rn^*\rn\subset\rn$ hence $MN^*N\in\rn$. Since $\rn^*\cdot\rn=\ra$
we get $MA\in\rn$ for all $A\in\ra$. Let $A_i$ be an approximate
identity for the $C^*$-algebra $\ra$. Since one can factorize $M=M'A'$
with $M'\in\mr$ and $A'\in\ra$ the sequence $MA_i=M'A'A_i$ converges
to $ M'A'=M$ in norm. Thus $M\in\rn$.
\qed

\begin{proposition}\label{pr:2ss}
Let $\ce,\cf,\ch$ be Hilbert spaces and let $\mr\subset L(\ch,\ce)$
and $\rn\subset L(\ch,\cf)$ be Hilbert $C^*$-submodules. Let $\ra$ be
a $C^*$-algebra of operators on $\ch$ such that $\mr^*\cdot\mr$ and
$\rn^*\cdot\rn$ are ideals of $\ra$ and let us view $\mr$ and $\rn$ as
Hilbert $\ra$-modules. Then $\ck(\mr,\rn)\cong\rn\cdot\mr^*$ the
isometric isomorphism being determined by the condition
$\ket{N}\bra{M}=NM^*$. 
\end{proposition}

\subsection{Graded Hilbert $C^*$-modules}
\label{ss:gf}

This is due to Georges Skandalis \cite{Sk} (see also Remark
\ref{re:squant}).

\begin{definition}\label{df:grm}
Let $\cs$ be a semilattice and $\ra$ an $\cs$-graded
$C^*$-algebra. A Hilbert $\ra$-module $\mr$ is an \emph{$\cs$-graded
  Hilbert $\ra$-module} if a linearly independent family
$\{\mr(\sigma)\}_{\sigma\in \cs}$ of closed subspaces of $\mr$ is
given such that $\sum_\sigma\mr(\sigma)$ is dense in $\mr$ and:
\begin{equation}\label{eq:grm}
\mr(\sigma)\ra(\tau)\subset\mr(\sigma\wedge\tau) 
\hspace{2mm}\text{and}\hspace{2mm}
\braket{\mr(\sigma)}{\mr(\tau)}\subset\ra(\sigma\wedge\tau)
\hspace{2mm} \text{for all } \sigma,\tau\in \cs.
\end{equation}
\end{definition}
Note that $\ra$ equipped with its canonical Hilbert $\ra$-module
structure is an $\cs$-graded Hilbert \mbox{$\ra$-module}. 
\eqref{eq:grm} implies that each $\mr(\sigma)$ is a
Hilbert $\ra(\sigma)$-module and if $\sigma\leq\tau$ then
$\mr(\sigma)$ is an $\ra(\tau)$-module.

From \eqref{eq:grm} we also see that \emph{the imprimitivity algebra
  $\ck(\mr(\sigma))$ of the Hilbert $\ra(\sigma)$-module
  $\mr(\sigma)$ is naturally identified with the clspan in
  $\ck(\mr)$ of the elements $MM^*$ with $M\in\mr(\sigma)$}. Thus
$\ck(\mr(\sigma))$ is identified with a $C^*$-subalgebra of
$\ck(\mr)$. We use this identification below.

\begin{theorem}\label{th:kghm}
If $\mr$ is a graded Hilbert $\ra$-module then $\ck(\mr)$ becomes a
graded $C^*$-algebra if we define
$\ck(\mr)(\sigma)=\ck(\mr(\sigma))$. If $M\in\mr(\sigma)$ and
$N\in\mr(\tau)$ then there are elements $M'$ and $N'$ in
$\mr(\sigma\wedge\tau)$ such that $MN^*=M'N'^*$;
in particular $MN^*\in\ck(\mr)(\sigma\wedge\tau)$.
\end{theorem}
\proof As explained before, $\ck(\mr)(\sigma)$ are $C^*$-subalgebras
of $\ck(\mr)$. To show that they are linearly independent, let
$T(\sigma)\in\ck(\mr)(\sigma)$ such that $T(\sigma)=0$ but for a
finite number of $\sigma$ and assume $\sum_\sigma T(\sigma)=0$. Then
for each $M\in\mr$ we have $\sum_\sigma T(\sigma)M=0$. Note that the
range of $T(\sigma)$ is included in $\mr(\sigma)$. Since the linear
spaces $\mr(\sigma)$ are linearly independent we get $T(\sigma)M=0$
for all $\sigma$ and $M$ hence $T(\sigma)=0$ for all $\sigma$.

We now prove the second assertion of the proposition. Since
$\mr(\sigma)$ is a Hilbert $\ra(\sigma)$-module there are
$M_1\in\mr(\sigma)$ and $S\in\ra(\sigma)$ such that $M=M_1S$, cf. the
Cohen-Hewitt theorem or Lemma 4.4 in \cite{La}. Similarly, $N=N_1T$
with $N_1\in\mr(\tau)$ and $T\in\ra(\tau)$. Then $MN^*=M_1(S
T^*)N_1^*$ and $S T^*\in \ra(\sigma\wedge\tau)$ so we may factorize it
as $S T^*=UV^*$ with $U,V\in \ra(\sigma\wedge\tau)$, hence
$MN^*=(M_1U)(N_1V)^*$.  By using \eqref{eq:grm} we see that $M'=M_1U$
and $N'=N_1V$ belong to $\mr(\sigma\wedge\tau)$.  In particular, we
have $MN^*\in\ck(\mr)(\sigma\wedge\tau)$ if $M\in\mr(\sigma)$ and
$N\in\mr(\tau)$.

Observe that the assertion we just proved implies that
$\sum_\sigma\ck(\mr)(\sigma)$ is dense in $\ck(\mr)$.  
It remains to see that
$\ck(\mr)(\sigma)\ck(\mr)(\tau)\subset\ck(\mr)(\sigma\wedge\tau)$.
For this it suffices that $M\braket{M}{N}N^*$ be in
$\ck(\mr)(\sigma\wedge\tau)$ if $M\in\mr(\sigma)$ and
$N\in\mr(\tau)$. Since $\braket{M}{N}\in\ra(\sigma\wedge\tau)$ we
may write $\braket{M}{N}=S T^*$ with $S,T\in\ra(\sigma\wedge\tau)$
so $M\braket{M}{N}N^*=(MS)(NT)^*\in\ck(\mr)(\sigma\wedge\tau)$ by
\eqref{eq:grm}. 
\qed

We recall that the direct sum of a family $\{\rM_i\}$ of Hilbert
$\ra$-modules is defined as follows: $\oplus_i\rM_i$ is the space of
elements $(M_i)_i\in\prod_i\rM_i$ such that the series
$\sum_i\braket{M_i}{M_i}$ converges in $\ra$ equipped with the natural
$\ra$-module structure and with the $\ra$-valued inner product defined
by
\begin{equation}\label{eq:sum}
\braket{(M_i)_i}{(N_i)_i} 
=\textstyle\sum_i\braket{M_i}{N_i}.
\end{equation} 
The algebraic direct sum of the $\ra$-modules $\rM_i$ is dense in
$\oplus_i\rM_i$.

It is easy to check that if each $\mr_i$ is graded and if we set
$\mr(\sigma)=\oplus_i\mr_i(\sigma)$ then $\mr$ becomes a graded
Hilbert $\ra$-module.  For example, if $\rn$ is a graded Hilbert
$\ra$-module then $\rn\oplus\ra$ is a graded Hilbert $\ra$-module and
so the \emph{linking algebra $\ck(\rn\oplus\ra)$ is equipped with a
graded algebra structure}. We recall \cite[p. 50-52]{RW} that we
have a natural identification
\begin{equation}\label{eq:link}
\ck(\rn\oplus\ra)=
\begin{pmatrix}
\ck(\rn)& \rn\\
\rn^*& \ra
\end{pmatrix}
\end{equation}
and by Theorem \ref{th:kghm} this is a graded algebra whose
$\sigma$-component is equal to
\begin{equation}\label{eq:links}
\ck(\rn(\sigma)\oplus\ra(\sigma))=
\begin{pmatrix}
\ck(\rn(\sigma))& \rn(\sigma)\\
\rn(\sigma)^*& \ra(\sigma)
\end{pmatrix}.
\end{equation}
If $\rn$ is a $C^*$-submodule of $L(\ce,\cf)$ and if we set
$\rn^*\cdot\rn=\ra,\rn\cdot\rn^*=\rb$ then the linking algebra
$\begin{pmatrix}\rb& \mr\\\mr^*& \ra\end{pmatrix}$ of $\mr$ is a
$C^*$-algebra of operators on $\cf\oplus\ce$.

Some of the graded Hilbert $C^*$-modules which we shall use later on
will be constructed as follows.

\begin{proposition}\label{pr:rhm}
Let $\ce,\cf$ be Hilbert spaces and let $\mr\subset L(\ce,\cf)$ be a
Hilbert $C^*$-submodule, so that $\ra\equiv\mr^*\cdot\mr\subset
L(\ce)$ is a $C^*$-algebra and $\mr$ is a full Hilbert
$\ra$-module. Let $\cc$ be a $C^*$-algebra of operators on $\ce$
graded by the family of $C^*$-subalgebras
$\{\cc(\sigma)\}_{\sigma\in\cs}$. Assume that we have
\begin{equation}\label{eq:tas}
\ra\cdot\cc(\sigma)=\cc(\sigma)\cdot\ra
\equiv\rc(\sigma)
\hspace{2mm} \text{for all } \sigma\in\cs
\end{equation}
and that the family $\{\rc(\sigma)\}$ of subspaces of $L(\cf)$ is
linearly independent. Then the $\rc(\sigma)$ are $C^*$-algebras of
operators on $\ce$ and $\rc=\sum^\rmc_\sigma\rc(\sigma)$ is a
$C^*$-algebra graded by the family $\{\rc(\sigma)\}$. If
$\rn(\sigma)\equiv\mr\cdot\cc(\sigma)$ then
$\rn=\sum^\rmc_\sigma\rn(\sigma)$ is a full Hilbert $\rc$-module
graded by $\{\rn(\sigma)\}$.
\end{proposition}
\proof
We have
$$
\rc(\sigma)\cdot\rc(\tau)=\ra\cdot\cc(\sigma)\cdot\ra\cdot\cc(\tau)
=\ra\cdot\ra\cdot\cc(\sigma)\cdot\cc(\tau)\subset
\ra\cdot\cc(\sigma\wedge\tau)=\rc(\sigma\wedge\tau).
$$ 
This proves that the $\rc(\sigma)$ are $C^*$-algebras and that
$\rc$ is $\cs$-graded. Then:
$$
\rn(\sigma)\cdot\rc(\tau)=\mr\cdot\cc(\sigma)\cdot\cc(\tau)\cdot\ra
\subset\mr\cdot\cc(\sigma\wedge\tau)\cdot\ra=
\mr\cdot\ra\cdot\cc(\sigma\wedge\tau)=
\mr\cdot\cc(\sigma\wedge\tau)=\rn(\sigma\wedge\tau)
$$
and
$$
\rn(\sigma)^*\cdot\rn(\tau)=
\cc(\sigma)\cdot\mr^*\cdot\mr\cdot\cc(\tau)=
\cc(\sigma)\cdot\ra\cdot\cc(\tau)=
\ra\cdot\cc(\sigma)\cdot\cc(\tau)\subset
\ra\cdot\cc(\sigma\wedge\tau)=\rc(\sigma\wedge\tau).
$$
Observe that this computation also gives
$\rn(\sigma)^*\cdot\rn(\sigma)=\rc(\sigma)$.  Then
$$
\big({\textstyle\sum_\sigma}\rn(\sigma)^*\big)
\big({\textstyle\sum_\sigma}\rn(\sigma)\big)=
{\textstyle\sum_{\sigma,\tau}}\rn(\sigma)^*\rn(\tau)\subset
{\textstyle\sum_{\sigma,\tau}}\rc(\sigma\wedge\tau)\subset
{\textstyle\sum_{\sigma}}\rc(\sigma)
$$ and by the preceding remark we get $\rn^*\cdot\rn=\rc$ so $\rn$
is a full Hilbert $\rc$-module. To show the grading property it
suffices to prove that the family of subspaces $\rn(\sigma)$ is
linearly independent. Assume that $\sum N(\sigma)=0$ with
$N(\sigma)\in\rn(\sigma)$ and $N(\sigma)=0$ for all but a finite
number of $\sigma$. Assuming that there are non-zero elements in
this sum, let $\tau$ be a maximal element of the set of $\sigma$
such that $N(\sigma)\neq0$. From
$\sum_{\sigma_1,\sigma_2}N(\sigma_1)^*N(\sigma_2)=0$ and since
$N(\sigma_1)^*N(\sigma_2)\in\rc(\sigma_1\wedge\sigma_2)$ we get
$\sum_{\sigma_1\wedge\sigma_2=\sigma}N(\sigma_1)^*N(\sigma_2)=0$ for
each $\sigma$. Take here $\sigma=\tau$ and observe that if
$\sigma_1\wedge\sigma_2=\tau$ and $\sigma_1>\tau$ or $\sigma_2>\tau$
then $N(\sigma_1)^*N(\sigma_2)=0$. Thus $N(\tau)^*N(\tau)=0$ so
$N(\tau)=0$. But this contradicts the choice of $\tau$, so
$N(\sigma)=0$ for all $\sigma$.  \qed

\subsection{Tensor products}
\label{ss:ha}

In this subsection we collect some facts concerning tensor products
which are useful in what follows.  We recall the definition of the
tensor product of a Hilbert space $\ce$ and a \mbox{$C^*$-algebra}
$\ra$ in the category of Hilbert $C^*$-modules, cf. \cite{La}. We
equip the algebraic tensor product $\ce\odot\ra$ with the obvious
right $\ra$-module structure and with the $\ra$-valued sesquilinear
map given by
\begin{equation}\label{eq:hoa}
\braket{{\textstyle\sum_{u\in\ce}}u\otimes
  A_u}{{\textstyle\sum_{v\in\ce}}v\otimes B_v} 
={\textstyle\sum_{u,v}}\braket{u}{v}A^*_uB_v
\end{equation}
where $A_u=B_u=0$ outside a finite set. Then the completion of
$\ce\odot\ra$ for the norm $\|M\|:=\|\braket{M}{M}\|^{1/2}$ is a
full Hilbert $\ra$-module denoted $\ce\otimes\ra$. Clearly its
imprimitivity algebra is
\begin{equation}\label{eq:cha}
\ck(\ce\otimes\ra)=K(\ce)\otimes\ra.
\end{equation}  
If $\ra$ is $\cs$-graded then $\ce\otimes\ra$ is equipped with an
obvious structure of $\cs$-graded Hilbert $\ra$-module.

If $\ra$ is realized on a Hilbert space $\cf$ then one has a natural
isometric embedding $\ce\otimes\ra \subset L(\cf,\ce\otimes\cf)$.
Indeed, there is a unique linear map $\ce\otimes\ra\to
L(\cf,\ce\otimes\cf)$ which associates to $u\otimes A$ the function
$f\mapsto u\otimes(Af)$ and due to \eqref{eq:hoa} this map is an
isometry.  Thus the Hilbert $\ra$-module $\ce\otimes\ra$ is realized
as a Hilbert $C^*$-submodule of $L(\cf,\ce\otimes\cf)$, the dual
module is realized as the set of adjoint operators
$(\ce\otimes\ra)^*\subset L(\ce\otimes\cf,\ce)$, and one clearly has
\begin{equation}\label{eq:etens}
(\ce\otimes\ra)^*\cdot (\ce\otimes\ra)=\ra, \hspace{2mm}
(\ce\otimes\ra)\cdot(\ce\otimes\ra)^*=K(\ce)\otimes\ra.
\end{equation}

If $X$ is a locally compact space equipped with a Radon measure then
$L^2(X)\otimes\ra$ is the completion of $\Cc(X;\ra)$ for the norm
$\|\int_X F(x)^*F(x) \rmd x\|^{1/2}$.  Note that $L^2(X;\ra)\subset
L^2(X)\otimes\ra$ strictly in general, cf.\ the example below.  If
$\ra\subset L(\cf)$ then the norm on $L^2(X)\otimes\ra$ is
\begin{equation}\label{eq:L2a}
\|{\textstyle\int_X} F(x)^*F(x) \rmd x\|^2=
{\textstyle\sup_{f\in\cf,\|f\|=1}} {\textstyle\int_X} 
\|F(x)f\|^2 \rmd x.
\end{equation}
If $Y$ is a locally compact space then
$\ce\otimes\Co(Y)\cong\Co(Y;\ce)$.  Hence $L^2(X)\otimes\Co(Y)$ is
the completion of $\Cc(X\times Y)$ for the norm $\sup_{y\in
  Y}(\int_X |F(x,y)|^2 \rmd x)^{1/2}$. Assume that $X=Y$ is a
locally compact abelian group and let $f\in L^\infty(X)$ with
compact support and $g\in L^2(X)$. It is easy to check that
$F(x,y)=f(x)g(x+y)$ is an element of
$\Co(X;L^2(X))=L^2(X)\otimes\Co(X)$ but if $F(x,\cdot)=f(x)U_x g$ is
not zero then it does not belong to $\Co(X)$ and is not even a bounded
function if $g$ is not. Thus the elements of $L^2(X)\otimes\ra$ can
not be realized as bounded operator valued (equivalence classes of)
functions on $X$.

More generally, if $\cf'$, $\cf''$ are Hilbert spaces and
$\mr\subset L(\cf',\cf'')$ is a closed subspace then we define
$L^2(X)\otimes\mr$ as the completion of the space $\Cc(X;\mr)$ for a
norm similar to \eqref{eq:L2a}. We clearly have
$L^2(X)\otimes\mr\subset L(\cf',L^2(X)\otimes\cf'')$ isometrically
and $L^2(X;\mr)\subset L^2(X)\otimes\mr$ continuously.

If $\ce,\cf,\cg,\ch$ are Hilbert spaces and $\mr\subset L(\ce,\cf)$
and $\rn\subset L(\cg,\ch)$ are closed linear subspaces then we
denote $\mr\otimes\rn$ the closure in
$L(\ce\otimes\cg,\cf\otimes\ch)$ of the algebraic tensor product of
$\mr$ and $\rn$. Now suppose that $\mr$ is a $C^*$-submodule of
$L(\ce,\cf)$ and that $\rn$ is a $C^*$-submodule of $L(\cg,\ch)$ and
let $\ra=\mr^*\cdot\mr$ and $\rb=\rn^*\cdot\rn$. Then $\mr$ is a
Hilbert $\ra$-module and $\rn$ is a Hilbert $\rb$-module hence the
exterior tensor product, denoted temporarily
$\mr\otimes_{\text{ext}}\rn$, is well defined in the category of
Hilbert $C^*$-modules \cite{La} and is a Hilbert
$\ra\otimes\rb$-module. On the other hand, it is easy to check that
$(\mr\otimes\rn)^*=\mr^*\otimes\rn^*$ and then that $\mr\otimes\rn$
is a Hilbert $C^*$-submodule of $L(\ce\otimes\cg,\cf\otimes\ch)$
such that
$(\mr\otimes\rn)^*\cdot(\mr\otimes\rn)=\ra\otimes\rb$. Finally, it
is clear that $L(\ce\otimes\cg,\cf\otimes\ch)$ and
$\mr\otimes_{\text{ext}}\rn$ induce the same $\ra\otimes\rb$-valued
inner product on the algebraic tensor product of $\mr$ and $\rn$.
Thus we we get a canonical isometric isomorphism
$\mr\otimes_{\text{ext}}\rn=\mr\otimes\rn$.

As an application we give now an abstract version of the "toy
models" described in Example \ref{ex:fried}. Let $\ce,\cf$ be
Hilbert spaces and let us define $\ch=(\ce\otimes\cf)\oplus\cf$.
Let $\ra$ and $\rb$ be $C^*$-algebras of operators on $\cf$ and
$\ce\otimes\cf$ respectively. We embed $\ce\otimes\ra \subset
L(\cf,\ce\otimes\cf)$ as above. We simplify notation and denote
$\ce^*\otimes\ra:=(\ce\otimes\ra)^*\subset L(\ce\otimes\cf,\cf)$ the
dual module.

\begin{proposition}\label{pr:toymodel}
  Let $\cs$ be a semilattice and $\ct$ an ideal of $\cs$. Assume
  that the $C^*$-algebras $\ra$ and $\rb$ are $\cs$-graded and that
  we have $\ra(\sigma)=\{0\}$ if $\sigma\nin\ct$ and $\rb(\tau)=
  K(\ce)\otimes \ra(\tau)$ for $\tau\in\ct$.  Then
\begin{equation}\label{e:toy}
\rc=
\begin{pmatrix}
\rb & \ce\otimes\ra\\
\ce^*\otimes\ra & \ra
\end{pmatrix}.
\end{equation}
is an $\cs$-graded $C^*$-algebra if we define its components as
follows:
\begin{equation}\label{e:gtoy}
\rc(\sigma)=
\begin{pmatrix}
\rb(\sigma) & \ce\otimes\ra(\sigma)\\
\ce^*\otimes\ra(\sigma) & \ra(\sigma)
\end{pmatrix} \quad \text{for all} \quad \sigma\in\cs.
\end{equation}
\end{proposition}
\proof
Observe that if we set $\ct'=\cs\setminus\ct$ then
\begin{equation}\label{e:imptoy}
\rc=
\begin{pmatrix}
K(\ce)\otimes\ra & \ce\otimes\ra\\
\ce^*\otimes\ra & \ra
\end{pmatrix} +
\begin{pmatrix}
\rb(\ct') &  0 \\
0 & 0
\end{pmatrix} =
\ck(\rn\oplus\ra) +
\begin{pmatrix}
\rb(\ct') &  0 \\
0 & 0
\end{pmatrix}
\end{equation}
where $\rn=\ce\otimes\ra$ is an $\cs$-graded Hilbert $\ra$-module,
cf. \eqref{eq:link} and \eqref{eq:cha}. It is easy to see that the
family $\{\rc(\sigma)\}$ is linearly independent and that $\rc$ is
the closure of its sum. By taking into account \eqref{eq:links} we
see that it suffices to show that
$\rc(\sigma)\rc(\tau)\subset\rc(\sigma\wedge\tau)$ if
$\sigma\in\ct'$ and $\tau\in\ct$. After computing the coefficients
of the matrices we see that it suffices to check that
$\rb(\sigma)\cdot \ce\otimes\ra(\tau)
\subset\ce\otimes\ra(\sigma\wedge\tau)$. 
But:
\begin{align*}
\rb(\sigma)\cdot\ce\otimes\ra(\tau) & =
\rb(\sigma)\cdot K(\ce)\otimes\ra(\tau)\cdot \ce\otimes\ra(\tau) = 
\rb(\sigma)\cdot\rb(\tau)\cdot \ce\otimes\ra(\tau) \\
& \subset \rb(\sigma\wedge\tau)\cdot \ce\otimes\ra(\tau)
=K(\ce)\otimes\ra(\sigma\wedge\tau)\cdot \ce\otimes\ra(\tau)
\subset \ce\otimes\ra(\sigma\wedge\tau)
\end{align*}
which finishes the proof.
\qed

The extension to an increasing family of ideals
$\ct_1\subset\ct_2\dots\subset\cs$ is straightforward.

\section{The many-body $C^*$-algebra }
\label{s:grass}
\protect\setcounter{equation}{0}

In this section we introduce the many-body $C^*$-algebra and
describe its main properties (in particular, we prove the theorems
\ref{th:C} and \ref{th:CG}). Subsection \ref{ss:hilbert} contains
some preparatory material on concrete realizations of Hilbert
$C^*$-modules which implement the Morita equivalence between some
crossed products.

\subsection{Notations}
\label{ss:group}

Let $X$ be a locally compact abelian group with operation denoted
additively equipped with a Haar measures $\rmd x$. We abbreviate
this by saying that \emph{$X$ is an lca group}.  We set $\rl_X\equiv
L(L^2(X))$ and $\rk_X\equiv K(L^2(X))$ and note that these are
$C^*$-algebras independent of the choice of the measure on $X$.  If
$Y$ is a second lca group we shall use the abbreviations
\begin{equation}\label{eq:lkxy}
\rl_{XY}=L(L^2(Y),L^2(X)) \quad\text{and}\quad
\rk_{XY}=K(L^2(Y),L^2(X)). 
\end{equation}
We denote by $\varphi(Q)$ the operator in $L^2(X)$ of multiplication
by a function $\varphi$ and if $X$ has to be explicitly specified we
set $Q=Q_X$.  The bounded uniformly continuous functions on $X$ form
a $C^*$-algebra $\Cbu(X)$ which contains the algebras $\Cc(X)$ and
$\Co(X)$.  The map $\varphi\mapsto\varphi(Q)$ is an embedding
$\Cbu(X)\subset\rl_X$.

The group $\cc^*$-algebra $\rt_X$ of $X$ is the closed linear
subspace of $\rl_X$ generated by the convolution operators of the
form $(\varphi*f)(x)=\int_X \varphi(x-y)f(y)\rmd y$ with
$\varphi\in\Cc(X)$.  Observe that $f\mapsto\varphi*f$ is equal to
$\int_X\varphi(-a)U_a\,\rmd a$ where $U_a$ is the unitary
translation operator on $L^2(X)$ defined by $(U_af)(x)=f(x+a)$.

Let $X^*$ be the group dual to $X$ with operation denoted
additively\symbolfootnote[2]{\ Then $(k+p)(x)=k(x)p(x)$, $0(x)=1$,
  and the element $-k$ of $X^*$ represents the function
  $\bar{k}$. In order to avoid such strange looking expressions one
  might use the notation $k(x)=[x,k]$.  }.  If $k\in X^*$ we define
a unitary operator $V_k$ on $L^2(X)$ by $(V_ku)(x)=k(x) u(x)$.  The
Fourier transform of an integrable measure $\mu$ on $X$ is defined
by $(F\mu)(k)=\int \bar{k}(x)\mu(\rmd x)$.  Then $F$ induces a
bijective map $L^2(X)\rarrow L^2(X^*)$ hence a canonical isomorphism
$S\mapsto F^{-1}S F$ of $\rl_{X^*}$ onto $\rl_X$.  If $\psi$ is a
function on $X^*$ we set $\psi(P)\equiv\psi(P_X)=F^{-1}M_\psi F$,
where $M_\psi=\psi(Q_{X^*})$ is the operator of multiplication by
$\psi$ on $L^2(X^*)$.  The map $\psi\mapsto\psi(P)$ gives an
isomorphism $\Co(X^*)\cong \rt_X$.

If $Y\subset X$ is a closed subgroup then $\pi_Y:X\to X/Y$ is the
canonical surjection.  We embed $\Cbu(X/Y)\subset\Cbu(X)$ with the
help of the injective morphism $\varphi\mapsto\varphi\circ\pi_Y$. So
$\Cbu(X/Y)$ is identified with the set of functions
$\varphi\in\Cbu(X)$ such that $\varphi(x+y)=\varphi(x)$ for all
$x\in X$ and $y\in Y$.

In particular, $\Co(X/Y)$ is identified with the set of continuous
functions $\varphi$ on $X$ such that $\varphi(x+y)=\varphi(x)$ for
all $x\in X$ and $y\in Y$ and such that for each $\varepsilon>0$
there is a compact $K\subset X$ such that $|\varphi(x)|<\varepsilon$
if $x\nin K+Y$. By $x/Y\to\infty$ we mean $\pi_Y(x)\to\infty$, so
the last condition is equivalent to $\varphi(x)\to0$ if
$x/Y\to\infty$.  For coherence with later notations we set
\begin{equation}\label{eq:coxy}
\cc_X(Y)=\Co(X/Y)
\end{equation}
Observe that to an element $y\in Y$ we may associate a translation
operator $U_y$ in $L^2(X)$ and another translation operator in
$L^2(Y)$. However, in order not to overcharge the writing we shall
denote the second operator also by $U_y$.  The restriction map
$k\mapsto k|_Y$ is a continuous surjective group morphism $X^*\to
Y^*$ with kernel equal to $Y^\perp=\{k\in X^*\mid
k(y)=1\hspace{1mm}\forall y\in Y\}$ which defines the canonical
identification $Y^*\cong X^*/Y^\perp$. We denote by the same symbol
$V_k$ the operator of multiplication by the character $k\in X^*$ in
$L^2(X)$ and by the character $k|_Y\in Y^*$ in $L^2(Y)$.

We shall write $X=Y\oplus Z$ if $X$ is the direct sum of the two
closed subgroups $Y,Z$ equipped with compatible Haar measures, in
the sense that $\rmd x=\rmd y\otimes \rmd z$.  Then
$L^2(X)=L^2(Y)\otimes L^2(Z)$ as Hilbert spaces and
$\rk_X=\rk_Y\otimes \rk_Z$ and $\cc_X(Y)=1\otimes\Co(Z)$ as
$C^*$-algebras.

Let $O=\{0\}$ be the trivial group equipped with the Haar measure of
total mass $1$. Then $L^2(O)=\mbc$.

\subsection{Crossed products}
\label{ss:nbcrp}

Let $X$ be a locally compact abelian group. A $C^*$-subalgebra
$\ca\subset\Cbu(X)$ stable under translations will be called
\emph{$X$-algebra}.  The \emph{crossed product of
  $\ca$ by the action of $X$} is an abstractly defined $C^*$-algebra
$\ca\rtimes X$ canonically identified with the $C^*$-algebra of
operators on $L^2(X)$ given by
\begin{equation}\label{eq:crp}
\ca\rtimes X\equiv\ca\cdot \rt_X=\rt_X\cdot\ca.
\end{equation}
Crossed products of the form $\cc_X(Y)\rtimes X$ where $Y$ is a
closed subgroup of $X$ play an important role in the many-body
problem. To simplify notations we set
\begin{equation}\label{eq:Cxy}
\rc_X(Y)=\cc_X(Y)\rtimes X=\cc_X(Y)\cdot\rt_X=
\rt_X\cdot\cc_X(Y).
\end{equation}
If $X=Y\oplus Z$ and if we identify $L^2(X)=L^2(Y)\otimes L^2(Z)$ then
$\rt_X=\rt_Y\otimes\rt_Z$ hence
\begin{equation}\label{eq:crxyz}
\rc_X(Y)=\rt_Y\otimes \rk_Z.
\end{equation}
A useful ``symmetric'' description of $\rc_X(Y)$ is contained in the
next lemma. Let $Y^{(2)}$ be the closed subgroup of $X^2\equiv X\oplus
X$ consisting of elements of the form $(y,y)$ with $y\in Y$.

\begin{lemma}\label{lm:sym}
$\rc_X(Y)$ is the closure of the
set of integral operators with kernels $\theta\in\Cc(X^2/Y^{(2)})$.
\end{lemma}
\proof Let $\rc$ be the norm closure of the set of integral
operators with kernels $\theta\in\Cbu(X^2)$ having the properties:
(1) $\theta(x+y,x'+y)=\theta(x,x')$ for all $x,x'\in X$ and $y\in
Y$; (2) $\supp\theta\subset K_\theta+Y$ for some compact
$K_\theta\subset X^2$. We show $\rc=\rc_X(Y)$.  Observe that the map
in $X^2$ defined by $(x,x')\mapsto(x-x',x')$ is a topological group
isomorphism with inverse $(x_1,x_2)\mapsto(x_1+x_2,x_2)$ and sends
the subgroup $Y^{(2)}$ onto the subgroup $\{0\}\oplus Y$. This map
induces an isomorphism $X^2/Y^{(2)}\simeq X\oplus(X/Y)$. Thus any
$\theta\in\Cc(X^2/Y^{(2)})$ is of the form
$\theta(x,x')=\wtilde\theta(x-x',x')$ for some
$\wtilde\theta\in\Cc(X\oplus(X/Y))$. Thus $\rc$ is the closure in
$\rl_X$ of the set of operators of the form
$(Tu)(x)=\int_X\wtilde\theta(x-x',x')u(x') \rmd x'$. Since we may
approximate $\wtilde\theta$ with linear combinations of functions of
the form $a\otimes b$ with $a\in\Cc(X), b\in\Cc(X/Y)$ we see that
$\rc$ is the clspan of the set of operators of the form
$(Tu)(x)=\int_X a(x-x')b(x')u(x') \rmd x'$. But this clspan is
$\rt_X\cdot\cc_X(Y)=\rc_X(Y)$.  \qed

\subsection{Compatible subgroups}
\label{ss:compat}

If $X,Y$ is an arbitrary pair of lca groups then $X\oplus Y$ is the
set $X\times Y$ equipped with the product topology and group
structure. If $X,Y$ are closed subgroups of an lca group $G$ and if
the map $Y\oplus Z\to Y+Z$ defined by $(y,z)\mapsto y+z$ is open, we
say that they are \emph{compatible subgroups of $G$}.  In this case
$Y+Z$ is a closed subgroup of $X$.

\begin{remark}\label{re:DP}{\rm
If $G$ is $\sigma$-compact then $X,Y$ are compatible if and only if
$X+Y$ is closed. Indeed, a continuous surjective morphism between
two locally compact $\sigma$-compact groups is open and a subgroup
$H$ of a locally compact group $G$ is closed if and only if $H$ is
locally compact for the induced topology, see Theorems 5.11 and 5.29
in \cite{HR}.  We thank Lo\"ic Dubois and Benoit Pausader for
enlightening discussions on this matter.
}\end{remark}

The importance of the compatibility condition in the context of
graded $C^*$-algebras has been pointed out in \cite[Lemma 6.1.1]{Ma}
and one may find there several descriptions of this condition (see
also Lemma 3.1 from \cite{Ma3}). We quote two of them. Let $X/Y$ be
the image of $X$ in $G/Y$ considered as a subgroup of $G/Y$ equipped
with the induced topology. The group $X/(X\cap Y)$ is equipped with
the locally compact quotient topology and we have a natural map
$X/(X\cap Y)\to X/Y$ which is a bijective continuous group morphism.
Then $X,Y$ are compatible if and only if the following equivalent
conditions are satisfied:
\begin{align}
& \text{the natural map} \hspace{2mm} X/(X\cap Y)\to X/Y \hspace{2mm}
\text{is a homeomorphism}, \label{eq:ma1} \\
&
\text{the natural map }
G/(X\cap Y)\to G/X\times G/Y \hspace{1mm} \text{is closed}.
\label{eq:ma2}
\end{align}

If $\ca$ is a $G$-algebra let $\ca|_X$ be the set of restrictions
to $X$ of the functions from $\ca$. This is an $X$-algebra.

\begin{lemma}\label{lm:reg}
If $X,Y$ are compatible subgroups of $G$ then
\begin{align}
& \cc_G(X)\cdot\cc_G(Y) = \cc_G(X\cap Y)  \label{eq:reg1}\\
& \cc_G(Y)|_X =  \cc_X(X\cap Y).
\label{eq:reg2}
\end{align}
The second relation remains valid for the subalgebras $\Cc$.
\end{lemma}
\proof The fact that the inclusion $\subset$ in \eqref{eq:reg1} is
equivalent to the compatibility of $X$ and $Y$ is shown in Lemma
6.1.1 from \cite{Ma}, so we only have to prove that the equality
holds. Let $E = (G/X) \times (G/Y)$. If $\varphi \in \Co(G/X)$ and
$\psi \in \Co(G/Y)$ then $\varphi \otimes \psi$ denotes the function
$(s, t) \longmapsto \varphi(s) \psi(t)$, which belongs to
$\Co(E)$. The subspace generated by the functions of the form
$\varphi \otimes \psi$ is dense in $\Co(E)$ by the Stone-Weierstrass
theorem. If $F$ is a closed subset of $E$ then, by the Tietze
extension theorem, each function in $\Cc(F)$ extends to a function
in $\Cc(E)$, so the restrictions $(\varphi \otimes \psi)|_F$
generate a dense linear subspace of $\Co(F)$.  Let us denote by
$\pi$ the map $x \mapsto (\pi_X(x), \pi_Y(x))$, so $\pi$ is a group
morphism from $G$ to $E$ with kernel $V=X\cap Y$.  Then by
\eqref{eq:ma2} the range $F$ of $\pi$ is closed and the quotient map
$\wtilde\pi : G/V \to F$ is a continuous and closed bijection, hence
is a homeomorphism. So $\theta \mapsto \theta \circ \tilde \pi$ is
an isometric isomorphism of $\Co(F)$ onto $\Co(G/V)$. Hence for
$\varphi \in \Co(G/X)$ and $\psi \in \Co(G/Y)$ the function $\theta
= (\varphi \otimes \psi) \circ \tilde \pi$ belongs to $\Co(G/V)$, it
has the property $\theta \circ \pi_V = \varphi \circ \pi_X \cdot
\psi \circ \pi_Y$, and the functions of this form generate a dense
linear subspace of $\Co(G/V)$.

Now we prove \eqref{eq:reg2}. Recall that we identify $\cc_G(Y)$
with a subset of $\Cbu(G)$ by using $\varphi\mapsto\varphi\circ\pi_Y$
so in terms of $\varphi$ the restriction map which defines
$\cc_G(Y)|_X$ is just $\varphi\mapsto\varphi|_{X/Y}$. Thus we have a
canonical embedding $\cc_G(Y)|_X\subset\Cbu(X/Y)$ for an arbitrary
pair $X,Y$ . Then the continuous bijective group morphism
$\theta:X/(X\cap Y)\to X/Y$ allows us to embed
$\cc_G(Y)|_X\subset\Cbu(X/(X\cap Y))$. That the range of this map is
not $\cc_X(X\cap Y)$ in general is clear from the example $G=\mbr,
X=\pi\mbz,Y=\mbz$. But if $X,Y$ are compatible then $X/Y$ is closed
in $G/Y$, so $\cc_G(Y)|_X=\Co(X/Y)$ by the Tietze extension theorem,
and $\theta$ is a homeomorphism, hence we get \eqref{eq:reg2}.  \qed

\begin{lemma}\label{lm:double}
If $X,Y$ are compatible subgroups of $G$ then $X^2=X\oplus X$ and
$Y^{(2)}=\{(y,y)\mid y\in Y\}$ is a compatible pair of closed
subgroups of $G^2=G\oplus G$.
\end{lemma}
\proof Let $D=X^2\cap Y^{(2)}=\{(x,x)\mid x\in X\cap Y\}$. Due to to
\eqref{eq:ma1} it suffices to show that the natural map
$Y^{(2)}/D\to Y^{(2)}/X^2$ is a homeomorphism.  Here $Y^{(2)}/X^2$
is the image of $Y^{(2)}$ in $G^2/X^2\cong (G/X)\oplus(G/X)$, more
precisely it is the subset of pairs $(a,a)$ with $a=\pi_X(z)$ and
$z\in Y$, equipped with the topology induced by
$(G/X)\oplus(G/X)$. Thus the natural map $Y/X\to Y^{(2)}/X^2$ is a
homeomorphism. On the other hand, the natural map $Y/(X\cap Y)\to
Y^{(2)}/D$ is clearly a homeomorphism.  To finish the proof note
that $Y/(X\cap Y)\to Y/X$ is a homeomorphism because $X,Y$ is a
regular pair.  \qed

\begin{lemma}\label{lm:regp}
  Let $X,Y$ be compatible subgroups of an lca group $G$ and let
  $X^\perp,Y^\perp$ be their orthogonals in $G^*$.  Then $(X\cap
  Y)^\perp=X^\perp+Y^\perp$ and the closed subgroups
  $X^\perp,Y^\perp$ of $G^*$ are compatible.
\end{lemma}
\proof $X+Y$ is closed and, since
$(x,y)\mapsto(x,-y)$ is a homeomorphism, the map $S:X\oplus Y\to
X+Y$ defined by $S(x,y)=x+y$ is an open surjective morphism. Then
from the Theorem 9.5, Chapter 2 of \cite{Gu} it follows that the
adjoint map $S^*$ is a homeomorphism between $(X+Y)^*$ and its
range. In particular its range is a locally compact subgroup for the
topology induced by $X^*\oplus Y^*$ hence is a closed subgroup of
$X^*\oplus Y^*$, see Remark \ref{re:DP}.  We
have $(X+Y)^\perp=X^\perp\cap Y^\perp$, cf. 23.29 in \cite{HR}. Thus
from $X^*\cong G^*/X^\perp$ and similar representations for $Y^*$
and $(X+Y)^*$ we see that
$$
S^*:G^*/(X^\perp \cap Y^\perp)\to G^*/X^\perp\oplus G^*/Y^\perp
$$ is a closed map. But $S^*$ is clearly the natural map involved in
\eqref{eq:ma2}, hence the pair $X^\perp,Y^\perp$ is
regular. Finally, note that $(X\cap Y)^\perp$ is always equal to the
closure of the subgroup $X^\perp+Y^\perp$, cf. 23.29 and 24.10 in
\cite{HR}, and in our case $X^\perp+Y^\perp$ is closed.
\qed

\subsection{Green Hilbert $C^*$-modules} 
\label{ss:hilbert} 

Let $X,Y$ be a compatible pair of closed subgroups of a locally
compact abelian group $G$. Then the subgroup $X+Y$ of $G$ generated
by $X\cup Y$ is also closed. If we identify $X\cap Y$ with the
closed subgroup $D$ of $X\oplus Y$ consisting of the elements of the
form $(z,z)$ with $z\in X\cap Y$ then the quotient group $X\uplus Y
\equiv (X\oplus Y)/(X\cap Y)$ is locally compact and the map
\begin{equation}\label{eq:nat}
\phi:X\oplus Y \to X+Y \hspace{2mm}\text{defined by}\hspace{2mm}
 \phi(x,y)=x-y
\end{equation}
is an open continuous surjective group morphism $X\oplus Y\to X+Y$
with $X\cap Y$ as kernel.  Hence the group morphism
$\phi^\circ:X\uplus Y\to X+Y$ induced by $\phi$ is a homeomorphism.

Since $\Cc(X\uplus Y)\subset \Cbu(X\oplus Y)$ the elements
$\theta\in\Cc(X\uplus Y)$ are functions $\theta:X\times Y\to\mbc$
and we may think of them as kernels of integral operators.

\begin{lemma}\label{lm:bound}
  If $\theta\in\Cc(X\uplus Y)$ then $(T_\theta
  u)(x)=\int_Y\theta(x,y)u(y) \rmd y$ defines an operator in
  $\rl_{XY}$ with norm $\|T_\theta\|\leq C\sup|\theta|$ where $C$
  depends only on a compact which contains the support of $\theta$.
\end{lemma}
\proof 
By the Schur test
$$
\|T_\theta\|^2\leq 
{\textstyle\sup_{x\in X}}\int_Y|\theta(x,y)\rmd  y \cdot
{\textstyle\sup_{y\in Y}}\int_X|\theta(x,y)\rmd x.
$$ 
Let $K\subset X$ and $L\subset Y$ be compact sets such that
$(K\times L) + D$ contains the support of $\theta$.
Thus if $\theta(x,y)\neq0$ then $x\in z+K$ and $y\in z+L$ for some 
$z\in X\cap Y$ hence $ \int_Y|\theta(x,y)\rmd y \leq
\sup|\theta| \lambda_Y(L).  $ Similarly $\int_X|\theta(x,y)\rmd x
\leq \sup|\theta| \lambda_X(K)$.  \qed

\begin{definition}\label{df:ryz}
$\rt_{XY}$ is the norm closure in $\rl_{XY}$ of the set of operators
$T_\theta$ as in Lemma \ref{lm:bound}.
\end{definition}

\begin{remark}\label{re:rief}
  If $X\supset Y$ then $\rt_{XY}$ is a ``concrete'' realization of
  the Hilbert $C^*$-module introduced by Rieffel in \cite{Ri} which
  implements the Morita equivalence between the group $C^*$-algebra
  $\cc^*(Y)$ and the crossed product $\Co(X/Y)\rtimes X$. More
  precisely, $\rt_{XY}$ is a Hilbert $\cc^*(Y)$-module and its
  imprimitivity algebra is canonically isomorphic with
  $\Co(X/Y)\rtimes X$.  If $X,Y$ is an arbitrary couple of
  compatible subgroups of $G$ then we defined $\rt_{XY}$ such that
  $\rt_{XY}=\rt_{XG}\cdot\rt_{GY}$. On the other hand, from
  \eqref{eq:factor} we get $\rt_{XY}=\rt_{XE}\cdot\rt_{EY}$ with
  $E=X\cap Y$, hence $\rt_{XY}$ is naturally a Hilbert
  $(\Co(X/E)\rtimes X,\Co(Y/E)\rtimes Y)$ imprimitivity bimodule.
  It has been noticed by Georges Skandalis that $\rt_{XY}$ is in
  fact a ``concrete'' realization of a Hilbert $C^*$-module
  introduced by Green to show the Morita equivalence of the
  $C^*$-algebras $\Co(Z/Y)\rtimes X$ and $\Co(Z/X)\rtimes Y$ where
  we take $Z=X+Y$, cf. \cite[Example 4.13]{Wi}.
\end{remark}

We give now an alternative definition of $\rt_{XY}$. If
$\varphi\in\Cc(G)$ we define $T_{XY}(\varphi):\Cc(Y)\to\Cc(X)$ by
\begin{equation}\label{eq:ryz}
(T_{XY}(\varphi)u)(x)=\int_Y\varphi(x-y)u(y)\rmd y.
\end{equation}
This operator depends only the restriction $\varphi|_{X+Y}$ hence,
by the Tietze extension theorem, we could take $\varphi\in\Cc(Z)$
instead of $\varphi\in\Cc(G)$, where $Z$ is any closed subgroup of
$G$ containing $X\cup Y$.

\begin{proposition}\label{pr:def2}
$T_{XY}(\varphi)$ extends to a bounded operator $L^2(Y)\to L^2(X)$,
also denoted $T_{XY}(\varphi)$, and for each compact $K\subset G$
there is a constant $C$ such that if $\supp\varphi\subset K$
\begin{equation}\label{eq:nyz}
\|T_{XY}(\varphi)\|\leq C \sup\nolimits_{x\in G}|\varphi(x)|.
\end{equation}
The adjoint operator is given by $T_{XY}(\varphi)^*=
T_{YX}(\varphi^*)$ where $\varphi^*(x)=\bar\varphi(-x)$.  The space
$\rt_{XY}$ coincides with the closure in $\rl_{XY}$ of the set of
operators of the from $T_{XY}(\varphi)$.
\end{proposition}
\proof The set $X+Y$ is closed in $G$ hence the restriction map
$\Cc(G)\to\Cc(X+Y)$ is surjective. On the other hand, the map
$\phi^\circ:X\uplus Y\to X+Y$, defined after \eqref{eq:nat}, is a
homeomorphism so it induces an isomorphism
$\varphi\to\varphi\circ\phi^\circ$ of $\Cc(X+Y)$ onto $\Cc(X\uplus
Y)$. Clearly $T_{XY}(\varphi)=T_\theta$ if $\theta=\varphi\circ\phi$,
so the proposition follows from Lemma \ref{lm:bound}.  
\qed

We discuss now some properties of the spaces $\rt_{XY}$.
We set $\rt_{XY}^*\equiv(\rt_{XY})^*\subset\rl_{YX}$. 

\begin{proposition}\label{pr:nyza}
We have  $\rt_{XX}=\rt_X$ and:
\begin{align}
& \rt_{XY}^* =\rt_{YX} \label{eq:rad} \\
& \rt_{XY}   =\rt_{XY}\cdot \rt_Y=\rt_X\cdot\rt_{XY}
\label{eq:cyzc}  \\
& \ca|_X\cdot\rt_{XY} =\rt_{XY}\cdot\ca|_Y \label{eq:ayza}
\end{align}
where $\ca$ is an arbitrary  $G$-algebra.
\end{proposition}
\proof The relations $\rt_{XX}=\rt_X$ and \eqref{eq:rad} are
obvious.  Now we prove the first equality in \eqref{eq:cyzc} (then
the second one follows by taking adjoints). If $C(\eta)$ is the
operator of convolution in $L^2(Y)$ with $\eta\in \Cc(Y)$ then a
short computation gives
\begin{equation}\label{eq:yzc}
T_{XY}(\varphi)C(\eta)=T_{XY}(T_{G Y}(\varphi)\eta)
\end{equation}
for $\varphi\in\Cc(G)$. Since $T_{G Y}(\varphi)\eta\in\Cc(G)$ we get
$T_{XY}(\varphi)C(\eta)\in\rt_{GX}$, so $\rt_{XY}\cdot
\rt_Y\subset\rt_{XY}$. The converse follows by a standard
approximation argument.

Let $\varphi\in\Cc(G)$ and $\theta\in\ca$. We shall denote by
$\theta(Q_X)$ the operator of multiplication by $\theta|_X$ in
$L^2(X)$ and by $\theta(Q_Y)$ that of multiplication by $\theta|_Y$
in $L^2(Y)$.  Choose some $\varepsilon>0$ and let $V$ be a compact
neighborhood of the origin in $G$ such that
$|\theta(z)-\theta(z')|<\varepsilon$ if $z-z'\in V$. There are
functions $\alpha_k\in\Cc(G)$ with $0\leq\alpha_k\leq1$ such that
$\sum_k\alpha_k=1$ on the support of $\varphi$ and
$\supp\alpha_k\subset z_k+V$ for some points $z_k$.  Below we shall
prove:
\begin{equation}\label{eq:byza}
\|T_{XY}(\varphi)\theta(Q_Y)- 
{\textstyle\sum_k}\theta(Q_X-z_k)T_{XY}(\varphi\alpha_k)\| 
\leq
\varepsilon\|T_{XY}(|\varphi|)\|.
\end{equation}
This implies $\rt_{XY}\cdot\ca|_Y\subset\ca|_X\cdot\rt_{XY}$. If we
take adjoints, use \eqref{eq:rad} and interchange $X$ and $Y$ in the
final relation, we obtain $\ca|_X\cdot\rt_{XY}=\rt_{XY}\cdot\ca|_Y$
hence the proposition is proved.  For $u\in\Cc(X)$ we have:
\begin{align*}
(T_{XY}(\varphi)\theta(Q_Y)u)(x) &=
\int_Y\varphi(x-y)\theta(y)u(y)\rmd y 
=\sum_k\int_Y\varphi(x-y)\alpha_k(x-y)\theta(y)u(y)\rmd y \\
&=
\sum_k\int_Y\varphi(x-y)\alpha_k(x-y)\theta(x-z_k)u(y)\rmd y
+(Ru)(x)\\ 
&= 
\sum_k\left(\theta(Q_X-z_k)T_{XY}(\varphi\alpha_k)u\right)(x) 
+(Ru)(x).
\end{align*}
We can estimate the remainder as follows
$$
|(Ru)(x)|=\left|\sum_k\int_Y\varphi(x-y)\alpha_k(x-y)
[\theta(y)-\theta(x-z_k)]u(y)\rmd y \right|\leq
\varepsilon\int_Y|\varphi(x-y)u(y)|\rmd y.
$$ 
because $x-z_k-y\in V$.  This proves \eqref{eq:byza}. 
\qed

\begin{proposition}\label{pr:ryz}
 $\rt_{XY}$ is a Hilbert $C^*$-submodule of $\rl_{XY}$ and
\begin{equation}\label{eq:hyz}
\rt_{XY}^*\cdot\rt_{XY}=\rc_Y(X\cap Y), \hspace{2mm}
\rt_{XY}\cdot\rt_{XY}^*=\rc_X(X\cap Y).
\end{equation}
Thus $\rt_{XY}$ is a $(\rc_X(X\cap Y),\rc_Y(X\cap Y))$ imprimitivity
bimodule.
\end{proposition}
\proof Due to \eqref{eq:rad}, to prove the first relation in
\eqref{eq:hyz} we have to compute the clspan $\rc$ of the operators
$T_{XY}(\varphi)T_{YX}(\psi)$ with $\varphi,\psi$ in $\Cc(G)$.  We
recall the notation $G^2=G\oplus G$, this is a locally compact
abelian group and $X^2=X\oplus X$ is a closed subgroup. Let us
choose functions $\varphi_k,\psi_k\in\Cc(G)$ and let
$\Phi=\sum_k\varphi_k\otimes\psi_k\in\Cc(G^2)$. If
$\psi_k^\dag(x)=\psi_k(-x)$, then $\sum_k
T_{XY}(\varphi_k)T_{YX}(\psi_k^\dag)$ is an integral operator on
$L^2(X)$ with kernel $\theta_X=\theta|_{X^2}$ where $\theta:G^2\to
\mbc$ is given by
$$ 
\theta(x,x')= \int_Y\Phi(x+y,x'+y)\rmd y.
$$ Since the set of decomposable functions is dense in $\Cc(G^2)$ in
the inductive limit topology, an easy approximation argument shows
that $\rc$ contains all integral operators with kernels of the same
form as $\theta_X$ but with arbitrary $\Phi\in\Cc(G^2)$.  Let
$Y^{(2)}$ be the closed subgroup of $G^2$ consisting
of the elements $(y,y)$ with $y\in Y$. Then $K=\supp\Phi\subset G^2$
is a compact, $\theta$ is zero outside $K+Y^{(2)}$, and
$\theta(a+b)=\theta(a)$ for all $a\in G^2,b\in Y^{(2)}$. Thus
$\theta\in\Cc(G^2/Y^{(2)})$, with the usual identification
$\Cc(G^2/Y^{(2)})\subset\Cbu(G^2)$. From Proposition 2.48 in
\cite{Fo} it follows that reciprocally, any function $\theta$ in
$\Cc(G^2/Y^{(2)})$ can be represented in terms of some $\Phi$ in
$\Cc(G^2)$ as above. Thus $\rc$ is the closure of the set of
integral operators on $L^2(X)$ with kernels of the form $\theta_X$
with $\theta\in\Cc(G^2/Y^{(2)})$. According to Lemma
\ref{lm:double}, the pair of subgroups $X^2,Y^{(2)}$ is regular, so
we may apply Lemma \ref{lm:reg} to get
$\Cc(G^2/Y^{(2)})|_{X^2}=\Cc(X^2/D)$ where $D=X^2\cap
Y^{(2)}=\{(x,x)\mid x\in X\cap Y\}$.  But by Lemma \ref{lm:sym} the
norm closure in $\rl_X$ of the set of integral operators with
kernel in $\Cc(X^2/D)$ is $\rc_X/(X\cap Y)$. This proves
\eqref{eq:hyz}.

It remains to prove that $\rt_{XY}$ is a Hilbert $C^*$-submodule of
$\rl_{XY}$, i.e. that we have
\begin{equation}\label{eq:hyz1}
\rt_{XY}\cdot\rt_{XY}^*\cdot\rt_{XY}=\rt_{XY}.
\end{equation}
The first identity in \eqref{eq:hyz} and \eqref{eq:cyzc} imply
\begin{equation*}
\rt_{XY}\cdot\rt_{XY}^*\cdot\rt_{XY} = 
\rt_{XY}\cdot \rt_Y\cdot\cc_Y(X\cap Y)=
\rt_{XY}\cdot\cc_Y(X\cap Y).
\end{equation*}
From Lemma \ref{lm:reg} we get 
$$
\cc_Y(X\cap Y)=\cc_G(X\cap Y)|_Y =
\cc_G(X)|_Y \cdot \cc_G(Y)|_Y = \cc_G(X)|_Y 
$$ because $\cc_G(Y)|_Y=\mbc$.  Then by using Proposition
\ref{pr:nyza} we obtain
$$
\rt_{XY}\cdot\cc_Y(X\cap Y) = \rt_{XY}\cdot\cc_G(X)|_Y =
\cc_G(X)|_X \cdot\rt_{XY} =\rt_{XY}
$$
because $\cc_G(X)|_X=\mbc$. 
\qed

\begin{corollary}\label{co:txy}
We have
\begin{align}
\rt_{XY} &= \rt_{XY}\rt_Y=\rt_{XY}\cc_Y(X\cap Y) 
\label{eq:txy1}
\\
&= \rt_X\rt_{XY}=\cc_X(X\cap Y)\rt_{XY}.
\label{eq:txy2}
\end{align}
\end{corollary}
\proof If $\mr$ is a Hilbert $\ra$-module then $\mr=\mr\ra$ hence
$\rt_{XY}=\rt_{XY}\rc_Y(X\cap Y)$ by Proposition \ref{pr:ryz}. The
space $\rc_Y(X\cap Y)$ is a $\rt_Y$-bimodule and $\rc_Y(X\cap
Y)=\rc_Y(X\cap Y)\cdot \rt_Y$ by \eqref{eq:Cxy} hence we get
$\rc_Y(X\cap Y)=\rc_Y(X\cap Y)\rt_Y$ by the Cohen-Hewitt theorem.
This proves the first equality in \eqref{eq:txy1} and the other ones
are proved similarly.  \qed

If $\cg$ is a set of closed subgroups of $G$ then the
\emph{semilattice generated by $\cg$} is the set of finite
intersections of elements of $\cg$.

\begin{proposition}\label{pr:product}
Let $X,Y,Z$ be closed subgroups of $G$ such that any two subgroups
from the semilattice generated by the family $\{X,Y,Z\}$ are
compatible. Then:
\begin{align}\label{eq:product}
\rt_{XZ}\cdot\rt_{ZY} &=
\rt_{XY}\cdot\cc_Y(Y\cap Z)= \cc_X(X\cap Z)\cdot\rt_{XY} \\
&= \rt_{XY}\cdot\cc_Y(X\cap Y\cap Z)= 
\cc_X(X\cap Y\cap Z)\cdot\rt_{XY}.
\end{align} 
In particular, if $Z\supset X\cap Y$ then
\begin{equation}\label{eq:factor}
\rt_{XZ}\cdot\rt_{ZY}=\rt_{XY}.
\end{equation}
\end{proposition}
\proof We first prove \eqref{eq:factor} in the particular case
$Z=G$. As in the proof of Proposition \ref{pr:ryz} we see that
$\rt_{XG}\cdot\rt_{G Y}$ is the the closure in $\rl_{XY}$ of the
set of integral operators with kernels 
$\theta_{XY}=\theta|_{X\times Y}$ where $\theta:G^2\to \mbc$ is
given by 
$$ 
\theta(x,y)= \int_G\sum_k\varphi_k(x-z)\psi_k(z-y)\rmd z=
\int_G\sum_k\varphi_k(x-y-z)\psi_k(z)\rmd z\equiv\xi(x-y)
$$ where $\varphi_k,\psi_k\in\Cc(G)$ and $\xi=\sum_k\varphi_k*\psi_k$
convolution product on $G$. Since $\Cc(G)*\Cc(G)$ is dense in
$\Cc(G)$ in the inductive limit topology, the space
$\rt_{XG}\cdot\rt_{G Y}$ is the the closure of the set of integral
operators with kernels $\theta(x,y)=\xi(x-y)$ with $\xi\in\Cc(G)$.
By Proposition \ref{pr:def2} this is  $\rt_{XY}$. 

Now we prove \eqref{eq:product}. From \eqref{eq:factor} with $Z=G$
and \eqref{eq:hyz} we get:
\begin{equation*}
\rt_{XZ}\cdot\rt_{ZY} =
\rt_{XG}\cdot\rt_{G Z}\cdot\rt_{ZG}\cdot\rt_{G Y}
= \rt_{XG}\cdot\cc_G(Z)\cdot \rt_G\cdot\rt_{GY}.
\end{equation*}
Then from Proposition \eqref{pr:nyza} and Lemma \ref{lm:reg} we get:
$$
\cc_G(Z)\cdot \rt_G\cdot\rt_{GY}=\cc_G(Z)\cdot\rt_{G Y}=
\rt_{G Y}\cdot\cc_G(Z)|_Y= \rt_{G Y}\cdot\cc_Y(Y\cap Z).
$$ 
We obtain \eqref{eq:product} by using once again
\eqref{eq:factor} with $Z=G$ and taking adjoints. On the other hand,
the relation $\rt_{XY}=\rt_{XY}\cdot\cc_Y(X\cap Y)$ holds because
of \eqref{eq:txy1}, so we have
$$
\rt_{XY}\cdot\cc_Y(Y\cap Z)=
\rt_{XY}\cdot\cc_Y(X\cap Y)\cdot\cc_Y(Y\cap Z)=
\rt_{XY}\cdot\cc_Y(X\cap Y\cap Z)
$$ where we also used \eqref{eq:reg1} and the fact that $X\cap Y$,
$Z\cap Y$ are compatible. Finally, to get \eqref{eq:factor} for
$Z\supset X\cap Y$ we use once again \eqref{eq:hyz}.  \qed

The object of main interest for us is introduced in the next
definition. 

\smallskip

\begin{definition}\label{df:nxyz}
If $X,Y$ are compatible subgroups and $Z$ is a closed subgroup of
$X\cap Y$ then we set
\begin{equation}\label{eq:nxyz}
\rc_{XY}(Z):=\rt_{XY}\cdot\cc_Y(Z)=\cc_X(Z)\cdot\rt_{XY}.
\end{equation} 
\end{definition}
The equality above follows from \eqref{eq:ayza} with $\ca=\cc_G(Z)$.
We clearly have $\rc_{XY}(X\cap Y)=\rt_{XY}$ and
$\rc_{XX}(Y)=\rc_X(Y)$ if $X\supset Y$. Moreover
\begin{equation}\label{eq:nadj}
\rc_{XY}^*(Z):=\rc_{XY}(Z)^*=\rc_{YX}(Z)
\end{equation}
because of \eqref{eq:rad}.

\begin{theorem}\label{th:nxyz}
$\rc_{XY}(Z)$ is a Hilbert $C^*$-submodule of $\rl_{XY}$ such that
\begin{equation}\label{eq:nnz}
\rc_{XY}^*(Z)\cdot\rc_{XY}(Z)=\rc_Y(Z)
\hspace{2mm}\text{and}\hspace{2mm}
\rc_{XY}(Z)\cdot\rc_{XY}^*(Z)=\rc_X(Z).
\end{equation} 
In particular, $\rc_{XY}(Z)$ is a
$(\rc_X(Z),\rc_Y(Z))$ imprimitivity bimodule.
\end{theorem}
\proof 
By using \eqref{eq:nadj}, the definition \eqref{eq:nxyz}, and
\eqref{eq:reg1} we get
\begin{align*}
\rc_{XY}(Z)\cdot\rc_{YX}(Z) &=
\cc_X(Z)\cdot\rt_{XY}\cdot \rt_{YX}\cdot\cc_X(Z)\\
&=
\cc_X(Z)\cdot\cc_X(X\cap Y)\cdot \rt_X\cdot\cc_X(Z)\\
&=
\cc_X(Z)\cdot \rt_X\cdot\cc_X(Z)= \cc_X(Z)\cdot \rt_X
\end{align*}
which proves the second equality in \eqref{eq:nnz}.  The first one
follows by interchanging $X$ and $Y$.
\qed

\subsection{Many-body systems}
\label{ss:gaz} 

Here we give a formal definition of the notion of ``many-body
system'' then define and discuss the Hamiltonian algebra associated
to it.

Let $\rs$ be a set of locally compact abelian groups with the
following property: for any $X,Y\in\rs$ there is $Z\in\rs$ such that
$X$ and $Y$ are compatible subgroups of $Z$. Note that this implies
the following: if $Y\subset X$ then the topology and the group
structure of $Y$ coincide with those induced by $X$.

If $\rs$ is a set of $\sigma$-compact locally compact abelian groups
then the compatibility assumption is equivalent to the following
more explicit condition: for any $X,Y\in\rs$ there is $Z\in\rs$ such
that $X$ and $Y$ are closed subgroups of $Z$ and $X+Y$ is closed in
$Z$.

\begin{definition}\label{df:mb}
A \emph{many-body system} is a couple $(\cs,\lambda)$ where:
\begin{compactenum}
\item[(i)] $\cs\subset\rs$ is a subset such that 
$X,Y\in\cs\Rightarrow X\cap Y\in\cs$ and if
$X\supsetneq Y$ then $X/Y$ is not compact,
\item[(ii)] 
$\lambda$ is a map $X\mapsto\lambda_X$ which associates a Haar
measures $\lambda_X$ on $X$ to each $X\in\cs$. 
\end{compactenum}
\end{definition}

We identify $\cs=(\cs,\lambda)$ so the choice of Haar measures is
implicit. Note that the Hilbert space $\ch_\cs$ and the
$C^*$-algebra $\rc_\cs$ that we introduce below depend on $\lambda$
but different choices give isomorphic objects.  Each $X\in\cs$ is
equipped with a Haar measure so the Hilbert spaces $\ch_X= L^2(X)$
are well defined. If $Y\subset X$ are in $\cs$ then $X/Y$ is
equipped with the quotient measure so $\ch_{X/Y}= L^2(X/Y)$ is well
defined.

{\bf Example:} Let $\rs$ the set of all finite dimensional vector
subspaces of a vector space over an infinite locally compact field
and let $\cs$ be any subset of $\rs$ such that $X,Y\in\cs\Rightarrow
X\cap Y \in \cs$.

For each $X\in\cs$ let $\cs_X$ be the set of $Y\in\cs$ such that
$Y\subset X$.  This is an $N$-body system with $X$ as configuration
space in the sense of Definition \ref{df:nb}.  Then by Lemma
\ref{lm:reg} the space
\begin{equation}\label{eq:sax}
\cc_X := {\textstyle\sum^\rmc_{Y\in\cs_X}}\cc_X(Y)
\end{equation}
is an $X$-algebra so the crossed product $\cc_X\rtimes X$ is well
defined and we clearly have
\begin{equation}\label{eq:crsax}
\rc_X := \cc_X\rtimes X \equiv \cc_X\cdot\rt_X =
{\textstyle\sum^\rmc_{Y\in\cs_X}}\rc_X(Y).
\end{equation}
The $C^*$-algebra $\rc_X$ is realized on the Hilbert space $\ch_X$
and we think of it as the Hamiltonian algebra of the $N$-body system
determined by $\cs_X$.

\begin{theorem}\label{th:grsax}
The $C^*$-algebras $\cc_X$ and $\rc_X$ are $\cs_X$-graded by 
the decompositions \eqref{eq:sax} and \eqref{eq:crsax}. 
\end{theorem}

The theorem is a particular case of results due to A. Mageira,
cf. Propositions 6.1.2, 6.1.3 and 4.2.1 in \cite{Ma} (or see
\cite{Ma3}). We mention that the results in \cite{Ma,Ma3} are much
deeper since the groups are allowed to be noncommutative and the
treatment is so that the second part of condition (i) is not
needed. The case when $\cs$ consists of linear subspaces of a finite
dimensional real vector space has been considered in \cite{BG1,DG1}
and the corresponding version of Theorem \ref{th:grsax} is proved
there by elementary means.

\begin{definition}\label{df:main}
If $X,Y\in\cs$ then
$\rc_{XY} := \rt_{XY}\cdot\cc_Y=\cc_X\cdot\rt_{XY}$.
\end{definition}

In particular $\rc_{XX}=\rc_X$ is a $C^*$-algebra of operators on
$\ch_X$.  For $X\neq Y$ the space $\rc_{XY}$ is a closed linear
space of operators $\ch_Y\to \ch_X$ canonically associated to the
semilattice of groups $\cs_{X\cap Y}$, cf. \eqref{eq:main}. We call
these spaces \emph{coupling modules} because they are Hilbert
$C^*$-modules and determine the way the systems corresponding to $X$
and $Y$ are allowed to interact.

For each pair $X,Y\in\cs$ with $X\supset Y$ we set 
\begin{equation}\label{eq:saX}
\cc^Y_X :=
{\textstyle\sum^\rmc_{Z\in\cs_Y}}\cc_X(Z).
\end{equation}
This is also an $X$-algebra so we may define $\rc^Y_X=\cc^Y_X\rtimes
X$ and we have
\begin{equation}\label{eq:saX1}
\rc^Y_X := \cc^Y_X\rtimes X=
{\textstyle\sum^\rmc_{Z\in\cs_Y}}\rc_X(Z).
\end{equation}
If $X=Y\oplus Z$ then $\cc_X^Y\simeq\cc_Y\otimes 1$ and 
$\rc_X^Y\simeq\rc_Y\otimes \rt_Z$. 

\begin{lemma}\label{lm:xyprod}
Let $X\in\cs$ and $Y\in\cs_X$. Then
\begin{equation}\label{eq:xy1}
\cc_X^{Y}=\cc_X(Y)\cdot\cc_X \hspace{2mm}\text{and}\hspace{2mm}
\rc_X^{Y}=\cc_X(Y)\cdot\rc_X=\rc_X\cdot\cc_X(Y).
\end{equation}
Moreover, for all  $Y,Z\in\cs_X$ we have
\begin{equation}\label{eq:xy2}
\cc_X^Y\cdot\cc_X^Z=\cc_X^{Y\cap Z} 
\hspace{2mm}\text{and}\hspace{2mm} 
\rc_X^Y\cdot\rc_X^Z=\rc_X^{Y\cap Z}.
\end{equation}
\end{lemma}
\proof The abelian case follows from \eqref{eq:reg1} and a
straightforward computation. For the crossed product algebras we use
$\cc_X(Y)\cdot\rc_X=\cc_X(Y)\cdot\cc_X\cdot \rt_X$ and the first
relation in \eqref{eq:xy1} for example.  \qed

\begin{lemma}\label{lm:nxy}
For arbitrary $X,Y\in\cs$ we have
\begin{equation}\label{eq:main}
\cc_X\cdot\rt_{XY}=\rt_{XY}\cdot\cc_Y
=\rt_{XY}\cdot\cc_Y^{X\cap Y}=
\cc_X^{X\cap Y}\cdot\rt_{XY}.
\end{equation}
\end{lemma}
\proof
If $G\in\rs$ contains $X\cup Y$ then clearly
$$
\cc_X\cdot\rt_{XY}=
{\textstyle\sum^\rmc_{Z\in\cs_X} }\cc_X(Z)\cdot\rt_{XY}=
{\textstyle\sum^\rmc_{Z\in\cs_X} }\cc_G(Z)|_X\cdot\rt_{XY}.
$$
From \eqref{eq:ayza} and \eqref{eq:reg2} we get
$$
\cc_G(Z)|_X\cdot\rt_{XY}=\rt_{XY}\cdot\cc_Y(Y\cap Z).
$$ 
Since $Y\cap Z$ runs over $\cs_{X\cap Y}$ when $Z$ runs over
$\cs_X$ we obtain $\cc_X\cdot\rt_{XY}=\rt_{XY}\cdot\cc_Y^{X\cap Y}$.
Similarly $\rt_{XY}\cdot\cc_Y=\cc_X^{X\cap Y}\cdot\rt_{XY}$. 
On the other hand $\cc_X^{X\cap Y}=\cc_G^{X\cap Y}|_X$ and similarly
with $X,Y$ interchanged, hence 
$\cc_X^{X\cap Y}\cdot\rt_{XY}=\rt_{XY}\cdot\cc_Y^{X\cap Y}$
because of \eqref{eq:ayza}. 
\qed

\begin{proposition}\label{pr:mxyz}
Let $X,Y,Z\in\cs$. Then $\rc_{XY}^*=\rc_{YX}$ and
\begin{equation}\label{eq:mxyz}
\rc_{XZ}\cdot\rc_{ZY}=\rc_{XY}\cdot\cc_Y^{X\cap Y\cap Z}=
\cc_X^{X\cap Y\cap Z}\cdot\rc_{XY} \subset \rc_{XY}. 
\end{equation}
In particular $\rc_{XZ}\cdot\rc_{ZY}=\rc_{XY}$ if $Z\supset X\cap Y$.
\end{proposition}
\proof 
The first assertion follows from \eqref{eq:rad}.  From the
Definition \ref{df:main} and Proposition \ref{pr:product} we then get
\begin{align*}
\rc_{XZ}\cdot\rc_{ZY} &=
\cc_X\cdot\rt_{XZ}\cdot\rt_{ZY}\cdot\cc_Y =
\cc_X\cdot\rt_{XY}\cdot\cc_Y(X\cap Y\cap Z)\cdot\cc_Y \\ &=
\rt_{XY}\cdot\cc_Y\cdot\cc_Y(X\cap Y\cap Z)\cdot\cc_Y =
\rt_{XY}\cdot\cc_Y(X\cap Y\cap Z)\cdot\cc_Y.
\end{align*}
But $\cc_Y(X\cap Y\cap Z)\cdot\cc_Y=\cc_Y^{X\cap Y\cap Z}$ by Lemma
\ref{lm:xyprod}. For the last inclusion in \eqref{eq:mxyz} we use
the obvious relation $\cc_Y^{X\cap Y\cap Z}\cdot\cc_Y\subset\cc_Y$.
The last assertion of the proposition follows from \eqref{eq:main}. 
\qed

The following theorem is a consequence of the results obtained so
far.

\begin{theorem}\label{th:nmod}
$\rc_{XY}$ is a Hilbert $C^*$-submodule of $\rl_{XY}$ such that 
\begin{equation}\label{eq:nmod}
\rc_{XY}^*\cdot\rc_{XY}=\rc_Y^{X\cap Y} \text{ and }
\rc_{XY}\cdot\rc_{XY}^*=\rc_X^{X\cap Y}.
\end{equation}
In particular, $\rc_{XY}$ is a
$(\rc_X^{X\cap Y},\rc_Y^{X\cap Y})$ imprimitivity bimodule. 
\end{theorem}

We recall the conventions
\begin{align}
& X,Y\in\cs \text{ and } Y\not\subset X \Rightarrow 
\cc_X(Y)= \rc_X(Y)=\{0\}, \label{eq:convn} \\
& X,Y,Z\in\cs \text{ and } Z\not\subset X\cap Y \Rightarrow
\rc_{XY}(Z)=\{0\}.\label{eq:convn1}
\end{align}
From now on by ``graded'' we mean $\cs$-graded.  Then
$\rc_X=\sum^\rmc_{Y\in\cs}\rc_X(Y)$ is a graded
$C^*$-algebras supported by the ideal $\cs_X$ of $\cs$, in
particular it is a graded ideal in $\rc_X$.  With the notations of
Subsection \ref{ss:grca} the algebra $\rc^Y_X=\rc_X(\cs_Y)$ is a
graded ideal of $\rc_X$ supported by $\cs_Y$. Similarly for $\cc_X$
and $\cc_X^Y$.

Since $\rc_X^{X\cap Y}$ and $\rc_Y^{X\cap Y}$ are ideals in $\rc_X$
and $\rc_Y$ respectively, Theorem \ref{th:nmod} allows us to equip
$\rc_{XY}$ with (right) Hilbert $\rc_Y$-module and left Hilbert
$\rc_X$-module structures (which are not full in general).

\begin{theorem}\label{th:nmain}
The Hilbert $\rc_Y$-module $\rc_{XY}$ is graded by the family of
$C^*$-submodules  $\{\rc_{XY}(Z)\}_{Z\in\cs}$.
\end{theorem}
\proof We use Proposition \ref{pr:rhm} with $\mr=\rt_{XY}$ and
$\cc_Y(Z)$ as algebras $\cc(\sigma)$. Then $\ra=\rc_Y(X\cap Y)$ by
\eqref{eq:hyz} hence $\ra\cdot\cc_Y(Z)=\rc_Y(Z)$ and the conditions
of the proposition are satisfied.  \qed

\begin{remark}\label{re:precise}
The following more precise statement is a consequence of the Theorem
\ref{th:nmain}: the Hilbert $\rc_Y^{X\cap Y}$-module $\rc_{XY}$ is
$\cs_{X\cap Y}$-graded by the family of $C^*$-submodules
$\{\rc_{XY}(Z)\}_{Z\in\cs_{X\cap Y}}$.
\end{remark}

Finally, we may construct the $C^*$-algebra $\rc$ which is of main
interest for us, the many-body Hamiltonian algebra. We shall
describe it as an algebra of operators on the Hilbert space
\begin{equation}\label{eq:bigh}
\ch\equiv\ch_\cs={\textstyle\oplus_{X\in\cs}} \ch_X
\end{equation}
which is a kind of Boltzmann-Fock space (without symmetrization or
anti-symmetrization) determined by the semilattice $\cs$. Note that
if the zero group $O=\{0\}$ belongs to $\cs$ then $\ch$ contains
$\ch_O=\mbc$ as a subspace, this is the vacuum sector.  Let
$\Pi_{X}$ be the orthogonal projection of $\ch$ onto $\ch_X$ and let
us think of its adjoint $\Pi_{X}^*$ as the natural embedding
$\ch_X\subset\ch$. Then for any pair $X,Y\in\cs$ we identify
\begin{equation}\label{eq:identc}
\rc_{XY}\equiv\Pi^*_{X}\rc_{XY}\Pi_{Y} \subset L(\ch).
\end{equation}
Thus we realize $\{\rc_{XY}\}_{X,Y\in\cs}$ as a linearly independent
family of closed subspaces of $L(\ch)$ such that
$\rc_{XY}^*=\rc_{YX}$ and $\rc_{XZ}\rc_{Z'Y}\subset\rc_{XY}$ for all
$X,Y,Z,Z'\in\cs$. Then by what we proved before, especially
Proposition \ref{pr:mxyz}, the space
$\sum\nolimits_{X,Y\in\cs}\rc_{XY}$ is a $*$-subalgebra of $L(\ch)$
hence its closure
\begin{equation}\label{eq:bigco}
\rc\equiv\rc_\cs= {\textstyle\sum^\rmc_{X,Y\in\cs}}\rc_{XY}.
\end{equation}
is a $C^*$-algebra of operators on $\ch$.  Note that one may view
$\rc$ as a matrix $(\rc_{XY})_{X,Y\in\cs}$.  

In a similar way one may associate to the spaces $\rt_{XY}$ a
closed self-adjoint subspace $\rt\subset L(\ch)$.  It is also useful
to define a new subspace $\rt^\circ\subset L(\ch)$ by
$\rt^\circ_{XY}=\rt_{XY}$ if $X\sim Y$ and $\rt^\circ=\{0\}$ if
$X\not\sim Y$. Here $X\sim Y$ means $X\subset Y$ or $Y\subset X$
. Clearly $\rt^\circ$ is a closed self-adjoint linear subspace of
$\rt$.  Finally, let $\cc$ be the diagonal $C^*$-algebra
$\cc\equiv\oplus_X\cc_X$ of operators on $\ch$.

\begin{theorem}\label{th:tc}
We have $\rc=\rt\cdot\cc=\cc\cdot\rt=\rt\cdot\rt=
\rt^\circ\cdot\rt^\circ$.
\end{theorem}
\proof The first two equalities are an immediate consequence of the
Definition \ref{df:main}. To prove the third equality we use
Proposition \ref{pr:product}, more precisely the relation
\[
\rt_{XZ}\cdot\rt_{ZY}=\rt_{XY}\cdot\cc_Y(X\cap Y\cap Z)=
\rc_{XY}(X\cap Y\cap Z)
\]
which holds for any $X,Y,Z$. Then
\[
{\textstyle\sum^\rmc_Z}\rt_{XZ}\cdot\rt_{ZY}=
{\textstyle\sum^\rmc_Z}\rc_{XY}(X\cap Y\cap Z)=
{\textstyle\sum^\rmc_Z}\rc_{XY}(Z)=\rc_{XY}
\]
which is equivalent to $\rt\cdot\rt=\rc$. Now we prove the last
equality in the proposition. We have
\[
{\textstyle\sum^\rmc_Z} \rt^\circ_{XZ}\cdot\rt^\circ_{ZY}= 
\text{ closure of the sum }
{\textstyle\sum_{\substack{Z\sim X\\ Z\sim Y}}}
\rt_{XZ}\cdot\rt_{ZY}. 
\]
In the last sum we have four possibilities: $Z\supset X\cup Y$,
$X\supset Z\supset Y$, $Y\supset Z\supset X$, and $Z\subset X\cap
Y$. In the first three cases we have $Z\supset X\cap Y$ hence
$\rt_{XZ}\cdot\rt_{ZY}=\rt_{XY}$ by \eqref{eq:factor}. In the last
case we have $\rt_{XZ}\cdot\rt_{ZY}=\rt_{XY}\cdot\cc_Y(Z)$ by
\eqref{eq:product}. This proves $\rt^\circ\cdot\rt^\circ=\rc$.
\qed

Finally, we are able to equip $\rc$ with an $\cs$-graded
$C^*$-algebra structure.

\begin{theorem}\label{th:cgrad}
For each $Z\in\cs$ the space $\rc(Z):=
\sum^\rmc_{X,Y\in\cs}\rc_{XY}(Z)$ is a $C^*$-subalgebra of
$\rc$. The family $\{\rc(Z)\}_{Z\in\cs}$ defines a graded
$C^*$-algebra structure on $\rc$.
\end{theorem}
\proof 
We first prove the following relation:
\begin{equation}\label{eq:xyzef}
\rc_{XZ}(E)\cdot\rc_{ZY}(F)=\rc_{XY}(E\cap F)
\quad \text{if } X,Y,Z\in\cs \text{ and } E\subset{X\cap Z},
F\subset{Y\cap Z}.
\end{equation}
From Definition \ref{df:nxyz}, Proposition \ref{pr:product},
relations \eqref{eq:reg1} and \eqref{eq:ayza}, and 
$F\subset Y\cap Z$, we get
\begin{align*}
\rc_{XZ}(E)\cdot\rc_{ZY}(F) &=
\cc_X(E)\cdot\rt_{XZ}\cdot\rt_{ZY}\cdot\cc_Y(F) \\
&= \cc_X(E)\cdot\rt_{XY}\cdot\cc_{Y}(Y\cap Z)\cdot\cc_Y(F) \\
&= \cc_X(E)\cdot\rt_{XY}\cdot\cc_{Y}(F) \\
&= \rt_{XY}\cdot\cc_Y(Y\cap E)\cdot\cc_{Y}(F) \\
&= \rt_{XY}\cdot\cc_Y(Y\cap E\cap F).
\end{align*}
At the next to last step we used $\cc_X(E)=\cc_G(E)|_X$ for some
$G\in\rs$ containing both $X$ and $Y$ and then \eqref{eq:ayza},
\eqref{eq:reg2}. Finally, we use $\cc_Y(Y\cap E\cap F)=\cc_Y(E\cap
F)$ and the Definition \ref{df:nxyz}. This proves \eqref{eq:xyzef}.
Due to the conventions \eqref{eq:convn}, \eqref{eq:convn1} we now
get from \eqref{eq:xyzef} for $E,F\in\cs$
\[
{\textstyle\sum_{Z\in\cs}}\rc_{XZ}(E)\cdot\rc_{ZY}(F)=
\rc_{XY}(E\cap F).
\]
Thus $\rc(E)\rc(F)\subset\rc(E\cap F)$, in particular $\rc(E)$ is a
$C^*$-algebra. It remains to be shown that the family of
$C^*$-algebras $\{\rc(E)\}_{E\in\cs}$ is linearly independent. Let
$A(E)\in\rc(E)$ such that $A(E)=0$ but for a finite number of $E$
and assume that $\sum_E A(E)=0$. Then for all $X,Y\in\cs$
we have $\sum_E \Pi_X A(E) \Pi_Y^* =0$. Clearly 
$\Pi_X A(E) \Pi_Y^*\in\rc_{XY}(E)$ hence from Theorem \ref{th:nmain}
we get $\Pi_X A(E) \Pi_Y^*=0$ for all $X,Y$ so $A(E)=0$ for all $E$.
\qed

\subsection{Subsystems}
\label{ss:T} 

We now point out some interesting subalgebras of
$\rc$. If $\ct\subset\cs$ is any subset let
\begin{equation}\label{eq:t}
\rc^\ct_\cs\equiv{\textstyle\sum_{X,Y\in\ct}^\rmc}\rc_{XY} \quad
\text{and} \quad \ch_\ct\equiv\oplus_{X\in\ct}\ch_X.
\end{equation}
Note that the sum defining $\rc^\ct_\cs$ is already closed if $\ct$ is
finite and that $\rc^\ct_\cs$ is a $C^*$-algebra which lives on the
subspace $\ch_\ct$ of $\ch$. In fact, if $\Pi_\ct$ is the orthogonal
projection of $\ch$ onto $\ch_\ct$ then
\begin{equation}\label{eq:tt}
\rc^\ct_\cs=\Pi_\ct\rc_\cs\Pi_\ct
\end{equation}
and this is a $C^*$-algebra because $\rc\Pi_\ct\rc\subset\rc$ by
Proposition \ref{pr:mxyz}. 
It is easy to check that $\rc^\ct_\cs$ is a graded $C^*$-subalgebra of
$\rc$ supported by the ideal $\ccup_{X\in\ct}\cs_X$ generated by
$\ct$ in $\cs$. Indeed, we have
\[
\rc^\ct_\cs\,\ccap\,\rc(E)=
\left({\textstyle\sum_{X,Y\in\ct}^\rmc}\rc_{XY}\right) \ccap
\left({\textstyle\sum_{X,Y\in\cs}^\rmc}\rc_{XY}(E)\right) =
{\textstyle\sum_{X,Y\in\ct}^\rmc}\rc_{XY}(E).
\]
It is clear that $\rc$ is the inductive limit of the increasing
family of $C^*$-algebras $\rc^\ct_\cs$ with finite $\ct$.

If $\ct=\{X\}$ then $\rc^\ct_\cs$ is just $\rc_X$. If
$\ct=\{X,Y\}$ with distinct $X,Y$ we get a simple but nontrivial
situation. Indeed, we shall have $\ch_\ct=\ch_X\oplus\ch_Y$ and
$\rc^\ct_\cs$ may be thought as a matrix
\[
\rc^\ct_\cs=
\begin{pmatrix}
\rc_X & \rc_{XY}\\
\rc_{YX} & \rc_Y
\end{pmatrix}.
\]
The grading is now explicitly defined as follows: 
\begin{compactenum}
\item
If  $E\subset X\cap Y$ then
\[
\rc^\ct_\cs(E)=
\begin{pmatrix}
\rc_X(E) & \rc_{XY}(E)\\
\rc_{YX}(E) & \rc_Y(E)
\end{pmatrix}.
\]
\item   \label{p:2ex}
If  $E\subset X$ and $E\not\subset Y$ then
\[
\rc^\ct_\cs(E)=
\begin{pmatrix}
\rc_X(E) & 0\\
0 & 0
\end{pmatrix}.
\]
\item
If  $E\not\subset X$ and $E\subset Y$ then
\[
\rc^\ct_\cs(E)=
\begin{pmatrix}
0 & 0\\
0 & \rc_Y(E)
\end{pmatrix}.
\]
\end{compactenum}

The case when $\ct$ is of the form $\cs_X$ for some $X\in\cs$ is
especially interesting.  We denote $\rc_X^\#\equiv\rc_{\cs_X}$ and
we say that the $\cs_X$-graded $C^*$-algebra is the
\emph{unfolding} of the algebra $\rc_X$.  More explicitly
\begin{equation}\label{eq:xc}
\rc_X^\#\equiv{\textstyle\sum^\rmc_{Y,Z\in\cs_X}}\rc_{YZ}.
\end{equation}
The self-adjoint operators affiliated to $\rc_X$ live on the Hilbert
space $\ch_X$ and are (an abstract version of) Hamiltonians of an
$N$-particle system $\rs$ with a fixed $N$ (the configuration space
is $X$ and $N$ is the number of levels of the semilattice
$\cs_X$). The unfolding $\rc_X^\#$ lives on the ``Boltzmann-Fock
space'' $\ch_{\cs_X}$ and is obtained by adding interactions which
couple the subsystems of $\cs$ which have the groups $Y\in\cs_X$ as
configuration spaces and $\rc_Y$ as Hamiltonian algebras.

Clearly $\rc_X^\#\subset\rc_Y^\#$ if $X\subset Y$ and $\rc$ is the
inductive limit of the algebras $\rc_X^\#$.  Below we give an
interesting alternative description of $\rc_X^\#$.

\begin{theorem}\label{th:mor}
Let $\rn_X=\oplus_{Y\in\cs_X}\rc_{YX}$ be the direct sum of the
Hilbert $\rc_X$-modules $\rc_{YX}$ equipped with the direct sum graded
structure. Then $\ck(\rn_X) \cong \rc_X^\#$ the isomorphism being such
that the graded structure on $\ck(\rn_X)$ defined in Theorem
\ref{th:kghm} is transported into that of $\rc_X^\#$.  In other terms,
$\rc_X^\#$ is the imprimitivity algebra of the full Hilbert
$\rc_X$-module $\rn_X$ and $\rc_X$ and $\rc_X^\#$ are Morita
equivalent.
\end{theorem}
\proof If $Y\subset X$ then $\rc^*_{YX}\cdot\rc_{YX}=\rc_X^Y$ and
$\rc_{YX}$ is a full Hilbert $\rc_X^Y$-module. Since the $\rc_X^Y$
are ideals in $\rc_X$ and their sum over $Y\in\cs_X$ is equal to
$\rc_X$ we see that $\rn_X$ becomes a full Hilbert graded
$\rc_X$-module supported by $\cs_X$, cf.  Section \ref{s:grad}. By
Theorem \ref{th:kghm} the imprimitivity $C^*$-algebra $\ck(\rn_X)$
is equipped with a canonical $\cs_X$-graded structure.

We shall make a comment on $\ck(\mr)$ in the more general the case
when $\mr=\oplus_i\mr_i$ is a direct sum of Hilbert $\ra$-modules
$\mr_i$, cf. \S\ref{ss:gf}. First, it is clear that we have 
\[
\ck(\mr)={\textstyle\sum^\rmc_{ij}}\ck(\mr_j,\mr_i)\cong
(\ck(\mr_j,\mr_i))_{ij}.
\]
Now assume that $\ce,\ce_i$ are Hilbert spaces such that $\ra$ is a
$C^*$-algebra of operators on $\ce$ and $\mr_i$ is a Hilbert
$C^*$-submodule of $L(\ce,\ce_i)$ such that
$\ra_i\equiv\mr_i^*\cdot\mr_i$ is an ideal of $\ra$. 
Then by Proposition \ref{pr:2ss} we have
$\ck(\mr_j,\mr_i)\cong\mr_i\cdot\mr_j^*\subset L(\ce_j,\ce_i)$. 

In our case we take 
\[
i=Y\in\cs_X,\quad \mr_i=\rc_{YX}, \quad \ra=\rc_X,\quad
\ce=\ch_X,\quad  \ce_i=\ch_Y,\quad   \ra_i=\rc_X^Y.
\]
Then we get
\[
\ck(\mr_j,\mr_i)\equiv\ck(\rc_{ZX},\rc_{YX})\cong
\rc_{YX}\cdot\rc_{ZX}^*=\rc_{YX}\cdot\rc_{XZ}=\rc_{YZ}
\]
by Proposition \ref{pr:mxyz}.
\qed

\begin{remark}\label{re:squant}
  We understood the role in our work of the imprimitivity algebra of
  a Hilbert $C^*$-module thanks to a discussion with Georges
  Skandalis: he recognized (a particular case of) the main
  $C^*$-algebra $\rc$ we have constructed as the imprimitivity
  algebra of a certain Hilbert $C^*$-module. Theorem \ref{th:mor} is
  a reformulation of his observation and of his abstract
  construction of graded Hilbert $C^*$-modules in the present
  framework (at the time of the discussion our definition of $\rc$
  was rather different because we were working in a tensor product
  formalism).  More generally, if $\mr$ is a full Hilbert
  $\ra$-module then the imprimitivity $C^*$-algebra $\ck(\mr)$ could
  also be interpreted as Hamiltonian algebra of a system related in
  some natural way to the initial one.  For example, this is a
  natural method of ``second quantizing'' \mbox{$N$-body} systems,
  i.e.  introducing interactions which couple subsystems
  corresponding to different cluster decompositions of the $N$-body
  systems.  This is clear in the physical $N$-body situation
  discussed in \S\ref{ss:cexample}
\end{remark}

\section{An intrinsic description}
\label{s:id}
\protect\setcounter{equation}{0}

We begin with some preliminary facts on crossed products.  Let $X$
be a locally compact abelian group.  The next result, due to
Landstad \cite{Ld}, gives an ``intrinsic'' characterization of
crossed products of \mbox{$X$-algebras} by the action of $X$.  We
follow the presentation from \cite[Theorem 3.7]{GI4} which takes
advantage of the fact that $X$ is abelian.

\begin{theorem}\label{th:land}
A $C^*$-algebra $\ra\subset \rl_X$ is a crossed product
if and only for each $A\in\ra$ we have:
\begin{itemize}
\vspace{-2mm} 
\item
if $k\in X^*$ then $V_k^*AV_k\in\ra$ and
$\lim_{k\rarrow0}\|V_k^*AV_k-A\|=0$,
\item
if $x\in X$ then $U_xA\in\ra$ and $\lim_{x\rarrow0}\|(U_x-1)A\|=0$.
\vspace{-2mm}
\end{itemize} 
In this case one has $\ra=\ca\rtimes X$ for a unique $X$-algebra
$\ca\subset\Cbu(X)$  and this algebra is given by
\begin{equation}\label{eq:land}
\ca =\{\varphi\in\Cbu(X)\mid 
\varphi(Q)S \in {\ra} \hspace{1mm}\text{and}\hspace{1mm}
\bar\varphi(Q)S \in {\ra}
\hspace{1mm}\text{for all}\hspace{1mm} S \in \rt_X\}.
\end{equation} 
\end{theorem}
Note that the second condition above is equivalent
to $\rt_X\cdot\ra=\ra$, cf. Lemma \ref{lm:help}.

The following consequence of Landstad's theorem is an intrinsic 
description of $\rc_X(Y)$.

\begin{theorem}\label{th:cxy} 
$\rc_X(Y)$ is the set of $A\in \rl_X$ such that $U_y^*AU_y=A$ for all
$y\in Y$ and:
\begin{enumerate}
\item[{\rm(1)}]
$\|U_x^*AU_x-A\|\to 0$ if $x\to 0$ in $X$ and 
$\|V_k^*AV_k-A\|\to 0$ if $k\to 0$ in $X^*$,
\item[{\rm(2)}]
$\|(U_x-1)A\|\to 0$ if $x\to 0$ in $X$ and $\|(V_k-1)A\|\to 0$ if
$k\to 0$ in $Y^\perp$. 
\end{enumerate}   
\end{theorem}

By ``$k\to 0$ in $Y^\perp$'' we mean: $k\in Y^\perp$ and $k\to 0$.
Note that the second condition above is equivalent to:
\begin{equation}\label{eq:cxy}
\text{there are } \theta\in\rt_X,\ \psi\in\cc_X(Y)
\text{ and }  B,C\in\rl_X \text{ such that } 
A=\theta(P)B=\psi(Q)C.
\end{equation}
For the proof, use $Y^\perp\cong (X/Y)^*$ and apply Lemma
\ref{lm:help}. In particular, the last factorization shows that for
each $\varepsilon>$ there is a compact set $M\subset X$ such that
$\|\cchi_V(Q)A\|<\varepsilon$, where $V=X\setminus(M+Y)$.

\noindent{\bf Proof of Theorem \ref{th:cxy}:} Let $\ra\subset \rl_X$
be the set of operators $A$ satisfying the conditions from the
statement of the theorem.  We first prove that $\ra$ satisfies the
two conditions of Theorem \ref{th:land}. Let $A\in\ra$.  We have to
show that $A_p\equiv V_p^*AV_p\in\ra$ and $\|V_p^*AV_p-A\|\to0$ as
$p\to0$.  From the commutation relations $U_xV_p=p(x)V_pU_x$ we get
$\|(U_x-1)A_p\|=\|(U_x-p(x))A\|\to0$ if $x\to0$ and the second part
of condition 1 of the theorem is obviously satisfied by $A_p$. Then
for $y\in Y$
$$
U_y^*A_pU_y=U_y^*V_p^*AV_pU_y=V_p^*U_y^*AU_yV_p=V_p^*AV_p=A_p.
$$ Condition 2 is clear so we have $A_p\in\ra$ and the fact that
$\|V_p^*AV_p-A\|\to0$ as $p\to0$ is obvious. That $A$ satisfies the
second Landstad condition, namely that for each $a\in X$ we have
$U_aA\in\ra$ and $\|(U_a-1)A\|\to0$ as $a\to0$, is also clear because
$\|[U_a,V_k]\|\to0$ as $k\to0$.

Now we have to find the algebra $\ca$ defined by \eqref{eq:land}.
Assume that $\varphi\in\Cbu(X)$ satisfies $\varphi(Q)S\in\ra$ for all
$S\in \rt_X$. Since $U_y^*\varphi(Q)U_y=\varphi(Q-y)$ we get
$(\varphi(Q)-\varphi(Q-y))S=0$ for all such $S$ and all $y\in Y$,
hence $\varphi(Q)-\varphi(Q-y)=0$ which means $\varphi\in\Cbu(X/Y)$.
We shall prove that $\varphi\in\cc_X(Y)$ by reductio ad absurdum. 

If $\varphi\nin\cc_X(Y)$ then there is $\mu>0$ and there is a
sequence of points $x_n\in X$ such that $x_n/Y\to\infty$ and
$|\varphi(x_n)|>2\mu$. From the uniform continuity of $\varphi$ we
see that there is a compact neighborhood $K$ of zero in $X$ such
that $|\varphi|>\mu$ on $\bigcup_n(x_n+K)$.  Let $K'$ be a compact
neighborhood of zero such that $K'+K'\subset K$ and let us choose
two positive not zero functions $\psi,f\in\Cc(K')$. We define $S\in
\rt_X$ by $Su=\psi*u$ and recall that $\supp Su\subset\supp
\psi+\supp u$. Thus $\supp SU_{x_n}^*f\subset K'+x_n+K'\subset
x_n+K$. Now let $V$ be as in the remarks after \eqref{eq:cxy}. Since
$\pi_Y(x_n)\to\infty$ we have $x_n+K\subset V$ for $n$ large enough,
hence
$$
\|\cchi_V(Q)\varphi(Q)SU_{x_n}^*f\|\geq\mu\|SU_{x_n}^*f\|=
\mu\|Sf\| >0.
$$
On the other hand, for each $\varepsilon>0$ one can choose $V$ such
that $\|\cchi_V(Q)\varphi(Q)S\|<\varepsilon$.  Then we shall have
$\|\cchi_V(Q)\varphi(Q)SU_{x_n}^*f\|\leq\varepsilon\|f\|$ so
$\mu\|Sf\|\leq\varepsilon\|f\|$ for all $\varepsilon>0$ which is
absurd.  \qed

We now give a similar characterization of $\rc_{XY}(Z)$ where $X,Y$
is a compatible pair of closed subgroups of an lca group $G$.

\begin{theorem}\label{th:yzintr}
  $\rc_{XY}(Z)$ is the set of $T\in\rl_{XY}$ satisfying the
  following conditions:
\begin{enumerate} \vspace{-2mm}
\item[{\rm(1)}] $U_z^*T U_z=T$ if $z\in Z$ and $\|V^*_k T V_k-T\|\to
  0$ if $k\to 0$ in $(X+Y)^*$
\item[{\rm(2)}]
$\|(U_x-1)T\|\to 0$ if $x\to 0$ in $X$ and 
$\|T(U_y-1)\|\to 0$ if $y\to 0$ in $Y$, 
\item[{\rm(3)}]
$\|(V_k-1)T\|\to 0$ if $k\to 0$ in $(X/Z)^*$ and 
$\|T(V_k-1)\|\to 0$ if $k\to 0$ in $(Y/Z)^*$.
\end{enumerate}
\end{theorem}

Before the proof we make some preliminary comments. We think of
$X+Y$ as a closed subgroup of $G\in\rs$ which contains $X$ and $Y$
as closed subgroups.  Each character $k\in(X+Y)^*$ defines by
restriction a character $k|_X\in X^*$ and the map $k\mapsto k|_X$ is
a continuous open surjection. And similarly if $X$ is replaced by
$Y$. In (1) the operator $V_k$ acts in $L^2(X)$ as multiplication by
$k|_X$ and in $L^2(Y)$ as multiplication by $k|_Y$. In the first
part of (3) we take $k\in X^*$ and identify $(X/Z)^*$ with the
orthogonal of $Z$ in $X^*$ and similarly for the second part.

Assumptions (2) and (3) of Theorem \ref{th:yzintr} are decay
conditions in certain directions in $P$ and $Q$ space. Indeed, by
Lemma \ref{lm:help} condition (2) is equivalent to:
\begin{equation}\label{eq:cond1}
\text{there are } S_1\in \rt_X, S_2\in \rt_Y \text{ and }
R_1,R_2\in\rl_{XY} \text{ such that } T=S_1R_1=R_2S_2.
\end{equation}
Recall that $\rt_X\cong\Co(X^*)$ for example. Then
condition (3) is equivalent to:
\begin{equation}\label{eq:cond2}
\text{there are } S_1\in \cc_X(Z), S_2\in \cc_Y(Z) \text{ and }
R_1,R_2\in\rl_{XY} \text{ such that } T=S_1R_1=R_2S_2.
\end{equation}

\noindent{\bf Proof of Theorem \ref{th:yzintr}:} The set $\rc$ of all
the operators satisfying the conditions of the theorem is clearly a
closed subspace of $\rl_{XY}$.  We have $\rc_{X,Y}(Z)\subset\rc$
because \eqref{eq:cond1}, \eqref{eq:cond2} are satisfied by any
$T\in\rc_{XY}(Z)$ as a consequence of Theorem \ref{th:nxyz}.  Then we
get:
$$
\rc_Y(Z)= 
\rc_{XY}^*(Z)\cdot\rc_{XY}(Z)\subset \rc^*\cdot\rc, 
\hspace{1mm}
\rc_X(Z)=
\rc_{XY}(Z)\cdot\rc_{XY}^*(Z)\subset \rc\cdot\rc^*.
$$ 
We prove that equality holds in both these relations. We show, for
example, that $A\equiv TT^*$ belongs to $\rc_X(Z)$ if $T\in\rc$ and
for this we shall use Theorem \ref{th:cxy} with $Y$ replaced by
$Z$. That $U_z^*AU_z=A$ for $z\in Z$ is clear. From \eqref{eq:cond1}
we get $A=S_1R_1R_1^*S_1^*$ with $S_1\in\rt_X$ hence
$\|(U_x-1)A\|\to 0$ and $\|A(U_x-1)\|\to 0$ as $x\to0$ in $X$ are
obvious and imply $\|U_x^*AU_x-A\|\to 0$. Then \eqref{eq:cond2}
implies $A=\psi(Q)C$ with $\psi\in\cc_X(Z)$ and bounded $C$ hence
\eqref{eq:cxy} is satisfied.

That $\rc\rc_Y(Z)\subset\rc$ is easily proven because $T=SA$ has the
properties \eqref{eq:cond1} and \eqref{eq:cond2} if $S$ belongs to
$\rc$ and $A$ to $\rc_Y(Z)$, cf. Theorem \ref{th:cxy}. From what we
have shown above we get $\rc\rc^*\rc\subset\rc\rc_Y(Z)\subset\rc$ so
$\rc$ is a Hilbert $C^*$-submodule of $\rl_{XY}$. On the other hand,
$\rc_{XY}(Z)$ is a Hilbert $C^*$-submodule of $\rl_{XY}$ such that
$\rc_{XY}^*(Z)\cdot\rc_{XY}(Z)=\rc^*\cdot\rc$ and
$\rc_{XY}(Z)\cdot\rc_{XY}^*(Z)=\rc\cdot\rc^*$. Since
$\rc_{XY}(Z)\subset\rc$ we get $\rc=\rc_{XY}(Z)$ from Proposition
\ref{pr:clsubmod}.  \qed

If $Z=X\cap Y$ then Theorem \ref{th:yzintr} gives an intrinsic
description of the space $\rt_{XY}$. For example:

\begin{corollary}\label{co:txyintr}
If $X\supset Y$ then $\rt_{XY}$ is the set of $T\in\rl_{XY}$
satisfying $U_y^*T U_y=T$    if $y\in Y$ and such that:
$U_xT\to T$ if $x\to 0$ in $X$, 
$V^*_k T V_k\to T$ if $k\to 0$ in $X^*$ and 
$V_kT\to T$ if $k\to 0$ in $Y^\perp$.
\end{corollary}

In the rest of this section we describe the structure of the objects
introduced in Section \ref{s:grass} when the subgroups are
complemented, e.g. if $\cs$ consists of finite dimensional vector
spaces.

We say that \emph{$Z$ is complemented in $X$} if $X=Z\oplus E$ for
some closed subgroup $E$ of $X$. If $X,Z$ are equipped with Haar
measures then $X/Z$ is equipped with the quotient Haar measure and we
have $E\simeq X/Z$.  If $Z$ is complemented in $X$ and $Y$ then
$\rc_{XY}(Z)$ can be expressed as a tensor product.

\begin{proposition}\label{pr:def3}
If $Z$ is complemented in $X$ and $Y$ then
\begin{equation}\label{eq:ryzsum}
\rc_{XY}(Z)\simeq \rt_Z\otimes \rk_{X/Z,Y/Z}.
\end{equation}
If $Y\subset X$ then $\rt_{XY}\simeq \rt_Y\otimes L^2(X/Y)$ tensor
product of Hilbert $C^*$-modules.
\end{proposition} 
\proof Note first that the tensor product in \eqref{eq:ryzsum} is
interpreted as the exterior tensor product of the Hilbert
$C^*$-modules $\rt_Z$ and $\rk_{X/Z,Y/Z}$. Let $X=Z\oplus E$ and
$Y=Z\oplus F$ for some closed subgroups $E,F$. Then, as explained in
\S\ref{ss:ha}, we may also view the tensor product as the norm closure
in the space of continuous operators from $L^2(Y)\simeq L^2(Z)\otimes
L^2(F)$ to $L^2(X)\simeq L^2(Z)\otimes L^2(E)$ of the linear space
generated by the operators of the form $T\otimes K$ with $T\in
\rt_Z$ and $K\in \rk_{EF}$.

We now show that under the conditions of the proposition $X+Y\simeq
Z\oplus E\oplus F$ algebraically and topologically. The natural map
$\theta:Z\oplus E\oplus F\to Z+E+F=X+Y$ is a continuous bijective
morphism, we have to prove that it is open.  Since $X,Y$ are
compatible, the  map \eqref{eq:nat} is a continuous open
surjection.  If we represent $X\oplus Y\simeq Z\oplus Z\oplus E\oplus
F$ then this map becomes $\phi(a,b,c,d)=(a-b)+c+d$.  Let
$\psi=\xi\oplus{\rm id}_E\oplus{\rm id}_F$ where $\xi:Z\oplus Z\to Z$
is given by $\xi(a,b)=a-b$. Then $\xi$ is continuous surjective and
open because if $U$ is an open neighborhood of zero in $Z$ then
$U-U$ is also an open neighborhood of zero. Thus
$\psi:(Z\oplus Z)\oplus E\oplus F \to Z\oplus E\oplus F$ is a
continuous open surjection and $\phi=\theta\circ\psi$. So if $V$ is
open in $Z\oplus E\oplus F$ then there is an open 
$U\subset Z\oplus Z\oplus E\oplus F$ such that $V=\psi(U)$ and then
$\theta(V)=\theta\circ\psi(U)=\phi(U)$ is open in $Z+E+F$.

Thus we may identify $L^2(Y)\simeq L^2(Z)\otimes L^2(F)$ and
$L^2(X)\simeq L^2(Z)\otimes L^2(E)$ and we must describe the norm
closure of the set of operators $T_{XY}(\varphi)\psi(Q)$ with
$\varphi\in\Cc(X+Y)$ (cf. the remark after \eqref{eq:ryz} and the fact
that $X+Y$ is closed) and $\psi\in\Co(Y/Z)$. Since $X+Y\simeq Z\oplus
E\oplus F$ and $Y=Z\oplus F$ it suffices to describe the clspan of the
operators $T_{XY}(\varphi)\psi(Q)$ with
$\varphi=\varphi_Z\otimes\varphi_E\otimes\varphi_F$ and
$\varphi_Z,\varphi_E,\varphi_F$ continuous functions with compact
support on $Z,E,F$ respectively and $\psi=1\otimes\eta$ where $1$ is
the function identically equal to $1$ on $Z$ and
$\eta\in\Co(F)$. Then, if $x=(a,c)\in Z\times E$ and $y=(b,d)\in
Z\times F$, we get:
$$
(T_{XY}(\varphi)\psi(Q)u)(a,c)=\int_{Z\times F} \varphi_Z(a-b)
\varphi_E(c)
\varphi_F(d)  \eta(d) u(b,d) \rmd b \rmd d. 
$$ But this is just
$C(\varphi_Z)\otimes\ket{\varphi_E}\bra{\bar\eta\bar\varphi_F}$ where
$\ket{\varphi_E}\bra{\bar\eta\bar\varphi_F}$ is a rank one operator
$L^2(F)\to L^2(E)$ and $C(\varphi_Z)$ is the operator of convolution
by $\varphi_Z$ on $L^2(Z)$.  \qed

If $X\cap Y$ is complemented in $X$ and $Y$ then $\rc_{XY}$ can be
expressed (non canonically) as a tensor product.

\begin{proposition}\label{pr:xytens}
If $X\cap Y$ is complemented in $X$ and $Y$ then
\[
\rc_{XY}\simeq \rc_{X\cap Y} \otimes \rk_{X/Y,Y/X}.
\] 
In particular,
if $X\supset Y$ then $\rc_{XY}\simeq \rc_{Y} \otimes \ch_{X/Y}$.
\end{proposition}
\proof If $X=(X\cap Y)\oplus E$ and $Y=(X\cap Y)\oplus F$ then we have
to show that $\rc_{XY}\simeq \rc_{X\cap Y} \otimes \rk_{EF}$ where the
tensor product may be interpreted either as the exterior tensor
product of the Hilbert $C^*$-modules $\rc_{X\cap Y}$ and $\rk_{EF}$ or
as the norm closure in the space of continuous operators from
$L^2(Y)\simeq L^2(X\cap Y)\otimes L^2(F)$ to $L^2(X)\simeq L^2(X\cap
Y)\otimes L^2(E)$ of the algebraic tensor product of $\rc_{X\cap Y}$
and $\rk_{EF}$.  From Proposition \ref{pr:def3} with $Z=X\cap Y$ we
get $\rt_{XY}\simeq\rt_{X\cap Y}\otimes\rk_{EF}$. The relations
\eqref{eq:main} and the Definition \ref{df:main} imply
$\rc_{XY}=\rt_{XY}\cdot\cc_Y^{X\cap Y}$ and we clearly have
\[
\cc_Y^{X\cap Y}={\textstyle\sum_{Z\subset X\cap Y}^\rmc} \cc_Y(Z)
\simeq {\textstyle\sum_{Z\subset X\cap Y}^\rmc} 
\cc_{X\cap Y}(Z)\otimes \Co(F)
\simeq \cc_{X\cap Y}\otimes \Co(F).
\]     
Then we get
\[
\rc_{XY}\simeq \rt_{X\cap Y}\otimes\rk_{EF}\cdot 
\cc_{X\cap Y}\otimes \Co(F)=
\big(\rt_{X\cap Y}\cdot\cc_{X\cap Y} \big)\otimes
\big(\rk_{EF}\cdot\Co(F)\big)
\]
and this is $\rc_{X\cap Y} \otimes \rk_{EF}$.
\qed

If $Z$ is complemented in $X$ and $Y$ then Theorem \ref{th:yzintr}
can be improved. We shall describe this improvement only in the
Euclidean case which will be useful in our treatment of
nonrelativistic Hamiltonians. Thus below we assume that $X,Y$ are
subspaces of an Euclidean space (see \S\ref{ss:mouint} for
notations). Note that $V_k$ is the operator of multiplication by the
function $x\mapsto\rme^{i\braket{x}{k}}$ where the scalar product
$\braket{x}{k}$ is well defined for any $x,k$ in the ambient space
$\cx$. 

\begin{theorem}\label{th:xyzeintr}
$\rc_{XY}(Z)$ is the set of $T\in\rl_{XY}$ satisfying:
\begin{enumerate} 
\item[{\rm(1)}] 
$U_z^*T U_z=T$ for $z\in Z$ and
$\|V^*_z T V_z-T\|\to 0$ if $z\to 0$ in $Z$,
\item[{\rm(2)}] 
$\|(U_x-1)T\|\to 0$ if $x\to 0$ in $X$ and $\|(V_k-1)T\|\to 0$ if
$k\to 0$ in $X/Z$.
\end{enumerate}
\end{theorem}

\begin{remark}\label{re:xyzeintr}
Condition 2 may be replaced by
\begin{compactenum}
\item[{\rm(2$'$)}]
$\|T(U_y-1)\|\to 0$ if $y\to 0$ in $Y$ and $\|T(V_k-1)\|\to 0$ if
$k\to 0$ in $Y/Z$.
\end{compactenum}
This will be clear from the next proof.
\end{remark}
\proof Let $\cf\equiv\cf_Z$ be the Fourier transformation in the
space $Z$, this is a unitary operator in the space $L^2(Z)$ which
interchanges the position and momentum observables $Q_Z,P_Z$. We
denote also by $\cf$ the operators $\cf\otimes1_{\ch_{X/Z}}$ and
$\cf\otimes1_{\ch_{Y/Z}}$ which are unitary operators in the spaces
$\ch_X$ and $\ch_Y$ due to \eqref{eq:xyzint}.  If $S=\cf T
\cf^{-1}$ then $S$ satisfies the following conditions:
\begin{enumerate} 
\item[(i)]
$V_z^*S V_z=S$ for $z\in Z$, $\|(V_z-1)S\|\to 0$ if $z\to 0$ in $Z$,
and $\|U_z S U^*_z-S\|\to 0$ if $z\to 0$ in $Z$;
\item[(ii)] 
$\|(U_x-1)S\|\to 0$ and $\|(V_x-1)S\|\to 0$ if $x\to 0$ in $X/Z$.
\end{enumerate}
For the proof, observe that the first part of condition (2) may be
written as the conjunction of the two relations $\|(U_z-1)T\|\to 0$
if $z\to 0$ in $Z$ and $\|(U_x-1)T\|\to 0$ if $x\to 0$ in $X/Z$.  We
shall work in the representations
\begin{equation}\label{eq:fiber}
\ch_X= L^2(Z;\ch_{X/Z}) \quad \text{and} \quad
\ch_Y= L^2(Z;\ch_{Y/Z}).
\end{equation} 
From the relation $V_z^*S V_z=S$ for all $z\in Z$ it follows that
there is a bounded weakly measurable function
$S(\cdot):Z\to\rl_{X/Z,Y/Z}$ such that in the representations
\eqref{eq:fiber} $S$ is the operator of multiplication by
$S(\cdot)$. Then $\|U_z S U^*_z-S\|\to 0$ if $z\to 0$ in $Z$ means
that the function $S(\cdot)$ is uniformly continuous. And clearly
$\|(V_z-1)S\|\to 0$ if $z\to 0$ in $Z$ is equivalent to the fact
that $S(\cdot)$ tends to zero at infinity. Thus we see that
$S(\cdot)\in\Co(Z;\rl_{X/Z,Y/Z})$.
The condition (ii) can now be written
\[
\sup_{z\in Z}\big(\|(U_x-1)S(z)\|+ \|(V_x-1)S(z)\|\big)\to 0
\quad \text{if } x\to 0 \text{ in } X/Z.
\]
From the Riesz-Kolmogorov theorem it follows that each $S(z)$ is a
compact operator. Thus we have $S(\cdot)\in\Co(Z;\rk_{X/Z,Y/Z})$
which implies $T\in\rc_{XY}(Z)$ by Proposition \ref{pr:def3}.  \qed

\begin{remark}\label{re:half}
Since $S(\cdot)$ is continuous and tends to zero at infinity, for
each $\varepsilon>0$ there are points $z_1,\dots,z_n\in Z$ and
complex functions $\varphi_1,\dots,\varphi_n\in\Cc(Z)$ such that
\[
\|S(z)-{\textstyle\sum_k}\varphi_k(z)S(z_k)\|\leq\varepsilon\quad
\forall z\in Z.
\]
The operators $S(z_k)$ being compact, applying once again the
Riesz-Kolmogorov theorem we get
\[
\sup_{z\in Z}\big(\|S(z)(U_y-1)\|+ \|S(z)(V_y-1)\|\big)\to 0 
\quad \text{if } y\to 0 \text{ in } Y/Z.
\]
This explains why the second parts of conditions (2) and (3) of
Theorem \ref{th:yzintr} is not needed. 
\end{remark}

\section{Affiliated operators}
\label{s:af}
\protect\setcounter{equation}{0}

In this section we give examples of self-adjoint operators
affiliated to the algebra $\rc$ constructed in Section \ref{s:grass}
and then we give a formula for their essential spectrum. We refer to
\S\ref{ss:grca} for terminology and basic results related to the
notion of affiliation that we use and to \cite{ABG,GI1,DG3} for
details.

We recall that a self-adjoint operator $H$ on a Hilbert space $\ch$
is \emph{strictly affiliated} to a $C^*$-algebra of operators $\ra$
on $\ch$ if $(H+i)^{-1}\in\ra$ (then $\varphi(H)\in\ra$ for all
$\varphi\in\Co(\mbr)$) and if $\ra$ is the clspan of the elements
$\varphi(H)A$ with $\varphi\in\Co(\mbr)$ and $A\in\ra$. This class
of operators has the advantage that each time $\ra$ is
non-degenerately represented on a Hilbert space $\ch'$ with the help
of a morphism $\rp:\ra\to L(\ch')$, the observable $\rp H$ is
represented by a usual (densely defined) self-adjoint operator on
$\ch'$.

The diagonal algebra
\begin{equation}\label{eq:d}
\rt_{\rmd}\equiv(\rt_{\cs})_\rmd=\oplus_{X\in\cs} \rt_X
\end{equation}
has a simple physical interpretation: this is the $C^*$-algebra
generated by the kinetic energy operators.  Since
$\rc_{XX}=\rc_X\supset\rc_{X}(X)= \rt_X$ we see that $\rt_\rmd$
is a $C^*$-subalgebra of $\rc$. From \eqref{eq:nxyz},
\eqref{eq:txy1}, \eqref{eq:txy2} and the Cohen-Hewitt theorem we get
\begin{equation}\label{eq:ed}
\rc(Z)\rt_\rmd=\rt_\rmd\rc(Z)=\rc(Z)\quad \forall Z\in\cs \quad 
\text{and}\hspace{2mm} \rc \rt_\rmd=\rt_\rmd\rc=\rc.
\end{equation} 
In other terms, $\rt_\rmd$ acts
non-degenerately\symbolfootnote[2]{\ Note that if $\cs$ has a
  largest element $\cx$ then the algebra $\rc(\cx)$ acts on each
  $\rc(Z)$ but this action is degenerate.} 
on each $\rc(Z)$ and on $\rc$. It follows that a self-adjoint
operator strictly affiliated to $\rt_\rmd$ is also strictly
affiliated to $\rc$.

For each $X\in\cs$ let $h_X:X^*\to\mbr$ be a continuous function
such that $|h_X(k)|\to\infty$ if $k\to\infty$ in $X^*$. Then the
self-adjoint operator $K_X\equiv h_X(P)$ on $\ch_X$ is strictly
affiliated to $\rt_X$ and the norm of $(K_X+i)^{-1}$ is equal to
$\sup_k(h^2_X(k)+1)^{-1/2}$. Let $K\equiv\bigoplus_{X\in\cs}K_X$,
this is a self-adjoint operator $\ch$. Clearly $K$ is affiliated to 
$\rt_\rmd$ if and only if 
\begin{equation}\label{eq:kin}
\lim_{X\to\infty}\sup\nolimits_{k}(h^2_X(k)+1)^{-1/2}=0
\end{equation}
and then $K$ is strictly affiliated to $\rt_\rmd$ (the set $\cs$ is
equipped with the discrete topology). If the functions $h_X$ are
positive this means that $\min h_X$ tends to infinity when
$X\to\infty$. One could avoid such a condition by considering an
algebra larger then $\rc$ such as to contain 
$\prod_{X\in\cs} \rt_X$, but we shall not develop this here.

Now let $H=K+I$ with $I\in\rc$ (or in the multiplier algebra) a
symmetric element. Then
\begin{equation}\label{eq:res}
(\lambda-H)^{-1}=(\lambda-K)^{-1}\left(1-I(\lambda-K)^{-1}\right)^{-1}
\end{equation}
if $\lambda\nin\sp(H)\cup\sp(K)$ . Thus $H$ is strictly affiliated
to $\rc$.  We interpret $H$ as the Hamiltonian of our system of
particles when the kinetic energy is $K$ and the interactions
between particles are described by $I$. Even in the simple case
$I\in\rc$ these interactions are of a very general nature being a
mixture of $N$-body and quantum field type interactions (which
involve creation and annihilation operators so the number of
particles is not preserved).

We shall now use Theorem \ref{th:gas} in order to compute the
essential spectrum of an operator like $H$. The case of unbounded
interactions will be treated later on. Let $\rc_{\geq E}$ be the
$C^*$-subalgebra of $\rc$ determined by $E\in\cs$ according to the
rules of $\S\ref{ss:grca}$. More explicitly, we set
\begin{equation}\label{eq:geqe}
\rc_{\geq E}={\textstyle\sum^\rmc_{F\supset E}\rc(F)}\cong 
\big({\textstyle\sum^\rmc_{F\supset E}}
\rc_{XY}(F)\big)_{X\cap Y\supset E}
\end{equation}
and note that $\rc_{\geq E}$ lives on the subspace $\ch_{\geq
  E}=\bigoplus_{X\supset E}\ch_X$ of $\ch$. Since in the second sum
from \eqref{eq:geqe} the group $F$ is such that $E\subset F\subset
X\cap Y$ the algebra $\rc_{\geq E}$ is strictly included in the
algebra $\rc_\ct$ obtained by taking $\ct=\{F\in\cs \mid F\supset
E\}$ in \eqref{eq:t}. 

Let $\rp_{\geq E}$ be the canonical idempotent morphism of $\rc$
onto $\rc_{\geq E}$ introduced in \S\ref{ss:grca}. We consider
the self-adjoint operator on the Hilbert space $\ch_{\geq E}$
defined as follows:
\begin{equation}\label{eq:post}
 H_{\geq E}=K_{\geq E}+I_{\geq E} \quad \text{where} \quad 
K_{\geq E}= \oplus_{X\geq E} K_X \hspace{2mm} \text{and} 
\hspace{2mm} I_{\geq E}=\rp_{\geq E}I.
\end{equation}
Then $H_{\geq E}$ is strictly affiliated to $\rc_{\geq E}$ and it
follows easily from \eqref{eq:res} that
\begin{equation}\label{eq:pre}
\rp_{\geq E}\varphi(H)=\varphi(H_{\geq E}) \quad \forall
\varphi\in\Co(\mbr).
\end{equation} 
Now let us assume that the group $O=\{0\}$ belongs to $\cs$. Then we
have
\begin{equation}\label{eq:O}
\rc(O)=K(\ch).
\end{equation} 
Indeed, from \eqref{eq:nxyz} we get
$\rc_{XY}(O)=\rt_{XY}\cdot\Co(Y)=\rk_{XY}$ which implies the
preceding relation. If we also assume that $\cs$ is atomic and we
denote $\cp(\cs)$ its set of atoms, then from Theorem \ref{th:ga} we
get a canonical embedding
\begin{equation}\label{eq:quotc}
\rc/K(\ch)\subset\pprod\nolimits_{E\in\cp(\cs)}\rc_{\geq E}
\end{equation} 
defined by the morphism $\rp\equiv(\rp_{\geq E})_{E\in\cp(\cs)}$.
Then from \eqref{eq:es2} we obtain:
\begin{equation}\label{eq:ess1}
\spe(H)=\overline{\ccup}_{E\in\cp(\cs)}\sp(H_{\geq E}).
\end{equation}
Our next purpose is to prove a similar formula for a certain class
of unbounded interactions $I$.

Let $\cg\equiv\cg_\cs=D(|K|^{1/2})$ be the form domain of $K$
equipped with the graph topology. Then $\cg\subset\ch$ continuously
and densely so after the Riesz identification of $\ch$ with its
adjoint space $\ch^*$ we get the usual scale
$\cg\subset\ch\subset\cg^*$ with continuous and dense embeddings.
Let us denote
\begin{equation}\label{eq:jap}
\jap{K} =|K+i|=\sqrt{K^2+1}.
\end{equation}
Then $\jap{K}^{1/2}$ is a self-adjoint operator on $\ch$ with domain
$\cg$ and $\jap{K}$ induces an isomorphism $\cg\to\cg^*$.  The
following result is a straightforward consequence of Theorem 2.8 and
Lemma 2.9 from \cite{DG3}. 

\begin{theorem}\label{th:af}
Let $I:\cg\to\cg^*$ be a continuous symmetric operator and let us
assume that there are real numbers $\mu,a$ with $0<\mu<1$ such that
one of the following conditions is satisfied:
\begin{compactenum}[(i)]
\item
$\pm I \leq\mu|K+ia|,$
\item
$K$ is bounded from below and $ I \geq -\mu|K+ia|.$
\end{compactenum}
Let $H=K+I$ be the form sum of $K$ and $I$, so $H$ has as domain the
set  of $u\in\cg$ such that $Ku+Iu\in\ch$ and acts as $Hu=Ku+Iu$. 
Then $H$ is a self-adjoint operator on $\ch$. If there is
$\alpha>1/2$ such that 
$\langle K\rangle^{-1/2}I\langle K\rangle^{-\alpha}\in\rc$ then $H$
is strictly affiliated to $\rc$. 
If $O\in\cs$ and the semilattice $\cs$ is atomic then
\begin{equation}\label{eq:ess2}
\spe(H)=\overline{\ccup}_{E\in\cp(\cs)}\sp(H_{\geq E}).
\end{equation}
\end{theorem}

The last assertion of the theorem follows immediately from Theorem
\ref{th:gas} and is a general version of the HVZ theorem.  In order
to have a more explicit description of the observables $H_{\geq
  E}\equiv\rp_{\geq E}H$ we now prove an analog of Theorem 3.5 from
\cite{DG3}. We cannot use that theorem in our context for three
reasons: first we did not suppose that $\cs$ has a maximal element,
then even if $\cs$ has a maximal element $\cx$ the action of the
corresponding algebra $\rc(\cx)$ on the algebras $\rc(E)$ is
degenerate, and finally our ``free'' operator $K$ is not affiliated
to $\rc(\cx)$.

\begin{theorem}\label{th:afi}
For each $E\in\cs$ let $I(E)\in L(\cg,\cg^*)$ be a symmetric
operator such that:
\begin{compactenum}[(i)]
\item
$\jap{K}^{-1/2}I(E)\jap{K}^{-\alpha}\in\rc(E)$ for some
$\alpha> 1/2$ independent of $E$,
\item
there are real positive numbers $\mu_E,a$ such that either $\pm
I(E) \leq\mu_E|K+ia|$ for all $E$ or $K$ is bounded from below and
$ I(E) \geq -\mu_E|K+ia|$ for all $E$,
\item
we have $\sum_E\mu_E\equiv\mu<1$ and the series $\sum_E I(E)\equiv I$
is norm summable in $L(\cg,\cg^*)$.
\end{compactenum}
Let us set $I_{\geq E}=\sum_{F\geq E}I(F)$. Define the self-adjoint
operator $H=K+I$ on $\ch$ as in Theorem \ref{th:af} and define
similarly the self-adjoint operator $H_{\geq E}=K_{\geq E}+I_{\geq
  E}$ on $\ch_{\geq E}$.  Then the operator $H$ is strictly
affiliated to $\rc$, the operator $H_{\geq E}$ is strictly
affiliated to $\rc_{\geq E}$, and we have $\rp_{\geq E}H=H_{\geq E}$.
\end{theorem}
\proof  We shall consider only the case when $\pm I(E)
\leq\mu_E|K+ia|$ for all $E$. The more singular situation when $K$
is bounded from below but there is no restriction on the positive
part of the operators $I(E)$ (besides summability) is more difficult
but the main idea has been explained in \cite{DG3}.

We first make some comments to clarify the definition of the
operators $H$ and $H_{\geq E}$.  Observe that our assumptions imply
$\pm I\leq\mu|K+ia|$ hence if we set
\[
\Lambda\equiv|K+ia|^{-1/2}=(K^2+a^2)^{-1/4}\in \rt_\rmd
\]
then  we obtain
\[
\pm\braket{u}{Iu}\leq\mu\braket{u}{|K+ia|u}=
\mu\| |K+ia|^{1/2}u\| = \mu\| \Lambda^{-1} u\|
\] 
which is equivalent to $\pm\Lambda I\Lambda\leq\mu$ or $\|\Lambda
I\Lambda\|\leq\mu$. In particular we may use Theorem \ref{th:af}
in order to define the self-adjoint operator $H$. Moreover, we have
\[
\jap{K}^{-1/2}I\jap{K}^{-\alpha}={\textstyle\sum_E}
\jap{K}^{-1/2}I(E)\jap{K}^{-\alpha}\in\rc
\]
because the series is norm summable in $L(\ch)$. Thus $H$ is
strictly affiliated to $\rc$.  

In order to define $H_{\geq E}$ we first make a remark on 
$I_{\geq E}$.  If we set $\cg_X=D(|K_X|^{-1/2})$ and if we equip
$\cg$ and $\cg_X$ with the norms  \label{p:formd}
\[
\|u\|_\cg=\|\jap{K}^{1/2}u\|_\ch \quad \text{and} \quad
\|u\|_{\cg_X}=\|\jap{K_X}^{1/2}u\|_{\ch_X}
\]
respectively then clearly $\cg=\oplus_X\cg_X$ and
$\cg^*=\oplus_X\cg^*_X $ where the sums are Hilbertian direct sums
and $\cg^*$ and $\cg_X^*$ are equipped with the dual norms.  Then
each $I(F)$ may be represented as a matrix
$I(F)=(I_{XY}(F))_{X,Y\in\cs}$ of continuous operators
$I_{XY}(E):\cg_Y\to\cg^*_X$. Clearly
\[
\jap{K}^{-1/2}I(F)\jap{K}^{-\alpha}=
\left(\jap{K_X}^{-1/2}I_{XY}(F)\jap{K_Y}^{-\alpha}\right)_{X,Y\in\cs}
\] 
and since by assumption (i) this belongs to $\rc(F)$ we see that
$I_{XY}(F)=0$ if $X\not\supset F$ or $Y\not\supset F$. Now fix $E$
and let $F\supset E$. Then, when viewed as a sesquilinear form,
$I(F)$ is supported by the subspace $\ch_{\geq E}$ and has domain
$\cg_{\geq E}= D(|K_{\geq E}|^{1/2}$.  It follows that $I_{\geq E}$ 
is a sesquilinear form with domain $\cg_{\geq E}$ supported by the
subspace $\ch_{\geq E}$ and may be thought as an element of
$L(\cg_{\geq E},\cg^*_{\geq E})$ such that
$\pm I_{\geq E}\leq \mu |K_{\geq E}+ia|$ because 
$\sum_{F\supset E}\mu_F\leq \mu$. To conclude, we may now define 
$H_{\geq E}=K_{\geq E}+I_{\geq E}$ exactly as in the case of $H$ and
get a self-adjoint operator on $\ch_{\geq E}$ strictly affiliated to
$\rc_{\geq E}$. Note that this argument also gives
\begin{equation}\label{eq:ek}
\jap{K}^{-1/2} I(F) \jap{K}^{-1/2}=
\jap{K_{\geq E}}^{-1/2} I(F) \jap{K_{\geq E}}^{-1/2}.
\end{equation}
It remains to be shown that $\rp_{\geq E}H=H_{\geq E}$. If we set
$R\equiv(ia-H)^{-1}$ and $R_{\geq E}\equiv(ia-H_{\geq E})^{-1}$ then
this is equivalent to $\rp_{\geq E}R=R_{\geq E}$.  Let us set
\[
U=|ia-K|(ia-K)^{-1}=\Lambda^{-2}(ia-K)^{-1}, \quad
J=\Lambda I\Lambda U.
\]
Then $U$ is a unitary operator and $\|J\|<1$, so we get a norm
convergent series expansion
\[
R=(ia-K-I)^{-1}=
\Lambda U(1-\Lambda I\Lambda U)^{-1}\Lambda =
{\textstyle\sum_{n\geq0}}\Lambda U J^n\Lambda
\]
which implies
\[
\rp_{\geq E} (R)=
{\textstyle\sum_{n\geq0}}
\rp_{\geq E}\big(\Lambda U J^{n}\Lambda\big)
\]
the series being norm convergent. Thus it suffices to prove that for
each $n\geq0$ 
\begin{equation}\label{eq:ekk}
\rp_{\geq E}\big(\Lambda U J^{n}\Lambda\big)=
\Lambda_{\geq E} (J_{\geq E})^{n}\Lambda_{\geq E}
\end{equation}
where $J_{\geq E}=\Lambda_{\geq E} I_{\geq E}\Lambda_{\geq E}
U_{\geq E}$.  Here $\Lambda_{\geq E}$ and $U_{\geq E}$ are
associated to $K_{\geq E}$ in the same way $\Lambda$ and $K$ are
associated to $K$. For $n=0$ this is obvious because 
$\rp_{\geq E}K=K_{\geq E}$. If $n=1$ this is easy because
\begin{align}\label{eq:e}
\Lambda U J\Lambda &= \Lambda U \Lambda I \Lambda U\Lambda=
(ia-K)^{-1} I (ia-K)^{-1} \\
&=
[(ia-K)^{-1}\jap{K}^{1/2}] \cdot 
[\jap{K}^{-1/2} I \jap{K}^{-\alpha}] \cdot
[\jap{K}^{\alpha} (ia-K)^{-1}] \nonumber
\end{align}
and it suffices to note that 
$\rp_{\geq E}(\jap{K}^{-1/2} I(F) \jap{K}^{-\alpha})=0$ if
$F\not\supset E$ and to use \eqref{eq:ek} for $F\supset E$. 

To treat the general case we make some preliminary remarks. 
If $J(F)=\Lambda I(F) \Lambda U$ then $J=\sum_F J(F)$ where the
convergence holds in norm on $\ch$ because of the condition (iii). 
Then we have a norm convergent expansion
\[
\Lambda U J^n \Lambda ={\textstyle\sum_{F_1,\dots,F_n\in\cs}}
\Lambda U J(F_1)\dots J(F_n) \Lambda.
\]
Assume that we have shown $\Lambda U J(F_1)\dots
J(F_n)\Lambda\in\rc(F_1\cap\dots\cap F_n)$.  Then we get
\begin{equation}\label{eq:ekkk}
\rp_{\geq E}(\Lambda U J^n \Lambda)=
{\textstyle\sum_{F_1\geq E,\dots,F_n\geq E}}
\Lambda U J(F_1)\dots J(F_n) \Lambda
\end{equation}
because if one $F_k$ does not contain $E$ then the intersection
$F_1\cap\dots\cap F_n$ does not contain $E$ hence $\rp_{\geq E}$
applied to the corresponding term gives $0$. Because of
\eqref{eq:ek} we have $J(F)=\Lambda_{\geq E} I(F) \Lambda_{\geq E}
U_{\geq E}$ if $F\supset E$ and we may replace everywhere in the
right hand side of \eqref{eq:ekkk} $\Lambda$ and $U$ by
$\Lambda_{\geq E}$ and $U_{\geq E}$. This clearly proves
\eqref{eq:ekk}. 

Now we prove the stronger fact 
$\Lambda U J(F_1)\dots J(F_n)\in\rc(F_1\cap\dots\cap F_n)$.  
If $n=1$ this follows from a slight modification of
\eqref{eq:e}: the last factor on the right hand side of \eqref{eq:e}
is missing but is not needed. Assume that the assertion holds for
some   $n$. Since $K$ is strictly affiliated to $\rt_\rmd$ and
$\rt_\rmd$ acts non-degenerately on each $\rc(F)$ we may use the
Cohen-Hewitt theorem to deduce that there is $\varphi\in\Co(\mbr)$
such that
$\Lambda U J(F_1)\dots J(F_n)=T\varphi(K)$ 
for some $T\in\rc(F_1\cap\dots\cap F_n)$. 
Then
\[
\Lambda U J(F_1)\dots J(F_n)J(F_{n+1})=T\varphi(K)J(F_{n+1})
\]
hence it suffices to prove that $\varphi(K)J(F)\in\rc(F)$ for any
$F\in\cs$ and any $\varphi\in\Co(\mbr)$. But the set of $\varphi$
which have this property is a closed subspace of $\Co(\mbr)$ which
clearly contains the functions $\varphi(\lambda)=(\lambda -z)^{-1}$
if $z$ is not real hence is equal to $\Co(\mbr)$.  \qed

\begin{remark}\label{re:alpha}
Choosing $\alpha>1/2$ allows one to consider perturbations of $K$
which are of the same order as $K$, e.g. in the $N$-body situations
one may add to the Laplacian $\Delta$ on operator like $\nabla^*
M\nabla$ where the function $M$ is bounded measurable and has the
structure of an $N$-body type potential, cf. \cite{DG3,DerI}.  
\end{remark}

The only assumption of Theorem \ref{th:afi} which is really relevant
is $\jap{K}^{-1/2}I(E)\jap{K}^{-\alpha}\in\rc(E)$.  We shall give
below more explicit conditions which imply it.  If we change
notation $E\to Z$ and use the formalism introduced in the proof of
Theorem \ref{th:afi} we have
\begin{equation}\label{eq:xye}
I(Z)=(I_{XY}(Z))_{X,Y\in\cs} \quad \text{with} \quad
I_{XY}(Z):\cg_Y\to\cg^*_X \text{ continuous}.
\end{equation}
We are interested in conditions on $I_{XY}(Z)$ which imply
\begin{equation}\label{eq:xye1}
\jap{K_X}^{-1/2}I_{XY}(Z)\jap{K_X}^{-\alpha} \in \rc_{XY}(Z).
\end{equation}
For this we shall use Theorem \ref{th:yzintr} which gives a simple
intrinsic characterization of $\rc_{XY}(Z)$.

The construction which follows is interesting only if $X$ is not a
discrete group, otherwise $X^*$ is compact and many conditions are
trivially satisfied. We shall use weights only in order to avoid
imposing on the functions $h_X$ regularity conditions stronger than
continuity.

A positive function $w$ on $X^*$ is a \emph{weight} if
$\lim_{k\to\infty} w(k)=\infty$ and $w(k+p)\leq\omega(k)w(p)$ for
some function $\omega$ on $X^*$ and all $k,p$. We say that $w$ is
\emph{regular} if one may choose $\omega$ such that
$\lim_{k\to0}\omega(k)=1$.  The example one should have in mind when
$X$ is an Euclidean space is $w(k)=\jap{k}^s$ for some $s>0$. Note
that we have $\omega(-k)^{-1}\leq w(k+p)w(p)^{-1} \leq \omega(k)$
hence if $w$ is a regular weight then \label{p:regw}
\begin{equation}\label{eq:regw}
\theta(k)\equiv \sup_{p\in X^*}\frac{|w(k+p)-w(p)|}{w(p)} 
\Longrightarrow
\lim_{k\to0}\theta(k)=0.
\end{equation}
It is clear that if $w$ is a regular weight and $\sigma\geq 0$ is a
real number then $w^\sigma$ is also a regular weight.

We say that two functions $f,g$ defined on a neighborhood of 
infinity of $X^*$ are \emph{equivalent} and we write $f\sim g$ if
there are numbers $a,b$ such that $a|f(k)|\leq|g(k)|\leq
b|f(k)|$. Then $|f|^\sigma\sim|g|^\sigma$ for all $\sigma>0$.

We denote $\cg^\sigma_X=D(|K_X|^{\sigma/2})$ and
$\cg^{-\sigma}_X\equiv(\cg^{\sigma}_X)^*$ with $\sigma \geq 1$.  In
particular $\cg^1_X=\cg_X$ and $\cg^{-1}_X=\cg^*_X$.

\begin{proposition}\label{pr:tex}
  Assume that $h_X,h_Y$ are equivalent to regular weights. Let
  $Z\subset X\cap Y$ and let $I_{XY}(Z)$ be a continuous map
  $\cg_Y\to\cg^*_X$ such that
\begin{enumerate}
\item
$U_z I_{XY}(Z)=I_{XY}(Z) U_z$ if $z\in Z$ and
$V^*_k I_{XY}(Z) V_k\to I_{XY}(Z)$ if $k\to 0$ in $(X+Y)^*$,
\item 
$(U_x-1)I_{XY}(Z)\to 0$ if $x\to 0$ in $X$ and
$(V_k-1) I_{XY}(Z)\to 0$ if $k\to 0$ in $(X/Z)^*$,
\end{enumerate}
where the limits hold in norm in $L(\cg^{\sigma}_Y,\cg^{-1}_X)$ for
some $\sigma\geq1$. Then \eqref{eq:xye1} holds with
$\alpha=\sigma/2$.
\end{proposition}
\proof We begin with some general comments on weights. Let $w$ be a
regular weight and let $\cg_X$ be the domain of the operator $w(P)$
in $\ch_X$ equipped with the norm $\|w(P)u\|$. Then $\cg_X$ is a
Hilbert space and if $\cg^*_X$ is its adjoint space then we get a
scale of Hilbert spaces $\cg_X\subset\ch_X\subset\cg^*_X$ with
continuous and dense embeddings. Since $U_x$ commutes with $w(P)$ it
is clear that $\{U_x\}_{x\in X}$ induces strongly continuous unitary
representation of $X$ on $\cg_X$ and $\cg^*_X$.  Then
\[
\|V_k u\|_{\cg_X}=\|w(k+P)u\|\leq\omega(k)\|u\|_{\cg_X}
\]
from which it follows that $\{V_k\}_{k\in X^*}$ induces by
restriction and extension strongly continuous representations of
$X^*$ in $\cg_X$ and $\cg^*_X$. Moreover, as operators on $\ch_X$
we have \label{p:RK}
\begin{align}  
|V_k^*w(P)^{-1}V_k-w(P)^{-1}| 
&=|w(k+P)^{-1}-w(P)^{-1}| 
= |w(k+P)^{-1}(w(P)-w(k+P))w(P)^{-1}|  \nonumber \\
& \leq \omega(-k)|(w(P)-w(k+P))w(P)^{-2}|
\leq \omega(-k)\theta(k) w(P)^{-1}. \label{eq:refw} 
\end{align}
Now let $w_X,w_Y$ be regular weights equivalent to
$|h_X|^{1/2},|h_Y|^{1/2}$ and let us set $S=I_{XY}(Z)$. Then
\[
\jap{K_X}^{-1/2}S\jap{K_Y}^{-\alpha}=
\jap{K_X}^{-1/2}w_X(P)\cdot 
w_X(P)S w_Y(P)^{-2\alpha} \cdot
w_Y(P)^{2\alpha}\jap{K_Y}^{-\alpha}
\]
and $\jap{h_X}^{-1/2}w_X$, $\jap{h_Y}^{-\alpha}w_Y^{2\alpha}$ and
their inverses are bounded continuous functions on $X,Y$. Since
$\rc_{XY}(Z)$ is a non-degenerate left $\rt_X$-module and right
$\rt_Y$-module we may use the Cohen-Hewitt theorem to deduce that
\eqref{eq:xye1} is equivalent to
\begin{equation}\label{eq:xye2}
w_X(P)^{-1}I_{XY}(Z) w_Y(P)^{-\sigma} \in \rc_{XY}(Z)
\end{equation}
where $\sigma=2\alpha$.  To simplify notations we set $W_X=w_X(P),
W_Y=w^\sigma_Y(P)$.  We also omit the index $X$ or $Y$ for the
operators $W_X,W_Y$ since their value is obvious from the
context. In order to show $W^{-1}SW^{-1}\in \rc_{XY}(Z)$ we check
the conditions of Theorem \ref{th:yzintr} with $T=W^{-1}SW^{-1}$.
The first part of condition (2) of the theorem is verified by the
hypothesis (2) of the present proposition.  We may assume $\sigma>1$
and then hence the second part of condition (2) of the theorem
follows from
\[
\|T(U_y-1)\|\leq \|W^{-1}I_{XY}(Z)w_Y^{-1}(P)\|\to 0 
\|(U_y-1)w_Y^{1-\sigma}(P)\| \quad \text{if } y\to0.
\]
To check the second part of condition (1) of the theorem set
$W_k=V_k^*WV_k$ and $S_k=V_k^* SV_k$ and write
\begin{align*}
V_k^*TV_k-T 
&= W_k^{-1} S_k W_k^{-1}-W^{-1}SW^{-1}\\
&= (W_k^{-1}-W^{-1})S_kW_k^{-1} + W^{-1}(S_k-S)W_k^{-1}
+W^{-1}S(W_k^{-1} -W^{-1}).
\end{align*}
Now if we use \eqref{eq:refw} and set $\xi(k)=\omega(-k)\theta(k)$
we get
\begin{align*}
\|V_k^*TV_k-T\| &\leq 
\xi(k)\|W^{-1}S_kW_k^{-1}\| + \|W^{-1}(S_k-S)W^{-1}\|\|W W_k^{-1}\|
+\xi(k)\|W^{-1}SW^{-1}\|
\end{align*}
which clearly tends to zero if $k\to0$. Condition
(3) of Theorem \ref{th:yzintr} follows by a similar argument.
\qed

Now let $H$ be defined according to the algorithm of \S\ref{ss:ex}.
Then condition (i) of Theorem \ref{th:afi} will be satisfied for all
$\alpha >1/2$. Indeed, from Proposition \ref{pr:tex} we get
$\jap{K}^{-1/2}\Pi_\ct I(Z)\Pi_\ct\jap{K}^{-\alpha}\in\rc(Z)$ for
any finite $\ct$ and this operator converges in norm to
$\jap{K}^{-1/2} I(Z)\jap{K}^{-\alpha}$.  Thus all conditions of
Theorem \ref{th:afi} are fulfilled by the Hamiltonian $H=K+I$ and so
$H$ is strictly affiliated to $\rc$.  \label{p:algor}

\section{The Mourre estimate} 
\label{s:mou}
\protect\setcounter{equation}{0}

\subsection{Proof of the Mourre estimate}
\label{ss:mest}

From now on we work in the framework of the second part of Section
\ref{s:euclid}, so we assume that $\cs$ is a \emph{finite}
semilattice of finite dimensional subspaces of an Euclidean
space. In this subsection we prove the Mourre estimate for
nonrelativistic Hamiltonians.  The strategy of the proof is that
introduced in \cite{BG2} and further developed in \cite{ABG,DG2}
(graded $C^*$-algebras over infinite semilattices and dispersive
Hamiltonians are considered in Section 5 from \cite{DG2}).  We
choose the generator $D$ of the dilation group $W_\tau$ in $\ch$ as
conjugate operator for reasons explained below. For special types of
interactions, similar to those occurring in quantum field models,
which are allowed by our formalism, better choices can be made, but
at a technical level there is nothing new in that with respect to
\cite{Geo} (these special interactions correspond to distributive
semilattices $\cs$).

The dilations implement a group of automorphisms of the
$C^*$-algebra $\rc$ which is compatible with the grading, i.e. it
leaves invariant each component $\rc(Z)$ of $\rc$.  In fact, it is
clear that $W_\tau^*\rc_{XY}(Z)W_\tau=\rc_{XY}(Z)$ for all $X,Y,Z$
hence $W_\tau^*\rc(Z)W_\tau=\rc(Z)$.  This fact plays a fundamental
role in the proof of the Mourre estimate for operators affiliated to
$\rc$ and explains the choice of $D$ as conjugate operator.
Moreover, for each $T\in\rc$ the map $\tau\mapsto W^*_\tau T W_\tau$
is norm continuous. We can compute explicitly the function
$\what\rho_H$ thanks to the relation
\begin{equation}\label{eq:dlap}
W^*_\tau \Delta_X W_\tau = \rme^\tau\Delta_X \quad \text{or}\quad
[\Delta_X, i D]=\Delta_X 
\end{equation}
We say that a self-adjoint operator $H$ \emph{is of class $C^1(D)$}
or \emph{of class $C^1_\rmu(D)$} if $W^*_\tau RW_\tau$ as a function
of $\tau$ is of class $C^1$ strongly or in norm respectively.  Here
$R=(H-z)^{-1}$ for some $z$ outside the spectrum of $H$.  The formal
relation
\begin{equation}\label{eq:dres}
[D,R]= R[H,D] R 
\end{equation}
can be given a rigorous meaning as follows. If $H$ is of class
$C^1(D)$ then the intersection $\rd$ of the domains of the operators
$H$ and $D$ is dense in $D(H)$ and the sesquilinear form with domain
$\rd$ associated to the formal expression $HD-DH$ is continuous for
the topology of $D(H)$ so extends uniquely to a continuous
sesquilinear form on the domain of $H$ which is denoted
$[H,D]$. This defines the right hand side of \eqref{eq:dres}.  The
left hand side can be defined for example as 
$i\frac{d}{d\tau}W_\tau^*RW_\tau|_{\tau=0}$. 

For Hamiltonians as those considered here it is easy to decide that
$H$ is of class $C^1(D)$ in terms of properties of the commutator
$[H, D]$.  Moreover, the following is easy to prove: \emph{if $H$
  is affiliated to $\rc$ then $H$ is of class $C^1_\rmu(D)$ if and
  only if $H$ is of class $C^1(D)$ and $[R,D]\in\rc$}.

Let $H$ be of class $C^1(D)$ and $\lambda\in\mbr$. Then for each
$\theta\in\Cc(\mbr)$ with $\theta(\lambda)\neq0$ one may find a real
number $a$ and a compact operator $K$ such that 
\begin{equation}\label{eq:must}
\theta(H)^*[H,iD]\theta(H)\geq a|\theta(H)|^2+K.
\end{equation}

\begin{definition}\label{df:must}
The upper bound $\what\rho_H(\lambda)$ of the numbers $a$ for which
such an estimate holds is \emph{the best constant in the Mourre
  estimate for $H$ at $\lambda$}.  The \emph{threshold set} of $H$
(relative to $D$) is the closed real set
\begin{equation}\label{eq:thr0}
\tau(H)=\{\lambda \mid \what\rho_H(\lambda)\leq0\}
\end{equation}
One says that $D$ is \emph{conjugate to} $H$ at $\lambda$ if 
$\what\rho_H(\lambda)>0$. 
\end{definition}

The set $\tau(H)$ is closed because the function
$\what\rho_H:\mbr\to]-\infty,\infty]$ is lower semicontinuous.

To each closed real set $A$ we associate the function
$N_A:\mbr\to[-\infty,\infty[$ defined by
\begin{equation}\label{eq:na}
N_A(\lambda)=\sup\{ x\in A \mid x\leq\lambda\}.
\end{equation}
We make the convention $\sup\emptyset=-\infty$. Thus $N_A$ may take
the value $-\infty$ if and only if $A$ is bounded from below and then
$N_A(\lambda)=-\infty$ if and only if $\lambda<\min A$. The function
$N_A$ is further discussed during the proof of Lemma \ref{lm:nab}.

Nonrelativistic many-body Hamiltonians have been introduced in
Definition \ref{df:NR}.  Let $\mathrm{ev}(T)$ be the set of
eigenvalues of an operator $T$.

\begin{theorem}\label{th:thr}
  Let $\cs$ be finite and let $H=H_\cs$ be a nonrelativistic
  many-body Hamiltonian of class $C^1_\rmu(D)$. Then
  $\what\rho_H(\lambda)=\lambda-N_{\tau(H)}(\lambda)$ for all real
  $\lambda$ and $\tau(H)$ is a closed \emph{countable} real set
  given by
\begin{equation}\label{eq:thr}
\tau(H)=\ccup_{X\neq O}\mathrm{ev}(H_{\cs/X}).
\end{equation} 
\end{theorem}
\proof We first treat the case $O\in\cs$. We need some facts which
are discussed in detail in Sections 7.2, 8.3 and 8.4 from \cite{ABG}
(see pages 51--61 in \cite{BG2} for a shorter presentation).

(i) For each real $\lambda$ let $\rho_H(\lambda)$ be the upper bound
of the numbers $a$ for which an estimate like \eqref{eq:must} but
with $K=0$ holds. This defines a lower semicontinuous function
$\rho_H:\mbr\to ]-\infty,\infty] $ hence the set
$\varkappa(H)=\{\lambda \mid \rho_H(\lambda)\leq 0\}$ is a closed
real set called \emph{critical set} of $H$ (relative to $D$). We
clearly have $\rho_H\leq\what\rho_H$ and so
$\tau(H)\subset\varkappa(H)$.

(ii) Let $\mu(H)$ be the set of eigenvalues of $H$ such that
$\what\rho_H(\lambda)>0$. Then $\mu(H)$ is a discrete subset of
$\mathrm{ev}(H)$ consisting of eigenvalues of finite
multiplicity. This is essentially the virial theorem.

(iii) There is a simple and rather unexpected relation between the
functions $\rho_H$ and $\what\rho_H$: they are ``almost'' equal. In
fact, $\rho_H(\lambda)=0$ if $\lambda\in\mu(H)$ and
$\rho_H(\lambda)=\what\rho_H(\lambda)$ otherwise. In particular
\begin{equation}\label{eq:tev}
\varkappa(H)=\tau(H)\cup \mathrm{ev}(H)=\tau(H)\sqcup\mu(H)
\end{equation}
where $\sqcup$ denotes disjoint union. 

(iv) This step is easy but rather abstract and the $C^*$-algebra
setting really comes into play. We assume that $H$ is affiliated to
our algebra $\rc$. The preceding arguments did not require more than
the $C^1(D)$ class. Now we require $H$ to be of class $C^1_\rmu(D)$.
Then the operators $H_{\geq X}$ are also of class $C^1_\rmu(D)$ and
we have the important relation (Theorem 8.4.3 in \cite{ABG} or
Theorem 4.4 in \cite{BG2})
\[
\what\rho_H=\min_{X\in\cp(\cs)}\rho_{H_{\geq X}}.
\]
To simplify notations we adopt the abbreviations $\rho_{H_{\geq
    X}}=\rho_{\geq X}$ and instead of $X\in\cp(\cs)$ we write
$X\gtrdot O$, which should be read ``$X$ covers $O$''. For coherence
with later notations we also set $\what\rho_H=\what\rho_\cs$.  So
\eqref{eq:mustq} may be written
\begin{equation}\label{eq:mustq}
\what\rho_\cs=\min_{X\gtrdot O}\rho_{\geq X}.
\end{equation}

(v) From \eqref{eq:dlap} and \eqref{eq:NR} we get
\[
H_{\geq X}=\Delta_X\otimes1+ 1\otimes H_{\cs/X}, \quad
[H_{\geq X},iD]=\Delta_X\otimes1+ 1\otimes [D,i H_{\cs/X}].
\]
Recall that we denote $D$ the generator of the dilation group
independently of the space in which it acts. We note that the formal
argument which gives the second relation above can easily be made
rigorous but this does not matter here. Indeed, since $H_{\geq X}$
is of class $C^1_\rmu(D)$ and by using the first relation above, one
can easily show that $H_{\cs/X}$ is also of class $C^1_\rmu(D)$ (see
the proof of Lemma 9.4.3 in \cite{ABG}). Now we may use Theorem
8.3.6 from \cite{ABG} to get
\[
\rho_{\geq X}(\lambda)=\inf_{\lambda_1+\lambda_2=\lambda}
\big(\rho_{\Delta_X}(\lambda_1) + \rho_{\cs/X}(\lambda_2) \big)
\]
where $\rho_{\cs/X}=\rho_{H_{\cs/X}}$. But clearly if $X\neq O$ we
have $\rho_{\Delta_X}(\lambda)=\infty$ if $\lambda<0$ and 
$\rho_{\Delta_X}(\lambda)=\lambda$ if $\lambda\geq0$. Thus we get
\begin{equation}\label{eq:mustt}
\rho_{\geq X}(\lambda)=\inf_{\mu\leq\lambda}
\big(\lambda-\mu + \rho_{\cs/X}(\mu) \big)
=\lambda-  \sup_{\mu\leq\lambda}\big(\mu - \rho_{\cs/X}(\mu) \big).
\end{equation}

(vi) Now from \eqref{eq:mustq} and \eqref{eq:mustt} we get
\begin{equation}\label{eq:musr}
\lambda-\what\rho_\cs(\lambda)=
\max_{X\gtrdot O}\sup_{\mu\leq\lambda}
\big(\mu - \rho_{\cs/X}(\mu)\big).
\end{equation}
Finally, we are able to prove the formula
$\what\rho_H(\lambda)=\lambda-N_{\tau(H)}(\lambda)$ by induction
over the semilattice $\cs$. In other terms, we assume that the
formula is correct if $H$ is replaced by $H_{\cs/X}$ for all $X\neq
O$ and we prove it for $H=H_{\cs/O}$. So we have to show that the
right hand side of \eqref{eq:musr} is equal to
$N_{\tau(H)}(\lambda)$.

According to step (iii) above we have $\rho_{\cs/X}(\mu)=0$ if
$\mu\in\mu(H_{\cs/X})$ and
$\rho_{\cs/X}(\mu)=\what\rho_{\cs/X}(\mu)$ otherwise. Since by the
explicit expression of $\what\rho_{\cs/X}$ this is a positive
function and since $\rho_H(\lambda)\leq0$ is always true if
$\lambda$ is an eigenvalue, we get $\mu-\rho_{\cs/X}(\mu)=\mu$ if
$\mu\in\mathrm{ev}(H_{\cs/X})$ and
\[
\mu-\rho_{\cs/X}(\mu)=\mu-\what\rho_{\cs/X}(\mu)=
N_{\tau(H_{\cs/X})}(\mu)
\]
otherwise. From the first part of Lemma \ref{lm:nab} below we get
\[
\sup_{\mu\leq\lambda}\big(\mu - \rho_{\cs/X}(\mu)\big)=
N_{\mathrm{ev}(H_{\cs/X}) \cup \tau(H_{\cs/X})}.
\]
If we use the second part of Lemma \ref{lm:nab} then we see that
\[
\max_{X\gtrdot O}\sup_{\mu\leq\lambda}\big(\mu -
\rho_{\cs/X}(\mu)\big)=
\max_{X\gtrdot O}N_{\mathrm{ev}(H_{\cs/X}) \cup \tau(H_{\cs/X})}
\]
is the $N$ function of the set 
\[
\bigcup_{X\gtrdot O}\big(\mathrm{ev}(H_{\cs/X}) 
\cup \tau(H_{\cs/X})\big)=
\bigcup_{X\gtrdot O}\left(\mathrm{ev}(H_{\cs/X}) 
\bigcup \bigcup_{Y>X}\mathrm{ev}(H_{\cs/Y})\right)=
\bigcup_{X>O}\mathrm{ev}(H_{\cs/X})
\]
which finishes the proof of
$\what\rho_H(\lambda)=\lambda-N_{\tau(H)}(\lambda)$ hence the proof
of Theorem \ref{th:thr} in the case $O\in\cs$.

No assume $O\nin\cs$ and let $E=\min\cs$. Then $O\in\cs/E$ so we may
use the preceding result for $H_{\cs/E}$. Moreover, we have
$H=\Delta_E\otimes 1 + 1\otimes H_{\cs/E}$. Thus
$\mathrm{ev}(H)=\emptyset$, $\what\rho_H=\rho_H$, and we may use a
relation similar to \eqref{eq:mustt} to get
\[
\lambda-\what\rho_H(\lambda)=
\sup_{\mu\leq\lambda}(\mu-\rho_{\cs/E}(\mu)).
\]
By what we have shown before we have
$\mu-\rho_{\cs/E}(\mu)=N_{\tau(H_{\cs/E})}(\mu)$ if
$\mu\nin\mu(H_{\cs/E})$ and otherwise $\mu-\rho_{\cs/E}(\mu)=\mu$.
From Lemma \ref{lm:nab} we get
$\lambda-\what\rho_H(\lambda)=N_{\tau(H_{\cs/E})\cup\mu(H_{\cs/E})}$.
But from \eqref{eq:tev} we get $\tau(H_{\cs/E})\cup\mu(H_{\cs/E})=
\tau(H_{\cs/E})\cup\mathrm{ev}(H_{\cs/E})$. From
\eqref{eq:thr} we get
\[
\tau(H_{\cs/E})=\ccup_{Y\in\cs/E, Y\neq O}\mathrm{ev}(H_{(\cs/E)/Y})
=\ccup_{X\in\cs, X\neq E} \mathrm{ev}(H_{\cs/X})
\] 
because if we write $Y=X/E$ with $X\in\cs, X\neq E$ then
$(\cs/E)/(X/E)=\cs/X$. Finally,
\[
\tau(H_{\cs/E})\,\ccup\,\mathrm{ev}(H_{\cs/E})=
\ccup_{X\in\cs} \mathrm{ev}(H_{\cs/X})
\]
which proves the Theorem in the case $O\nin\cs$.
\qed

It remains to show the following fact which was used above.

\begin{lemma}\label{lm:nab}
If $A$ and $A\cup B$ are closed and if $M$ is the function given by
$M(\mu)=N_A(\mu)$ for $\mu\nin B$ and $M(\mu)=\mu$ for $\mu\in B$
then $\sup_{\mu\leq\lambda}M(\mu)=N_{A\cup B}(\lambda)$.  If $A,B$
are closed then $\sup(N_A,N_B)=N_{A\cup B}$.
\end{lemma}
\proof The last assertion of the lemma is easy to check, we prove
the first one.  Observe first that the function $N_A$ has the
following properties:
\begin{compactenum}
\item[(i)]
$N_A$ is increasing and right-continuous,
\item[(ii)]
$N_A(\lambda)=\lambda$ if $\lambda\in A$,
\item[(iii)] $N_A$ is locally constant and $N_A(\lambda)<\lambda$ on
  $A^\rmc\equiv \mbr\setminus A$.
\end{compactenum} 
Indeed, the first assertion in (i) and assertion (ii) are obvious.
The second part of (i) follows from the more precise and easy to prove
fact 
\begin{equation}\label{eq:nae}
N_A(\lambda+\varepsilon)\leq N_A(\lambda)+\varepsilon \quad
\text{for all real } \lambda \text{ and } \varepsilon>0.
\end{equation}
A connected component of the open set $A^\rmc$ is necessarily an
open interval of one of the forms $]-\infty,y[$ or $]x,y[$ or
$]x,\infty[$ with $x,y\in A$. On the first interval (if such an
interval appears) $N_A$ is equal to $-\infty$ and on the second or
the third one it is clearly constant and equal to $N_A(x)$. We also
note that the function $N_A$ is characterized by the properties
(i)--(iii).

Thus, if we denote $N(\lambda)=\sup_{\mu\leq\lambda}M(\mu)$, then it
will suffices to show that the function $N$ satisfies the conditions
(i)--(iii) with $A$ replace by $A\cup B$. Observe that $M(\mu)\leq\mu$
and the equality holds if and only if $\mu\in A\cup B$.  Thus $N$ is
increasing, $N(\lambda)\leq\lambda$, and $N(\lambda)=\lambda$ if
$\lambda\in A\cup B$.

Now assume that $\lambda$ belongs to a bounded connected component
$]x,y[ $ of $A\cup B$ (the unbounded case is easier to treat). If
$x<\mu<y$ then $\mu\nin B$ so $M(\mu)=N_A(\mu)$ and $]x,y[$ is
included in a connected component of $A^\rmc$ hence $M(\mu)=N_A(x)$.
Then $N(\lambda)= \max(\sup_{\nu\leq x}M(\nu),N_A(x))$ hence $N$ is 
constant on $]x,y[$. Here we have $M(\nu)\leq\nu\leq x$ so if $x\in A$
then $N_A(x)=x$ and we get $N(\lambda)=x$. If $x\in B\setminus A$ then
$M(x)=x$ so $\sup_{\nu\leq x}M(\nu)=x$ and $N_A(x)<x$ hence 
$N(\lambda)=x$. Since $x\in A\cup B$ one of these two cases is 
certainly realized and the same argument gives $N(x)=x$. Thus the 
value of $N$ on $]x,y[$ is $N(x)$ so $N$ is right continuous on 
$[x,y[$.  Thus we proved that $N$ is locally constant and right 
continuous on the complement of $A\cup B$ and also that 
$N(\lambda)<\lambda$ there.

It remains to be shown that $N$ is right continuous at each point of
$\lambda\in A\cup B$. We show that \eqref{eq:nae} hold with $N_A$
replaced by $N$. If $\mu\leq\lambda$ then 
$M(\mu)\leq\mu\leq\lambda=M(\lambda)$ hence we have 
\[
N(\lambda+\varepsilon)=
\sup_{\lambda\leq\mu\leq\lambda+\varepsilon}M(\mu). 
\]
But $M(\mu)$ above is either $N_A(\mu)$ either $\mu$. In the second
case $\mu\leq\lambda+\varepsilon$ and  in the first case
\[
N_A(\mu)\leq N_A(\lambda+\varepsilon)\leq N_A(\lambda)+\varepsilon
\leq \lambda +\varepsilon. 
\]
Thus we certainly have $N(\lambda+\varepsilon)\leq\lambda+\varepsilon$
and $\lambda=N(\lambda)$ because $\lambda\in A\cup B$.
\qed

\subsection{A general class of interactions}
\label{ss:euc}

The rest of this section is devoted to some technical questions. Our
main purpose is to clarify the structure of the interactions in the
Euclidean case.

The following compactness criterion will be useful. This is a
consequence of the Riesz-Kolmogorov theorem and of the argument page
\pageref{p:RK} involving the regularity of the weight. Let $E,F$ be
arbitrary Euclidean space and $s,t\in\mbr$.

\begin{proposition}\label{pr:stsob}
  An operator $T\in L(\ch^s_E,\ch^t_F)$ is compact if and only if
  one of the next two equivalent conditions is satisfied,
  where $\|\cdot\|$ is the norm in $L(\ch^s_E,\ch^t_F)$:
\begin{compactenum}
\item[{\rm(i)}]
$\|(U_x-1)T\| + \|(V_x-1)T\| \to 0 \quad 
\text{if } x\to 0 \text{ in }F$,
\item[{\rm(ii)}]
$\|T(U_x-1)\| + \|T(V_x-1)\| \to 0 \quad 
\text{if } x\to 0 \text{ in }E$.
\end{compactenum}
\end{proposition}

We denote $L^\circ(\ch^s_E,\ch^{t}_F)$ the set of small at infinity
operators, cf.  Definition \ref{df:small}. Clearly
\emph{$L^\circ(\ch^s_E,\ch^{t}_F)$ is a closed subspace of
  $L(\ch^s_E,\ch^{t}_F)$}.

\begin{corollary}\label{co:small}
  An operator $T\in L(\ch^s_E,\ch^{t}_F)$ is small at infinity if
  and only if $\lim_{k\to0}T(V_k-1)= 0$ in norm in
  $L(\ch^{s+\varepsilon}_E,\ch^{t}_F)$ for some $\varepsilon>0$.
  Then this holds for all $\varepsilon>0$.
\end{corollary}

Indeed, the first part of condition (ii) of Proposition
\ref{pr:stsob} ($s$ replaced by $s+\varepsilon$) is
automatically satisfied.

We now give a Sobolev space version of Proposition \ref{pr:tex}
which uses the weights $\jap{\cdot}^s$ and is convenient in
applications.  By using Theorem \ref{th:xyzeintr} instead of Theorem
\ref{th:yzintr} in the proof of Proposition \ref{pr:tex} we get:

\begin{proposition}\label{pr:etex}
Let $s,t>0$ and $Z\subset X\cap Y$. Let 
$I_{XY}(Z)\in  L(\ch^t_Y,\ch^{-s}_X)$
such that the following relations hold in norm in
$ L(\ch^{t+\varepsilon}_Y,\ch^{-s}_X)$ for some $\varepsilon > 0$:
\begin{enumerate}
\item[{\rm(1)}]
$U_z I_{XY}(Z)=I_{XY}(Z) U_z$ if $z\in Z$ and
$V^*_z I_{XY}(Z) V_z\to I_{XY}(Z)$ if $z\to 0$ in $Z$,
\item[{\rm(2)}] 
$I_{XY}(Z)(V_y-1)\to 0$ if $y\to 0$ in $Y/Z$.
\end{enumerate}
If $h_X,h_Y$ are continuous real functions on $X,Y$ such that
$h_X(x)\sim\jap{x}^{2s}$ and $h_Y(y)\sim\jap{y}^{2t}$ and if we set
$K_X=h_X(P), K_Y=h_Y(P)$ then
$\jap{K_X}^{-1/2}I_{XY}(Z)\jap{K_Y}^{-\alpha}\in\rc_{XY}(Z)$ if
$\alpha>1/2$.
\end{proposition}

Our next purpose is to discuss in more detail the structure of the
operators $I_{XY}(Z)$ from Proposition \ref{pr:etex}. For this we
make a Fourier transformation $\cf_Z$ in the $Z$ variable as in the
proof of Theorem \ref{th:xyzeintr}.

We fix $X,Y,Z$ with $Z\subset X\cap Y$, use the tensor
factorizations \eqref{eq:xyzint} and make identifications like
$\ch_Z\otimes\ch_{X/Z}= L^2(Z;\ch_{X/Z})$. Thus
$\ch_X=\ch_Z\otimes\ch_{X/Z}$ and $\Delta_X=\Delta_Z\otimes 1 +
1\otimes \Delta_{X/Z}$ hence if $s\geq0$
\begin{equation}\label{eq:stens}
\ch^s(X)=\ch(Z;\ch^s(X/Z))\cap \ch^s(Z;\ch_{X/Z})=
\big(\ch_Z\otimes\ch^s(X/Z)\big)\cap 
\big(\ch^s(Z)\otimes\ch_{X/Z}\big)
\end{equation}
where our notations are extended to  vector-valued Sobolev spaces. 
Clearly
\begin{equation}\label{eq:lap}
\cf_Z \jap{P_X}^s \cf_Z^{-1} = 
\int_Z^\oplus (1+|k|^2+|P_{X/Z}|^2)^{s/2} \rmd k.
\end{equation}
Then from \eqref{eq:CZ} and $\rt_Z=\cf_Z^{-1}\Co(Z)\cf_Z$ we get
\begin{equation*}
\rc_{XY}(Z)=\rt_Z\otimes \rk_{X/Z,Y/Z}=
\cf_Z^{-1}\Co(Z;\rk_{X/Z,Y/Z})\cf_Z.
\end{equation*}
To each weakly measurable map $I_{XY}^Z:Z\to
L(\ch^t_{Y/Z},\ch^{-s}_{X/Z})$ such that
\begin{equation}\label{eq:Iest}
\sup\nolimits_k
\|(1+|k|+|P_{X/Z}|)^{-s}I_{XY}^Z(k)(1+|k|+|P_{Y/Z}|)^{-t}\| <\infty.
\end{equation} 
we associate a continuous operator 
$I_{XY}(Z):\ch^t_Y\to\ch^{-s}_X$ by the relation
\begin{equation}\label{eq:Ixyz}
\cf_Z I_{XY}(Z) \cf_Z^{-1} \equiv \int_Z^\oplus I_{XY}^Z(k) \rmd k.
\end{equation}
The following fact is known: a continuous operator
$T:\ch^t_Y\to\ch^{-s}_X$ is of the preceding form if and only if
$U_aT=TUa$ for all $a\in Z$. From the preceding results we get
(notations are as in Remark \ref{re:iaff}):

\begin{proposition}\label{pr:zxy}
  Let $X,Y,Z\in\cs$ with $Z\subset X\cap Y$ and assume that
  $\cg^1_X=\ch^s_X$ and $\cg^1_Y=\ch^t_Y$. An operator
  $I_{XY}(Z):\ch^t_Y\to\ch^{-s}_X$ satisfies the conditions of
  Remark \ref{re:iaff} if and only if it is of the form
  \eqref{eq:Ixyz} with a norm continuous function $I_{XY}^Z:Z\to
  L^\circ(\ch^t_{Y/Z},\ch^{-s}_{X/Z})$ satisfying \eqref{eq:Iest}.
\end{proposition}

\subsection{Auxiliary results}
\label{ss:lemma}

In this subsection we collect some useful technical results.  Let
$\ce,\cf,\cg,\ch$ be Hilbert spaces. Note that we have a canonical
identification (tensor products are discussed in \S\ref{ss:ha})
\begin{equation}\label{eq:comtens}
K(\ce,\cf)\otimes K(\cg,\ch)\cong K(\ce\otimes\cg,\cf\otimes\ch),
\hspace{2mm}\text{in particular}\hspace{2mm}
K(\ce,\cf\otimes\ch)\cong K(\ce,\cf)\otimes\ch.
\end{equation}
Assume that we have continuous injective embeddings $\ce\subset\cg$
and $\cf\subset\cg$. We equip $\ce\cap\cf$ with the intersection
topology defined by the norm $(\|g\|_\ce^2+\|g\|_\cf^2)^{1/2}$, hence
$\ce\cap\cf$ becomes a Hilbert space continuously embedded in $\cg$.

\begin{lemma}\label{lm:efgh}
  The map $K(\ce,\ch)\times K(\cf,\ch) \to K(\ce\cap\cf,\ch)$ which
  associates to $S\in K(\ce,\ch)$ and $T\in K(\cf,\ch)$ the operator
  $S|_{\ce\cap\cf}+T|_{\ce\cap\cf} \in K(\ce\cap\cf,\ch)$ is
  surjective. Thus, slightly formally,
\begin{equation}\label{eq:efgh}
K(\ce\cap\cf,\ch)=K(\ce,\ch) + K(\cf,\ch).
\end{equation}
\end{lemma}
\proof It is clear that the map is well defined. Let $R\in
K(\ce\cap\cf,\ch)$, we have to show that there are $S,T$ as in the
statement of the proposition such that $R=
S|_{\ce\cap\cf}+T|_{\ce\cap\cf}$.  Observe that the norm on
$\ce\cap\cf$ has been chosen such that the linear map
$g\mapsto(g,g)\in\ce\oplus\cf$ be an isometry with range a closed
linear subspace $\ci$. Consider $R$ as a linear map $\ci\to\ch$ and
extend it to the orthogonal of $\ci$ by zero. The so defined map
$\wtilde R:\ci\to\ch$ is clearly compact. Let $S,T$ be defined by
$Se=\wtilde R(e,0)$ and $Tf=\wtilde R(0,f)$. Clearly $S\in
K(\ce,\ch)$ and $T\in K(\cf,\ch)$ and if $g\in\ce\cap\cf$ then
\[
Sg+Tg=\wtilde R(g,0)+\wtilde R(0,g)=\wtilde R(g,g)=Rg
\]
which proves the lemma.
\qed

We give some applications. If $E,F$ are Euclidean spaces and $s>0$
is real then
\begin{equation}\label{eq:EFs}
\ch^s_{E\oplus F}=\big(\ch^s_E\otimes\ch_F\big)\cap
\big(\ch_E\otimes\ch^s_F\big)
\end{equation}
hence Lemma \ref{lm:efgh} gives for an arbitrary Hilbert space $\ch$
\begin{equation}\label{eq:EFH}
K(\ch^s_{E\oplus F},\ch)=
K(\ch^s_E\otimes\ch_F,\ch) + K(\ch_E\otimes\ch^s_F,\ch).
\end{equation}
If $\ch$ itself is a tensor product $\ch=\ch'\otimes\ch''$ then we
can combine this with  \eqref{eq:comtens} and get
\begin{equation}\label{eq:EFHEF}
K(\ch^s_{E\oplus F},\ch'\otimes\ch'')  =
K(\ch^s_E,\ch')\otimes K(\ch_F,\ch'')
 + K(\ch_E,\ch') \otimes K(\ch^s_F,\ch'').
\end{equation}
Consider now a triplet $X,Y,Z$ such that $Z\subset X\cap Y$ and
denote
\begin{equation}\label{eq:xyze}
E=(X\cap Y)/Z \hspace{2mm}\text{and}\hspace{2mm} 
X \boxplus Y = X/Y\times Y/X.
\end{equation}
Then $Y/Z=E\oplus(Y/X) \text{ and } X/Z=E\oplus(X/Y)$ hence by using
\eqref{eq:EFHEF} we get for example 
\begin{align}
\ch_{Y/Z} &=\ch_E\otimes\ch_{Y/X} \text{ and \ }
\ch_{X/Z} =\ch_E\otimes\ch_{X/Y} \label{eq:new} \\
\ch^2_{Y/Z} &=\big(\ch^2_E\otimes\ch_{Y/X}\big)\cap 
\big(\ch_E\otimes\ch^2_{Y/X}\big) \label{eq:klea} \\
\ch^{-2}_{X/Z} &= \ch^{-2}_E\otimes\ch_{X/Y}
+\ch_E\otimes\ch^{-2}_{X/Y}. \label{eq:klean}
\end{align}
By using once again \eqref{eq:EFHEF} and the notations introduced in
\eqref{eq:ikef}, we get
\begin{equation}\label{eq:kde3}
\rk^2_{X/Z,Y/Z} = \rk^2_E\otimes \rk_{X/Y,Y/X} + 
\rk_E\otimes\rk^2_{X/Y,Y/X}.
\end{equation}
We identify a Hilbert-Schmidt operator with its kernel, so $L^2(X
\boxplus Y)\subset \rk_{X/Y,Y/X}$ is the subspace of Hilbert-Schmidt
operators. The we have a strict inclusion
\begin{equation} \label{eq:kde9} 
L^2(X \boxplus Y;\rk^2_E) \subset \rk^2_E\otimes \rk_{X/Y,Y/X}
\end{equation}

\subsection{First order regularity conditions}
\label{ss:scs} 

In the next two subsections we consider interactions as in
Proposition \ref{pr:nrm} and give explicit conditions on the
$I_{XY}^Z$ such that $H$ be of class $C^1_\rmu(D)$. We recall that
the assumptions of Proposition \ref{pr:nrm} can be stated as
follows: for all $Z\subset X\cap Y$
\begin{align}
& I^Z_{XY}:\ch^2_{Y/Z}\to\ch_{X/Z} 
\hspace{2mm} \text{is compact and satisfies}\hspace{2mm} 
(I^{Z}_{XY})^*\supset I^Z_{YX}, \label{eq:A}\\
& [D,I^Z_{XY}]:\ch^2_{Y/Z}\to\ch^{-2}_{X/Z} \hspace{2mm}\text{is
compact}. \label{eq:B}
\end{align}
If \eqref{eq:A} is satisfied then by duality and interpolation we
get
\begin{equation}\label{eq:interpol}
I^Z_{XY}:\ch^\theta_{Y/Z}\to\ch^{\theta-2}_{X/Z} \quad 
\text{is a compact operator for all } 0\leq \theta \leq 2,
\end{equation} 
in particular the operator $[D,I^Z_{XY}]\equiv
D_{X/Z}I^Z_{XY}-I^Z_{XY}D_{Y/Z}$ restricted to the space of
functions in $\ch^2_{Y/Z}$ with compact support has values in the
space of functions locally in $\ch^{-1}_{X/Z}$.  We use, for
example, the relation $D_{X/Z}=D_E\otimes 1 + 1\otimes D_{X/Y}$
relatively to \eqref{eq:new} to decompose this operator as follows:
\begin{align}
[D,I^Z_{XY}] 
&=(D_E+D_{X/Y})I^Z_{XY}-I^Z_{XY}(D_E+D_{Y/X}) \nonumber \\
&=[D_E,I^Z_{XY}] +D_{X/Y}I^Z_{XY} -I^Z_{XY}D_{Y/X}. \label{eq:dec}
\end{align}
Since $I^Z_{XY}D_{Y/X}\subset (D_{Y/X}I^Z_{YX})^*$ if \eqref{eq:A}
is satisfied then condition \eqref{eq:B} follows from:
\begin{equation}\label{eq:dii}
[D_E,I^Z_{XY}] \text{ and } D_{X/Y}I^Z_{XY} \text{ are compact
  operators } \ch^2_{Y/Z}\to\ch^{-2}_{X/Z} \text{ for all } X,Y,Z.
\end{equation}
According to \eqref{eq:kde3} the first part of condition 
\eqref{eq:A} is equivalent to
\begin{equation}\label{eq:kde4}
I_{XY}^Z=J+J' \text{ for some }
J\in \rk^2_E\otimes \rk_{X/Y,Y/X} \text{ and }
J'\in\rk_E\otimes\rk^2_{X/Y,Y/X}. 
\end{equation}
As a particular case, from \eqref{eq:kde9} we obtain the example
discussed in \S\ref{ss:examples}. The compactness conditions
\eqref{eq:dii} are conditions on the kernels $[D_E,I^Z_{XY}(x',y')]$
and $x'\cdot\nabla_{x'}I_{XY}^Z(x',y')$ of the operators
$[D_E,I^Z_{XY}]$ and $D_{X/Y}I^Z_{XY}$.  Note that a condition on
$I^Z_{XY}D_{Y/X}$ is a requirement on the kernel
$y'\cdot\nabla_{y'}I_{XY}^Z(x',y')$.

\subsection{Creation-annihilation type interactions}
\label{ss:xsupy} 

To see the relation with the
creation-annihilation type interactions characteristic to quantum
field models we consider now the case when
$Y\subset X$ strictly. Then
\begin{equation}\label{eq:ysux}
\rc_{XY}=\rc_Y\otimes\ch_{X/Y}, \quad 
\rc_{XY}(Z)=\rc_Y(Z)\otimes\ch_{X/Y}, \quad
\ch_X=\ch_Y\otimes\ch_{X/Y}
\end{equation}
where the first two tensor product have to be interpreted as
explained in \S\ref{ss:ha}. In particular we have
\begin{equation}\label{eq:sux}
L^2(X/Y;\rc_Y)\subset\rc_{XY} \quad \text{and} \quad
L^2(X/Y;\rc_Y(Z))\subset\rc_{XY}(Z)
\quad \text{strictly}. 
\end{equation}
If $Z\subset Y$ then $X=Z\oplus(Y/Z)\oplus(X/Y)$
and $X/Z=(Y/Z)\oplus(X/Y)$ hence $\ch_{X/Z}=\ch_{Y/Z}\otimes\ch_{X/Y}$
and thus the operator $I^Z_{XY}$ is
just a compact operator
\begin{equation}\label{eq:ixyo}
I^Z_{XY} : \ch^2_{Y/Z}\to\ch_{Y/Z}\otimes\ch_{X/Y}.
\end{equation}
If $\ce,\cf,\cg$ are Hilbert spaces then $K(\ce,\cf\otimes\cg)\cong
K(\ce,\cf)\otimes\cg$. Hence \eqref{eq:ixyo} means
\begin{equation}\label{eq:ixyoo}
I^Z_{XY} \in \rk^2_{Y/Z}\otimes \ch_{X/Y}.
\end{equation}
Let $\ri_{XY}= {\textstyle\sum_{Z\subset X\cap Y}} 1_Z\otimes
\rk^2_{X/Z,Y/Z}$, where the sum is direct and closed in
$\rk^2_{XY}$.  A usual nonrelativistic $N$-body Hamiltonian
associated to the semilattice $\cs_X$ of subspaces of $X$ is of the
form $\Delta_X+I_X$ with $I_X\in\ri_{X}\equiv\ri_{XX}$. Thus the
interaction which couples the $X$ and $Y$ systems is of the form
\begin{equation}\label{eq:xyinter}
I_{XY}={\textstyle\sum_{Z\in\cs_Y}} 
1_Z\otimes I^Z_{XY} \in \ri_{Y}\otimes \ch_{X/Y}. 
\end{equation}
In particular we may take $I_{XY}\in L^2(X/Y;\ri_Y)$, but we stress
that the space $\ri_{Y}\otimes \ch_{X/Y}$ is much larger (see
\S\ref{ss:ha}).  More explicitly, a square integrable function
$I_{XY}:X/Y\to\ri_Y$ determines an operator $I_{XY}:\ch^2_Y\to\ch_X$
by the following rule: it associates to $u\in\ch^2(Y)$ the function
$y'\mapsto I_{XY}(y')u$ which belongs to $L^2(X/Y;\ch_{X/Y})=\ch_X$.
We may also write
\begin{equation}\label{eq:IV}
(I_{XY}u)(x)=(I_{XY}(y')u)(y) \quad \text{where }
x\in X=Y\oplus X/Y \text{ is written as } x=(y,y').
\end{equation}
We say that the operator valued function $I_{XY}$ is the kernel of
the operator $I_{XY}$. The adjoint $I_{YX}=I_{XY}^*$ is an integral
operator in the $y'$ variable (like an annihilation
operator). Indeed, if $v\in\ch_X$ is thought as a map $y'\mapsto
v(y')\in\ch_Y$ then we have $I_{YX}v=\int_{X/Y}
I^*_{XY}(y')v(y')\rmd y'$ at least formally.

The particular case when the function $I_{XY}$ is factorizable
clarifies the connection with the quantum field type interactions:
let $I_{XY}$ be a finite sum $I_{XY}=\sum_i V_Y^i\otimes\phi_i$
where $V^i_Y\in\ri_Y$ and $\phi_i\in \ch_{X/Y}$, then
\begin{equation}\label{eq:qft}
I_{XY}u={\textstyle\sum_i} (V_Y^i u)\otimes\phi_i \quad
\text{as an operator }  
I_{XY}:\ch^2_Y\to\ch_X=\ch_Y\otimes\ch_{X/Y}.
\end{equation}
This is a sum of $N$-body type interactions $V^i_Y$ tensorized with
operators which create particles in states~ $\phi_i$.

The conditions on the ``commutator'' $[D,I_{XY}]$ may be written
in terms of the kernel of $I_{XY}$. The
relation \eqref{eq:dec}  becomes $
[D,I_{XY}]=[D_Y,I_{XY}]+D_{X/Y}I_{XY}$. The operator $D_Y$ acts only
on the variable $y$ and $D_{X/Y}$ acts only on the variable
$y'$. Thus $[D_Y,I_{XY}]$ and $D_{X/Y}I_{XY}$ are operators of the
same nature as $I_{XY}$ but more singular: the kernel of
$[D_Y,I_{XY}]$ is the function $y'\mapsto [D_Y,I_{XY}(y')]$ and that
of $2iD_{X/Y}I_{XY}$ is the function $ y'\mapsto
(y'\cdot\nabla_{y'}+n/2) I_{XY}(y')$.  Thus, to get
\eqref{eq:B} it suffices to
require two conditions on the kernel $I_{XY}$, one on
$[D_Y,I_{XY}(y')]$ and a second one on
$y'\cdot\nabla_{y'}I_{XY}(y')$.

If we decompose $I_{XY}$ as in \eqref{eq:xyinter} with
$I^Z_{XY}:\ch^2_{Y/Z}\to\ch_{Y/Z}\otimes\ch_{X/Y}$ compact then the
(formal) kernel of $I^Z_{XY}$ is a $\rk^2_{Y/Z}$ valued map on
$X/Y$.  We require that $[D_{Y/Z},I^Z_{XY}]$ and $D_{X/Y}I^Z_{XY}$
be compact operators $\ch^2_{Y/Z}\to\ch^{-2}_{X/Z}$. From
\eqref{eq:stens} and $X/Z=(Y/Z)\oplus(X/Y)$ we get
\begin{equation*}
\ch^2_{X/Z} = \big(\ch_{Y/Z}\otimes\ch^2_{X/Y}\big)\cap
\big(\ch^2_{Y/Z}\otimes\ch_{X/Y}\big), \quad
\ch^{-2}_{X/Z} = \ch_{Y/Z}\otimes\ch^{-2}_{X/Y} +
\ch^{-2}_{Y/Z}\otimes\ch_{X/Y}
\end{equation*}  
which are helpful in checking these compactness requirements.

\subsection{Besov regularity classes}
\label{ss:scsc} 

We recall some facts concerning the Besov type regularity class $
C^{1,1}(D)$; we refer to \cite{ABG} for details on these
matters. Since the conjugate operator $D$ is fixed we shall not
indicate it in the notation from now on. An operator $T\in L(\ch)$
is of class $C^{1,1}$ if
\begin{equation}\label{eq:c11}
\int_0^1\|W^*_{2\varepsilon}T W_{2\varepsilon}-
2W^*_{\varepsilon}T W_{\varepsilon} +T\|
\frac{\rmd\varepsilon}{\varepsilon^2} \equiv
\int_0^1\|(\cw_\varepsilon-1)^2
T\|\frac{\rmd\varepsilon}{\varepsilon^2} <\infty
\end{equation}
where $\cw_\varepsilon$ is the automorphism of $L(\ch)$ defined by
$\cw_\varepsilon T=W_\varepsilon^*T W_\varepsilon$. The condition
\eqref{eq:c11} implies that $T$ is of class $C^1_\rmu$ and is
just slightly more than this. Indeed, $T$ is of class $C^1$ or
$C^1_\rmu$ if and only if the limit
\[
\lim_{\tau\to 0}\int_\tau^1 (\cw_\varepsilon-1)^2 T
\frac{\rmd\varepsilon}{\varepsilon^2}
\] 
exists strongly or in norm respectively. The following subclass of
$C^{1,1}$ is useful in applications:  $T$ is called of class
$C^{1+}$ if $T$ is of class $C^1$, so the commutator $[D,T]$ is a
bounded operator, and
\begin{equation}\label{eq:din}
\int_0^1\|W_\varepsilon^*[D,T]W_\varepsilon-[D,T]\|
\frac{\rmd\varepsilon}{\varepsilon} <\infty.
\end{equation}
Then $C^{1+}\subset C^{1,1}$. The class most frequently used in the
context of the Mourre theorem is $C^2$: this is the set of $T\in
C^1$ such that $[D,T]\in C^1$. But $[D,T]\in C^1$ if and only if
\[
\|W_\varepsilon^*[D,T] W_\varepsilon-[D,T]\| \leq 
C |\varepsilon| \quad \text{for some constant $C$ and all real }
\varepsilon 
\]
hence this condition is much stronger then the Dini type condition
\eqref{eq:din}.  A self-adjoint operator $H$ is of class $C^{1,1}$,
$C^{1+}$ or $C^2$ if its resolvent is of class $C^{1,1}$, $C^{1+}$
or $C^2$ respectively. 

We now consider a Hamiltonian as in Proposition \ref{pr:nrm} and
discuss conditions which ensure that $H$ is of class $C^{1,1}$. An
important point is that the domain $\ch^2$ of $H$ is stable under
the dilation group $W_\tau$.  Then Theorem 6.3.4 from \cite{ABG}
implies that $H$ is of class $C^{1,1}$ if and only if
\begin{equation}\label{eq:c11h}
\int_0^1\|(\cw_\varepsilon-1)^2H\|_{\ch^2\to\ch^{-2}}
\frac{\rmd\varepsilon}{\varepsilon^2} <\infty.
\end{equation}
As above $\cw_{\varepsilon} H=W^*_{\varepsilon}H W_{\varepsilon}$
hence $ (\cw_\varepsilon-1)^2H=W^*_{2\varepsilon}H W_{2\varepsilon}-
2W^*_{\varepsilon}H W_{\varepsilon} +H$.  We have $H=\Delta +I$ and
due to \eqref{eq:dlap} the relation \eqref{eq:c11h} is trivially
verified by the kinetic part $\Delta$ of $H$ hence we are only
interested in conditions on $I$ which ensure that \eqref{eq:c11h} is
satisfied with $H$ replaced by $I$. If this is the case, by a slight
abuse of language we say that $I$ is of class $C^{1,1}$. In terms of
the coefficients $I_{XY}$, this means
\begin{equation}\label{eq:c11xyz}
\int_0^1\|(\cw_\varepsilon-1)^2I_{XY}^Z\|_{\ch^2_{Y/Z}\to\ch^{-2}_{X/Z}}
\frac{\rmd\varepsilon}{\varepsilon^2} <\infty
\quad\text{for all } X,Y,Z.
\end{equation}
We recall one fact (see \cite[Ch. 5]{ABG}). Let $I:\ch^2\to\ch^{-2}$
be an arbitrary linear continuous operator. Then
$[D,I]:\ch^2_\rmc\to\ch^{-3}_{\mathrm{loc}}$ is well defined and $I$
is of class $C^1$ (in an obvious sense) if and only if this operator
is the restriction of a continuous map $\ch^2\to\ch^{-2}$, which
will be denoted also $[D,I]$. We say that $I$ is of class $C^{1+}$
if this condition is satisfied and 
\begin{equation}\label{eq:dini}
\int_0^1\|W_\varepsilon^*[D,I]W_\varepsilon-[D,I]\|_{\ch^2\to\ch^{-2}}
\frac{\rmd\varepsilon}{\varepsilon} <\infty.
\end{equation}
As before, if $I$ is of class $C^{1+}$ then it is of class
$C^{1,1}$. In terms of the coefficients $I_{XY}^Z$ this means
\begin{equation}\label{eq:dinix}
\int_0^1\|W_\varepsilon^*[D,I_{XY}^Z]W_\varepsilon-[D,I_{XY}^Z]
\|_{\ch^2_{Y/Z}\to\ch^{-2}_{X/Z}}
\frac{\rmd\varepsilon}{\varepsilon} <\infty.
\end{equation}
Such a condition should be imposed on each of the three terms in the
decomposition \eqref{eq:dec} separately.

The techniques developed in \S 7.5.3 and on pages 425--429 from
\cite{ABG} can be used to get more concrete conditions.  The only
new fact with respect to the $N$-body situation as treated there is
that $\cw_\tau$ when considered as an operator on $\rl_{X/Z,Y,Z}$
factorizes in a product of three commuting operators. Indeed, if we
write $\ch_{Y/Z}=\ch_E\otimes\ch_{Y/X}$ and
$\ch_{X/Z}=\ch_E\otimes\ch_{X/Y}$ then we get
$\cw_\tau(T)=W^{X/Y}_{-\tau}\cw^E_\tau(T)W^{Y/X}_\tau$ where this
time we indicated by an upper index the space to which the operator
is related and, for example, we identified $W^{Y/X}_\tau=1_E\otimes
W^{Y/X}_\tau$. To check the $C^{1,1}$ property in this context
one may use:

\begin{proposition}\label{pr:inter}
If $T\in\rl:= L(\ch^2_{Y/Z},\ch^{-2}_{X/Z})$ then 
$\int_0^1\|(\cw_\varepsilon-1)^2T\|_\rl
\rmd\varepsilon/\varepsilon^2<\infty$ follows from
\begin{equation}\label{eq:inter}
\int_0^1\left(
\|(W^{X/Y}_{\varepsilon}-1)^2T\|_\rl+ 
\|(\cw^E_\varepsilon-1)^2T\|_\rl+
\|T(W^{Y/X}_{\varepsilon}-1)^2\|_\rl
\right)
\frac{\rmd\varepsilon}{\varepsilon^2} < \infty.
\end{equation}
\end{proposition}
\proof We shall interpret $\int_0^1\|(\cw_\varepsilon-1)^2T\|_\rl
\rmd\varepsilon/\varepsilon^2<\infty$ in terms of real interpolation
theory. Let $L_\tau$ be the operator of left multiplication by
$W^{X/Y}_{-\tau}$ and $N_\tau$ the operator of right multiplication
by $W^{Y/X}_{\tau}$ on $\rl_{X/Z,Y/Z}$. If we also set
$M_\tau=\cw^E_\tau$ then we get three commuting operators
$L_\tau,M_\tau,N\tau$ on $\rl_{X/Z,Y/Z}$ such that $\cw_\tau=L_\tau
M_\tau N_\tau$.  Then it is easy to check a Dini type condition like
\eqref{eq:dinix} by using
\begin{equation}\label{eq:tmp}
\cw_\tau-1=(L_\tau-1)M_\tau N_\tau+(M_\tau -1)N_\tau+(N_\tau-1). 
\end{equation}
On the other hand, observe that $\cw_\tau,L_\tau,M_\tau,N_\tau$ are
one parameter groups of operators on the Banach space $\rl$.  These
groups are not continuous in the ordinary sense but this does not
really matter, in fact we are in the setting of \cite[Ch. 5]{ABG}.
The main point is that the integral
$\int_0^1\|(\cw_\varepsilon-1)^2T\|_\rl\rmd\varepsilon/\varepsilon^2$
is finite if and only if
$\int_0^1\|(\cw_\varepsilon-1)^6T\|_\rl\rmd\varepsilon/\varepsilon^2$
is finite (see Theorem 3.4.6 in \cite{ABG}; this is where real
interpolation comes into play).  Now by taking the sixth power of
\eqref{eq:tmp} and developing the right hand side we easily get the
result, cf. the formula on top of page 132 of \cite{ABG}.  \qed

The proof of Theorem \ref{th:BVR} is based on an extension of
Propositions 9.4.11 and 9.4.12 from \cite{ABG} to the present
context. Since the argument is very similar, we do not enter into
details. We mention only that the operator $D$ can be written as
$4D=P \cdot Q+Q \cdot P$ where $P=\oplus_X P_X$ and $Q=\oplus_X Q_X$
are suitably interpreted. The proofs in \cite{ABG} depend only on
this structure.

\section{Appendix: Hamiltonian algebras} 
\label{s:appb}

We prove here some results on $C^*$-algebras generated by certain
classes of ``elementary'' Hamiltonians.

\subsection{}\label{ss:a1}

Let $X$ be a locally compact abelian group and let
$\{U_x\}_{x\in X}$ be a strongly continuous unitary representation
of $X$ on a Hilbert space $\ch$. Then one can associate to it a
Borel regular spectral measure $E$ on $X^*$ with values projectors
on $\ch$ such that $U_x=\int_{X^*}k(x)E(\rmd k)$ and this allows us
to define for each Borel function $\psi:X^*\to\mbc$ a normal
operator on $\ch$ by the formula $\psi(P)=\int_{X^*} \psi(k)E(\rmd
k)$.  The set $\rt_X(\ch)$ of all the operators $\psi(P)$ with
$\psi\in\Co(X^*)$ is clearly a non-degenerate $C^*$-algebra of
operators on $\ch$. The following result, which will be useful in
several contexts, is an easy consequence of the Cohen-Hewitt
factorization theorem, see Lemma 3.8 from \cite{GI4}. Consider an
operator $A\in L(\ch)$.

\begin{lemma}\label{lm:help}
$\displaystyle{\lim_{x\to0}}\|(U_x-1)A\|=0$ if and
only if $A=\psi(P)B$ for some $\psi\in\Co(X^*)$ and $B\in L(\ch)$.
\end{lemma}

We say that an operator $S\in L(\ch)$ is of class $C^0(P)$ if the
map $x\mapsto U_xSU_x^*$ is norm continuous.

\begin{lemma}\label{lm:cop}
Let $S\in L(\ch)$ be of class $C^0(P)$ and let $T\in
\rt_X(\ch)$. Then for each $\varepsilon>0$ there is $Y\subset X$
finite and there are operators $T_y\in \rt_X(\ch)$ such that
$\|ST-\sum_{y\in Y} T_y U_{y}SU_{y}^*\|<\varepsilon$.
\end{lemma}
\proof It suffices to assume that $T=\psi(P)$ where $\psi$ has a
Fourier transform integrable on $X$, so that $T=\int_X U_x
\what\psi(x) \rmd x$, and then to use a partition of unity on $X$
and the uniform continuity of the map $x\mapsto U_xSU_x^*$ (see the
proof of Lemma 2.1 in \cite{DG1}).  \qed

We say that a subset $\cb$ of $L(\ch)$ is $X$-stable if
$U_xSU_x^*\in\cb$ whenever $S\in\cb$ and $x\in X$. From Lemma
\ref{lm:cop} we see that if $\cb$ is an $X$-stable real linear space
of operators of class $C^0(P)$ then
\[
\cb\cdot \rt_X(\ch)= \rt_X(\ch)\cdot\cb.
\] 
Since the $C^*$-algebra $\ca$ generated by $\cb$ is also $X$-stable
and consists of operators of class $C^0(P)$
\begin{equation}\label{eq:cop}
\ra\equiv\ca\cdot \rt_X(\ch)= \rt_X(\ch)\cdot\ca
\end{equation} 
is a $C^*$-algebra. The operators $U_x$ implement a norm continuous
action of $X$ by automorphisms of the algebra $\ca$ so the
$C^*$-algebra crossed product $\ca\rtimes X$ is well defined and the
algebra $\ra$ is a quotient of this crossed product.

A function $h$ on $X^*$ is called \emph{$p$-periodic} for some
non-zero $p\in X^*$ if $h(k+p)=h(k)$ for all $k\in X^*$.

\begin{proposition}\label{pr:cop}
Let $\cv$ be an $X$-stable set of symmetric bounded operators of
class $C^0(P)$ and such that $\lambda\cv\subset\cv$ if
$\lambda\in\mbr$. Denote $\ca$ the $C^*$-algebra generated by $\cv$
and define $\ra$ by \eqref{eq:cop}.  Let $h:X^*\to\mbr$ be
continuous, not $p$-periodic if $p\neq0$, and such that
$|h(k)|\to\infty$ as $k\to\infty$.  Then $\ra$ is the $C^*$-algebra
generated by the self-adjoint operators of the form $h(P+k)+V$ with
$k\in X^*$ and $V\in\cv$.
\end{proposition}
\proof Denote $K=h(P+k)$ and let $R_\lambda=(z-K-\lambda V)^{-1}$
with $z$ not real and $\lambda$ real. Let $\rc$ be the $C^*$-algebra
generated by such operators (with varying $k$ and $V$). By taking
$V=0$ we see that $\rc$ will contain the $C^*$-algebra generated by
the operators $R_0$. By the Stone-Weierstrass theorem this algebra
is $\rt_X(\ch)$ because the set of functions $p\to(z-h(p+k))^{-1}$
where $k$ runs over $X^*$ separates the points of $X^*$. The
derivative with respect to $\lambda$ at $\lambda=0$ of $R_\lambda$
exists in norm and is equal to $R_0VR_0$, so $R_0VR_0\in\rc$. Since
$\rt_X\subset\rc$ we get $\phi(P)V\psi(P)\in\rc$ for all
$\phi,\psi\in\Co(X^*)$ and all $V\in\cv$. Since $V$ is of class
$C^0(P)$ we have $(U_x-1)V\psi(P)\sim V(U_x-1)\psi(P)\to0$ in norm
as $x\to0$ from which we get $\phi(P)V\psi(P)\to S\psi(P)$ in norm
as $\phi\to1$ conveniently. Thus $V\psi(P)\in\rc$ for $V,\psi$ as
above. This implies $V_1\cdots V_n\psi(P)\in\rc$ for all
$V_1,\dots,V_n\in\cv$. Indeed, assuming $n=2$ for simplicity, we
write $\psi=\psi_1\psi_2$ with $\psi_i\in\Co(X^*)$ and then Lemma
\ref{lm:cop} allows us to approximate $V_2\psi_1(P)$ in norm with
linear combinations of operators of the form $\phi(P)V^x_2$ where
the $V^x_2$ are translates of $V_2$.  Since $\rc$ is an algebra we
get $V_1\phi(P) V^x_2\psi_2(P)\in\rc$ hence passing to the limit we
get $V_1V_2\psi(P)\in\rc$. Thus we proved $\ra\subset\rc$. The
converse inclusion follows from a series expansion of $R_\lambda$ in
powers of $V$.  \qed

The next two corollaries follow easily from Proposition
\ref{pr:cop}. We take $\ch=L^2(X)$ which is equipped with the usual
representations $U_x,V_k$ of $X$ and $X^*$ respectively. Let
$W_\xi=U_xV_k$ with $\xi=(x,k)$ be the phase space translation
operator, so that $\{W_\xi\}$ is a projective representation of the
phase space $\Xi=X\oplus X^*$.  Fix some classical kinetic energy
function $h$ as in the statement of Proposition \ref{pr:cop} and let
the classical potential $v:X\to\mbr$ be a bounded uniformly
continuous function. Then the quantum Hamiltonian will be
$H=h(P)+v(Q)\equiv K+V$. Since the origins in the configuration and
momentum spaces $X$ and $X^*$ have no special physical meaning one
may argue \cite{Be1,Be2} that $W_\xi H W^*_\xi=h(P-k)+v(Q+x)$ is a
Hamiltonian as good as $H$ for the description of the evolution of
the system. It is not clear to us whether the algebra generated by
such Hamiltonians (with $h$ and $v$ fixed) is in a natural way a
crossed product. On the other hand, it is natural to say that the
coupling constant in front of the potential is also a variable of
the system and so the Hamiltonians $H_\lambda=K+\lambda V$ with any
real $\lambda$ are as relevant as $H$. Then we may apply Proposition
\ref{pr:cop} with $\cv$ equal to the set of operators of the form
$\lambda v(Q+x)$. Thus:

\begin{corollary}\label{co:cop1}
Let $v\in\Cbu(X)$ real and let $\ca$ be the $C^*$-subalgebra of
$\Cbu(X)$ generated by the translates of $v$.  Let $h:X^*\to\mbr$ be
continuous, not $p$-periodic if $p\neq0$, and such that
$|h(k)|\to\infty$ as $k\to\infty$.  Then the $C^*$-algebra generated
by the self-adjoint operators of the form $W_\xi H_\lambda W^*_\xi$
with $\xi\in\Xi$ and real $\lambda$ is the crossed product
$\ca\rtimes X$.
\end{corollary}

Now let $\ct$ be a set of closed subgroups of $X$ such that the
semilattice $\cs$ generated by it (i.e. the set of finite
intersections of elements of $\ct$) consists of pairwise compatible
subgroups. Set $\cc_X(\cs)=\sum^\rmc_{Y\in\cs} \cc_X(Y)$. From
\eqref{eq:reg1} it follows that this is the $C^*$-algebra generated
by  $\sum_{Y\in\ct} \cc_X(Y)$.

\begin{corollary}\label{co:cop2}
Let $h$ be as in Corollary \ref{co:cop1}. Then the $C^*$-algebra
generated by the self-adjoint operators of the form $h(P+k)+v(Q)$
with $k\in X^*$ and $v\in\sum_{Y\in\ct} \cc_X(Y)$ is the
crossed product $\cc_X(\cs)\rtimes X$.
\end{corollary}

\begin{remark}\label{re:cop}
Proposition \ref{pr:cop} and Corollaries \ref{co:cop1} and
\ref{co:cop2} remain true and are easier to prove if we consider the
$C^*$-algebra generated by the operators $h(P)+V$ with all
$h:X^*\to\mbr$ continuous and such that $|h(k)|\to\infty$ as
$k\to\infty$. If in Proposition \ref{pr:cop} we take $\ch=L^2(X;E)$
with $E$ a finite dimensional Hilbert space (describing the spin
degrees of freedom) then the operators $H_0=h(P)$ with $h:X\to L(E)$
a continuous symmetric operator valued function such that
$\|(h(k)+i)^{-1}\|\to 0$ as $k\to\infty$ are affiliated to $\ra$
hence also their perturbations $H_0+V$ where $V$ satisfies the
criteria from \cite{DG3}, for example.
\end{remark}

\subsection{}
\label{ss:anbody} 

We consider the framework of \S\ref{ss:cexample} and use Corollary
\ref{co:cop2} to prove that the Hamiltonian algebra of a
nonrelativistic $N$-body system is generated in a natural way by the
operators of the form \eqref{eq:nonrel}. To state a precise result
it suffices to consider the reduced Hamiltonians (for which we keep
the notation $H$).

Let $\mathfrak{S}_2$ be the set of cluster decompositions which
contain only one nontrivial cluster which consists of exactly two
elements. This cluster is of the form $\{j,k\}$ for a unique pair of
numbers $1\leq j < k \leq N$ and we denote by $(jk)$ the
corresponding cluster decomposition. The map $x\mapsto x_j-x_k$
sends $X$ onto $\mbr^d$ and has $X_{(jk)}$ as kernel hence
$V_{jk}(x_j-x_k)=V_{(jk)}\circ\pi_{(jk)}(x)$ where
$V_{(jk)}:X/X_{(jk)}\to\mbr$ is continuous with compact support and
$\pi_{(jk)}:X\to X/X_{(jk)}$ is the canonical surjection.

Thus the reduced Hamiltonians corresponding to \eqref{eq:nonrel} are
the operators on $\ch_X$ of the form
\begin{equation}\label{eq:nonre}
\Delta_X+
{\textstyle{\sum_{\sigma\in\mathfrak{S}_2}}} V_\sigma\circ\pi_\sigma
\hspace{2mm}\text{with}\hspace{2mm} V_\sigma:X/X_\sigma\to\mbr
\text{ continuous with compact support}.
\end{equation} 
These operators must be affiliated to the Hamiltonian algebra of the
$N$-body system. On the other hand, if a Hamiltonian $h(P)+V$ is
considered as physically admissible then $h(P+k)+V$ should be
admissible too because the zero momentum $k=0$ should not play a
special role. In other terms, translations in momentum space should
leave invariant the set of admissible Hamiltonians. Hence it is
natural to consider \emph{the smallest $C^*$-algebra $\rc_X(\cs)$
  such that the operators \eqref{eq:nonre} are affiliated to it and
  which is stable under translations in momentum space. But this
  algebra is exactly the crossed product}
\[
\rc_X=\cc_X\rtimes X=\cc_X\cdot \rt_X 
\hspace{2mm}\text{with}\hspace{2mm}
\cc_X={\textstyle\sum_\sigma}\rc_X(X_\sigma).
\]
Indeed, it is clear that the semilattice generated by
$\mathfrak{S}_2$ is $\mathfrak{S}$ so we may apply Corollary
\ref{co:cop2}.

\subsection{}
\label{ss:amotiv} 

Here we prove Theorem \ref{th:motiv}.

Let $\rc'$ be the $C^*$-algebra generated by the operators of the
form $(z-K-\phi)^{-1}$ where $z$ is a not real number, $K$ is a
standard kinetic energy operator, and $\phi$ is a symmetric field
operator. With the notation \eqref{eq:d} we easily get
$\rt_\rmd\subset\rc'$.  If $\lambda\in\mbr$ then $\lambda\phi$ is
also a field operator so $(z-K-\lambda\phi)^{-1}\in\rc'$. By taking
the derivative with respect to $\lambda$ at $\lambda=0$ of this
operator we get $(z-K)^{-1}\phi (z-K)^{-1}\in\rc$. Since
$(z-K)^{-1}=\oplus_X(z-h_X(P))^{-1}$ (recall that $P$ is the
momentum observable independently of the group $X$) and since
$\rt_\rmd\subset\rc'$ we get $S\phi(\theta) T\in\rc'$ for all
$S,T\in \rt_\rmd$ and $\theta=(\theta_{XY})_{X\supset Y}$.

Let $\rc'_{XY}=\Pi_X\rc'\Pi_Y\subset\rl_{XY}$ be the components of
the algebra $\rc'$ and let us fix $X\supset Y$. Then we get
$\varphi(P) a^*(u) \psi(P)\in\rc'_{XY}$ for all
$\varphi\in\Co(X^*)$, $\psi\in\Co(Y^*)$, and $u\in\ch_{X/Y}$.  The
clspan of the operators $a^*(u) \psi(P)$ is $\rt_{XY}$, see
Proposition \ref{pr:def3} and the comments after \eqref{eq:L2a}, and
from \eqref{eq:cyzc} we have $\rt_X\cdot\rt_{XY}=\rt_{XY}$.  Thus
the clspan of the operators $\varphi(P) a^*(u) \psi(P)$ is
$\rt_{XY}$ for each $X\supset Y$ and then we get
$\rt_{XY}\subset\rc'_{XY}$.  By taking adjoints we get
$\rt_{XY}\subset\rc'_{XY}$ if $X\sim Y$.

Now recall that the subspace $\rt^\circ\subset L(\ch)$ defined by
$\rt^\circ_{XY}=\rt_{XY}$ if $X\sim Y$ and $\rt^\circ=\{0\}$ if
$X\not\sim Y$ is a closed self-adjoint linear subspace of $\rt$ and
that $\rt^\circ\cdot\rt^\circ=\rc$, cf. Theorem \ref{th:tc}. By
what we proved before we have $\rt^\circ\subset\rc'$ hence
$\rc\subset\rc'$. The converse inclusions is easy to prove. This
finishes the proof of Theorem \ref{th:motiv}.



\begin{thebibliography}{DeG3}


\bibitem[ABG]{ABG} Amrein, W., Boutet de Monvel, A., Georgescu, V.:
{\it $C_{0}$-groups, commutator methods and spectral theory of
$N$-body Hamiltonians}, Birkh{\"a}user, 1996.

\bibitem[Be1]{Be1} Bellissard, J.: K-Theory of $C^*$-algebras in
  solid state physics, in \emph{Statistical Mechanics and Field
    Theory: Mathematical Aspects}, T.C.\ Dorlas, N.M.\ Hugenholtz,
  M.\ Winnink (eds.), 1985.

\bibitem[Be2]{Be2} Bellissard, J.: Gap labeling theorems for
Schr\"odinger operators,  in \emph{From Number Theory to Physics},
Les Houches 1989, J.M.\ Luck, P.\ Moussa, M.\ Waldschmidt (eds.),
Springer Proceedings in Physics {\bf 47} (1993), 538--630.

\bibitem[Bla]{Bl} Blackadar, B.: \emph{Operator algebras}, Springer,
2006. 

\bibitem[BG1]{BG1} Boutet de Monvel, A., Georgescu, V.: Graded
C*-algebras in the $N$-body problem, J.\ Math.\ Phys.\
\textbf{32}  (1991), 3101--3110.

\bibitem[BG2]{BG2} Boutet de Monvel, A., Georgescu, V.: Graded
C*-algebras associated to symplectic spaces and spectral analysis of
many channel Hamiltonians, in {\em Dynamics of complex and
irregular systems (Bielefeld encounters in Mathematics and Physics
VIII, 1991)}, Ph.\ Blanchard, L.\ Streit, M.\ Sirugue-Collin, D.\
Testard (eds.), World Scientific, 1993, 22--66.

\bibitem[DaG1]{DG1} Damak, M., Georgescu, V.: $C^*$-crossed
products and a generalized quantum mechanical $N$-body problem, 
99--481 at http://www.ma.utexas.edu/mp\_arc/. 

\bibitem[DaG2]{DG2} Damak, M., Georgescu, V.:
C*-algebras related to the N-body problem and the self-adjoint
operators affiliated to them, 99--482 at
http://www.ma.utexas.edu/mp\_arc/.

\bibitem[DaG3]{DG3} Damak, M., Georgescu, V.: Self-adjoint operators
  affiliated to $C^*$-algebras, Rev.\ Math.\ Phys.\ {\bf16} (2004),
  257--280.

\bibitem[DaG4]{DG4} Damak, M., Georgescu, V.: Hilbert C*-modules and
  spectral analysis of many-body systems, preprint 08-105 at
  http://www.ma.utexas.edu/mp\_arc/ or arXiv:0806.0827v1 at
  http://arXiv.org.

\bibitem[Der1]{De1} Derezinski, J.: The Mourre Estimate For
  Dispersive N-Body Schr\"odinger Operators,
  Trans.\ AMS \textbf{317} (1990), 773--798.

\bibitem[Der2]{De2} Derezinski, J.: Asymptotic completeness in
  quantum field theory. A class of Galilee-covariant models, Rev.\
  Math.\ Phys.\ {\bf 10} (1998), 191--233 (97-256 at
  http://www.ma.utexas.edu/mp\_arc).

\bibitem[DeG1]{DeG1} Derezinski, J., G\'erard, C.: {\it Scattering
  theory of classical and quantum $N$-particle scattering},
  Springer, 1997.

\bibitem[DeG2]{DeG2} Derezinski, J., G{\'e}rard, C.: Spectral and
 scattering theory of spatially cut-off $P(\phi)_{2}$
 Hamiltonians, Comm.\ Math.\ Phys.\ {\bf 213} (2000), 39--125.

\bibitem[DeIf]{DerI}
Dermenjian, Y.,  Iftimie, V.:  M\'ethodes \`a $N$ corps pour un
probl\`eme  de milieux pluristratifi\'es perturb\'es,
Publications of RIMS, {\bf 35} (1999), 679--709.

\bibitem[FeD]{FD}
Fell, J.M.G., Doran, R.S.:
{\it Representations of $*$-algebras, locally compact groups,
and Banach $*$-algebraic bundles;
volume 1, Basic representation theory of groups and algebras},
Academic Press, 1988.

\bibitem[Foll]{Fo} Folland, G.B.: {\it A course in abstract harmonic
analysis}, CRC Press, 1995.

\bibitem[Geo]{Geo} Georgescu, V.: On the spectral analysis of quantum
  field Hamiltonians, J.\ Funct.\ Analysis {\bf 245} (2007),
  89--143 (and 	arXiv:math-ph/0604072v1 at http://arXiv.org).

\bibitem[GI1]{GI1} Georgescu, V., Iftimovici,  A.: Crossed
products of $C^*$-algebras and spectral analysis of quantum
Hamiltonians, Comm.\ Math.\ Phys.\ {\bf 228} (2002), 519--560
(and preprint 00--521 at http://www.ma.utexas.edu/mp\_arc).

\bibitem[GI2]{GI2} Georgescu, V., Iftimovici, A.: $C^*$-algebras of
quantum Hamiltonians, in {\em Operator Algebras and Mathematical
Physics, Constan\c{t}a (Romania), July 2-7 2001}, Conference
Proceedings, J.M.~Combes, J.\ Cuntz, G.\ A.\ Elliot, G.\ Nenciu,
H.\ Siedentop, \c{S}.\ Str\u atil\u a (eds.), Theta, Bucharest 2003,
123--169 (or 02--410 at http://www.ma.utexas.edu/mp\_arc).
 

\bibitem[GI3]{GI4} Georgescu, V., Iftimovici,  A.: Localizations at
infinity and essential spectrum of quantum Hamiltonians: I. General
theory,  Rev.\ Math.\ Phys.\ {\bf 18} (2006), 417--483
(and arXiv:math-ph/0506051 at http://arxiv.org).

\bibitem[Ger1]{Ger1} G\'erard, C.: The Mourre Estimate For Regular
  Dispersive Systems, Ann. Inst. H. Poincare,
  Phys. Theor. \textbf{54}  (1991), 59-88. 

\bibitem[Ger2]{Ger2} G\'erard, C.:
Asymptotic completeness for the spin-boson model with a particle
number cutoff, Rev.\ Math.\ Phys.\ {\bf8} (1996), 549--589.

\bibitem[Gur]{Gu}
Gurarii, V.P.: 
{\em Group methods in commutative harmonic analysis \/},
in Commutative Harmonic Analysis II, eds. V. P. Havin, N. K. Nikolski,
Springer, Encyclopedia of Mathematical Sciences, {\bf 25},  1998.

\bibitem[HRe]{HR}
Hewitt, E., Ross, K.A.: \emph{Abstract harmonic analysis I}, second
edition, Springer, 1979.


\bibitem[Lac]{La} Lance, C.: \emph{Hilbert $C^*$-modules}, Cambridge
  University Press, 1995.

\bibitem[Lad]{Ld}
Landstad, M.B.: Duality theory for covariant systems,
Trans. Amer. Math. Soc. {\bf248} (1979), 223--269.

\bibitem[Ma1]{Ma} Mageira, A.: $C^*$-alg\`ebres gradu\'ees par un
  semi-treillis, thesis University of Paris 7, February 2007,
and preprint  arXiv:0705.1961v1 at http://arxiv.org.

\bibitem[Ma2]{Ma2} Mageira, A.: Graded $C^*$-algebras, 
J.\ Funct.\ Analysis {\bf 254} (2008), 1683--1701.

\bibitem[Ma3]{Ma3} Mageira, A.: Some examples of graded
  $C^*$-algebras,  Math.\ Phys.\ Anal.\ Geom.\  {\bf 11}  (2008),
  381--398.  

\bibitem[RW]{RW}
Raeburn, I., Williams, D.P.: 
{\em Morita equivalence and continuous-trace C*-algebras\/},
American Mathematical Society, 1998.

\bibitem[Rie]{Ri} Rieffel, A.M.: Induced representations of
  $C^*$-algebras, Adv. Math. {\bf 13} (1974), 176--257. 

\bibitem[SSZ]{SSZ} Sigal, I.M., Soffer, A., Zielinski, L.: On the
  spectral properties of Hamiltonians without conservation of the
  particle number, J.\ Math.\ Phys.\ {\bf 43} (2002), 1844--1855
  (and preprint 02-32 at http://www.ma.utexas.edu/mp\_arc).

\bibitem[Ska]{Sk} Skandalis, G.: private communication, June 2007.

\bibitem[Wil]{Wi} Williams, D.P.: \emph{Crossed products of
  $C^*$-algebras}, American Mathematical Society, 2007.


\end{thebibliography}
\end{document}